\documentclass[aps,prd,showpacs,preprintnumbers,notitlepage,nofootinbib,11pt]{revtex4-1}

\usepackage{graphicx}% Include figure files
\usepackage{bm}% bold math
\usepackage{enumitem}
\usepackage{epsf,epsfig}
\usepackage{amscd}
\usepackage{amsmath}
\usepackage{amssymb}
\usepackage{hyperref}
\hypersetup{colorlinks=true,linkcolor=blue,citecolor=red,urlcolor=blue}

\usepackage{feynmp}
\newcommand{\be}{\begin{equation}}
\newcommand{\ee}{\end{equation}}
\newcommand{\bea}{\begin{eqnarray}}
\newcommand{\eea}{\end{eqnarray}}
\newcommand{\hf}{\frac12}
\newcommand{\nn}{\nonumber\\}
\def\eq#1{(\ref{#1})}
\def\lal{\left\langle}
\def\rar{\right\rangle}
\def\la{\langle}
\def\ra{\rangle}
\def\Tr{{\mathrm{Tr}}}
\def\tr{{\mathrm{tr}}}
\def\ord#1{{\cal O}\left(#1\right)}
\def\mr#1{{\mathrm{#1}}}
\def\vec#1{{\bm{#1}}}
\def\ih{\frac{i}\hbar}
\def\fd#1#2{\frac{\delta#1}{\delta#2}}

\def\fdd#1#2#3{\frac{\delta^2#1}{\delta#2\delta#3}}
\def\br{\hskip-6pt/}
\def\bre{\hskip-4pt/}
\def\ha{{\hat a}}
\def\hj{{\hat j}}
\def\hphi{\hat\phi}
\def\hs{\hat\sigma}
\def\hsi{\hat\psi}
\def\hsib{\hat{\bar\psi}}
\def\hA{\hat A}
\def\hD{\hat D}
\def\hF{\hat F}
\def\hG{\hat G}
\def\hJ{\hat J}

\def\psib{\bar\psi}

\def\CL{\mathcal{L}}

\def\CO{\mathcal{O}}
\def\CT{\mathcal{T}}

\begin{document}
\title{Dynamics of the electric current in an ideal electron gas: a sound mode inside the quasi-particles}
\author{Sa\v{s}o Grozdanov}
\email{grozdanov@lorentz.leidenuniv.nl}
\affiliation{Rudolf Peierls Centre for Theoretical Physics, University of Oxford, \\1 Keble Road, Oxford OX1 3NP, United Kingdom    \\and\\Instituut-Lorentz for Theoretical Physics, Leiden University,\\Niels Bohrweg 2, Leiden 2333 CA, The Netherlands }
\author{Janos Polonyi}
\email{polonyi@iphc.cnrs.fr}
\affiliation{Strasbourg University, CNRS-IPHC, \\23 rue du Loess, BP28 67037 Strasbourg Cedex 2, France}

\preprint{OUTP-14-16P}

\begin{abstract}
We study the equation of motion for the Noether current in an electron gas within the framework of the Schwinger-Keldysh Closed-Time-Path formalism. The equation is shown to be highly non-linear and irreversible even for a non-interacting, ideal gas of electrons at non-zero density. We truncate the linearised equation of motion, written as the Laurent series in Fourier space, so that the resulting expressions are local in time, both at zero and at small finite temperatures. Furthermore, we show that the one-loop Coulomb interactions only alter the physical picture quantitatively, while preserving the characteristics of the dynamics that the electric current exhibits in the absence of interactions. As a result of the composite nature of the Noether current, {\it composite sound} waves are found to be the dominant IR collective excitations at length scales between the inverse Fermi momentum and the mean free path that would exist in an interacting electron gas. We also discuss the difference and the transition between the hydrodynamical regime of an ideal gas, defined in this work, and the hydrodynamical regime in phenomenological hydrodynamics, which is normally used for the description of interacting gases.
\end{abstract}
\maketitle

\begingroup
\hypersetup{linkcolor=black}
\tableofcontents
\endgroup

\newpage
\section{Introduction and motivation}
Hydrodynamical description of many-body systems with collective excitations is a well-established and widely applied method, which successfully combines microscopic and phenomenological information about the physical system. Despite the numerous successes of this mixed scheme, the systematic, microscopic derivation of the hydrodynamical equations is not fully understood. Rather than deriving them directly, one usually calculates transport coefficients from the microscopic description and uses their values in the equations of phenomenological hydrodynamics, i.e. the gradient expanded constitutive relations \cite{landau2013fluid,forster,Kovtun:2012rj}. 

Much recent research has been focused on understanding hydrodynamics in terms of an effective theory of Goldstone modes \cite{Dubovsky:2005xd,Dubovsky:2011sj,Bhattacharya:2012zx,Haehl:2013hoa,Haehl:2013kra}, including works that proposed ways of incorporating dissipation into the effective description \cite{Grozdanov:2013dba,Endlich:2012vt,Kovtun:2014hpa,Galley:2014wla}. However, none of these approaches provides us with a detailed microscopic view of the origin of the effective field theory of hydrodynamics. This present work grew from a desire to bridge this gap and to gain new insight into the {\it ab initio} elements of a hydrodynamic system, derived directly from a microscopic quantum field theory, including dissipation. In essence, this paper is one of the simplest examples of a microscopically derived effective Schwinger-Keldysh field theory with the structure of the classical theory explored in \cite{Grozdanov:2013dba}. 

An important simplifying feature of this work is that our present goal is to understand the role of many-body correlations that characterise the hydrodynamic flow, independently of the interactions among the participating particles. The suspicion that highly non-trivial correlations might be present even in the {\it absence of interactions} comes from the simple fact that gauge-invariant (i.e. charge conserving) observables are composite operators, and hence the structure of their connected Green's functions is richer than for those of elementary operators. We will argue that similar types of phenomena are not restricted only to quantum systems but that they are present also in classical mechanics.

Our main goal in this paper is to find the equation of motion for the expectation value of the current operator, $\left\langle j^\mu \right\rangle$, in an ideal gas of electrons at non-zero density, both at vanishing and low temperatures. In the absence of any phenomenological input, the relevant effective theory must be such that $\left\langle j^\mu \right\rangle$ satisfies its variational equation. It is the calculation of expectation values that requires us to go beyond the conventional formalism of quantum field theory and perform the calculation in the Schwinger-Keldysh Closed-Time-Path (CTP) formalism \cite{schw,schwbooks,Bakshi:1962dv,Mahanthappa:1962ex,keldysh}. 

The CTP effective action will be calculated within the spirit of the Landau-Ginzburg double expansion, by organising each order of the expansion in powers of the amplitude. This expansion gives rise to a Laurent series and orders the terms in the equation of motion in increasing powers of the wave vector. This should be compared with the phenomenological approach to classical hydrodynamics where the equations, expressing the conservation of the energy-momentum and charges, are also constructed by using the traditional double expansion \cite{landau2013fluid}. These two approaches are fully analogous with the exception of three important differences. Firstly, the equations derived from microscopic physics are closed without the requirement of additional thermodynamical considerations. Local equilibrium in an infinite, homogeneous system is automatically ensured so long as the perturbations are weak and slowly varying. As a result, the usual difficulties of the phenomenological approach to zero temperature hydrodynamics appear as a problem in performing the Legendre transform in deriving the effective action. Secondly, a new element of this approach is the fact that the derivation reveals a spatial non-locality induced by the Fermi surface, which is reflected by the appearance of the negative powers of the  wave vector, $1/|\vec{q}|$, in the Laurent series. Such terms, generated by loop-integrals, are absent in the usual phenomenological approaches, which are based on the na\"{i}ve Taylor series expansion of classical physics, i.e. in $\omega$ and $\vec q$.\footnote{An example of the complexity of hydrodynamics beyond an analytic Taylor expansion in $\omega$ and $\vec q$, which has been well known and also arises in the computation of current-current correlation functions (away from large-$N$) are the so-called ``long-time tails'', which exhibit non-analytic behaviour in $\omega$ \cite{Pomeau:1974hg,Kovtun:2012rj,Kovtun:2003vj,CaronHuot:2009iq}.} The third difference concerns the definition of the hydrodynamical regime. In the phenomenological approach, the relevant regime is defined at length scales where the local density, current and thermodynamical potentials can be defined. In our approach, based solely on local expectation values, there is a well defined effective action, valid at any length scale, well beyond the minimal length, set by the UV cutoff of the underlying quantum field theory model. We define the hydrodynamical limit by the requirement that the equations of motion be local in time and display integro-differential structure in space.

The transport coefficients, entering into the phenomenological hydrodynamical equations, can be derived from quantum field theory by using Kubo's formulae \cite{kubo,Kovtun:2012rj,Moore:2010bu}. Our generating functional for the connected CTP Green's functions of the current, calculated at the quadratic order, reproduces the usual Kubo formulae. It is then the next step in our work that goes beyond the traditional treatment of the electron gas. We perform the functional Legendre transform and calculate the effective action for the current, which is what produces closed equations of motion without the need for any thermodynamical input. Even though the derivation of the standard Kubo formula for electrical resistivity \cite{kubo} is the first step in obtaining the dynamical equation for $\la j^\mu\ra$, this line of research has not been pursued in a systematic manner to our knowledge. 

The behaviour of a Fermi liquid is normally formulated by Landau's phenomenological theory \cite{landau1980statistical,baym2008landau,Polchinski:1992ed}, which is based on the idea of quasi-particles. However, the charged, local fields that represent the quasi-particle excitations around the Fermi surface are not Hermitian physical observables. This problem is circumvented in Landau's argument by carefully constructing the effective dynamics near local equilibrium in terms of the scattering processes of the quasi-particles. The complication in deriving such a theory from microscopic physics in a systematic manner is the necessity for using non-local transition amplitudes, which are given by the gauge invariant residues of the amputated, connected Green's functions. Instead, we wish to formulate an effective theory in terms of local observables to understand how phenomenological equations of hydrodynamics arise. We must therefore rely on neutral, composite operators. The obvious choice, the current $j^\mu(x)=\psib(x)\gamma^\mu\psi(x)$, makes our effective theory formally different from Landau's Fermi liquid theory.

The main observation we wish to present in this work is that a {\it non-interacting, ideal gas} of electrons displays complicated correlations and dynamics when one analyses the behaviour of the electric current. It exhibits features that are usually characteristic of interacting systems, such as the presence of a {\it sound mode}. To better understand the role of interactions, imagine a calculation of the AC electrical conductivity in an electron gas. This can be done by introducing an external plane-wave electric field with the amplitude $\vec{E} = (E,0,0)$ and then calculating the response of the system, namely the amplitude of the induced current $\vec{j}=(j,0,0)$. Perturbatively, the result has the form
\begin{align}
j =e\sum_{m=1}^\infty \sum_{n=0}^\infty c_{mn}\,\left(\frac{eE}{\epsilon_F}\right)^m\left(\frac{e^2}{\hbar c}\right)^n,
\end{align} 
expanded in terms of two small parameters, the QED coupling strength, $e$, and the ratio of the external perturbation, $eE$, to the Fermi energy, $\epsilon_F$. The double expansion reflects the double role the electromagnetic interactions play in this problem; on the one hand, $e$ is the {\it fundamental} interaction among the charges in the gas and on the other hand, it is the interaction used to diagnose the gas. The key point is that the dynamics of an out-of-equilibrium gas behaves differently depending on the order the two independent $e\to0$ and $eE/\epsilon_F\to0$ limits are taken. 

To characterise the gas itself, without the large interference of the external probe, one goes into the limit of the {\it linear-response regime}. Namely, one carries out the limit $e E /\epsilon_F \to 0$ first in such a way that the transport coefficient, i.e. the conductivity, becomes well defined and finite,
\begin{align}
\sigma\equiv \frac{j}{E} =  \frac{e^2}{\epsilon_F} \sum_{n=0}^\infty c_{1n}\, \left(\frac{e^2}{\hbar c}\right)^n + \ord{\frac{eE}{\epsilon_F}}.
\end{align} 
In this limit, the interactions between the gas and the observational probe are severely simplified. However, the interactions among the charges still play an important and complicated role in determining the coefficients $c_{1n}$. To simplify the contribution from the fundamental interactions, one usually considers the un-physical limit in which the QED interaction becomes extremely weak. The leading-order result, $\sigma/e^2\sim c_{10}/\epsilon_F$, is then sufficient for describing the dynamics of a (nearly) ideal gas in the presence of a weak external perturbation $e E$. 

Alternatively, one may be interested in the way the elementary interactions of QED dress the non-interacting, ideal gas in the presence of external perturbations. This scenario corresponds to the other order of limits where $e\to0$ is performed for finite $eE$. In this approach, the external perturbation has to be restricted to an AC form in order to avoid the instabilities of non-interacting particles subject to homogeneous external forces. This is the strategy followed in this paper in which we derive the linearised equation of motion for the electric current in the presence of an external perturbation, within the one-loop approximation in QED. The external source, $a_\mu(x)$, is fictitious in our scheme. It is used to drive the electron gas from its equilibrium state at $t_i=-\infty$ to a desired, non-trivial state at $t=0$ when the source is switched off, i.e. $a_\mu(x)=0$ for $t>0$. The equation of motion, satisfied in the absence of the source, is therefore well defined so long as it is local in time. Since it is only possible to find a local equation of motion for sufficiently slow motion of the gas, we will restrict our attention to external sources with slow time dependence.

The non-trivial correlations in a non-interacting, ideal gas appear due to the redistribution of the energy-momentum into its non-interacting normal modes, when the dynamics is diagnosed by composite operators. The initial energy-momentum is injected into the gas from the external source. Such a redistribution generates new collective modes with non-trivial dispersion relations, leading to new {\it composite sound} waves. Dissipation manifests itself in the spread of the flow pattern in space, as a function of time. As a result, this type of a sound wave is different from both the zero and first sound modes of the usual Fermi liquid. 

There are three major parts of this paper: the discussion of inertial forces that arise from non-linear coordinate transformations, the argument that effective theories have to be derived within the CTP formalism and the presentation of the dynamics of the electric current. 

In Section \ref{tpyms}, we point out that inertial forces show a surprising similarity with genuine interactions. The equations of motion become highly involved and lead to non-trivial dispersion relations for the collective excitations. For us, the non-linear transformation of interest in the electron gas is the field transformation from that of the electron, $\psi(x)$ and $\psib(x)$, to the Noether current, $j^\mu(x)=\psib(x)\gamma^\mu\psi(x)$. 

In Section \ref{ctps}, we proceed with an introduction of the CTP formalism. Special emphasis is placed on the structure of the CTP propagators. Physical phenomena that arise in the CTP formalism from the coupling between the two time axes, namely decoherence and irreversibility, are discussed in Section \ref{ctaxs}. We also discus why decoherence and irreversibility can be present in harmonic systems. 

The CTP scheme is then applied to the electron gas in Section \ref{cdegs} by working out the effective action for the current at the quadratic level and by using the one-loop approximation. The corresponding equations of motion are presented for longitudinal and transverse components of the current. Section \ref{hydrlims} is devoted to the discussion of the infrared, hydrodynamical limit of the solutions. We first consider the linearised equations of motion and decoherence in a dense ideal gas with vanishing temperature. We then couple it to a heat bath and finally turn on the Coulomb interactions. 

Finally, we present our conclusions in Section \ref{concls}. Three Appendices include a brief discussion of the linear response formalism, the structure of the free Schwinger-Keldysh propagators and the summary of the one-loop current-current Green's function calculation.

\section{Inertial forces and interactions}\label{tpyms}
We begin this paper by analysing the dynamics of physical systems in which the quantities of interest are non-linear combinations of the fundamental degrees of freedom.

\subsection{Particle dynamics}
To examine non-linear field transformations, consider first a simple harmonic oscillator,
\be
L=\frac{1}{2} m \dot x^2-\frac{1}{2} m\omega^2 x^2.
\ee
We can introduce a non-linear coordinate $y=x^2/2$, under which the Lagrangian becomes
\be\label{SHO2}
L=\frac{1}{2} \left(\frac{m}{2y}\right) \dot y^2- \frac{1}{2} \left(\frac{m}{2y}\right) \omega_y^2 y^2,
\ee
where $\omega_y = 2 \omega$. The system in Eq. \eqref{SHO2}, expressed in terms of $y$, appears to have a position-dependent effective mass, $m(y)=m/2y$, a periodically diverging speed and oscillations, which are unable to pass through the point $y=0$. Despite the fact that both systems describe the same physics, the inertial force exerted by the coordinate-dependent mass of the oscillator $y(t)$ hides the simplicity of the motion expressed in terms of $x(t)$. 

The lesson to be learned at this point is that inertial forces, arising from a non-linear change of the coordinates, may disguise the normal modes of a harmonic system beyond recognition. Let us suppose that our system, described by the coordinate $x$, obeys dynamics described by a quadratic action, $S[x]$, and that we are interested in the effective theory of some non-linear combination of the original coordinate, $y=y[x]$. The action, $S[x[y]]$, is then non-linear and contains terms that make the physical system appear to behave as an interacting system \cite{idgas}.

\subsection{Classical fields}
Let us increase the complexity of the model and consider a complex scalar field, $\phi(x)$, governed by a translationally invariant quadratic action,
\be\label{exfth}
S=\phi^*\CL \, \phi+j^*\phi+\phi^*j+k\Phi.
\ee
The products of the type $fg$ denote space-time integrations,
\be\label{scprod}
fg \equiv \frac1c\int d^4xf(x)g(x)=c\int\frac{d^4q}{(2\pi)^4}f(-q)g(q),
\ee
with the Fourier transform defined as\footnote{Throughout this work, we will be using the $\eta_{\mu\nu} = \text{diag}\{+1,-1,-1,-1\}$ sign convention for the Minkowski metric tensor.}
\be
f(q)=\frac1c\int d^4xe^{iq\cdot x}f(x).
\ee
The source $k(x)$ is coupled to the composite field $\Phi(x)=\phi^*(x)\phi(x)$ and the kernel ${\cal L}(i\partial_\mu)$ can be a polynomial in space-time derivatives. The normal modes of the model \eqref{exfth} with $j=k=0$ are plane waves with a wave-vector $\bar q^\mu=\left(\omega_0 (\vec{q}),\vec{q}\right)$, where ${\cal L}(\bar q_\mu)=0$. If we turn on a source $j$ with four-momentum $\bar q^\mu$, then all of the energy-momentum gets absorbed exclusively by the normal mode $\bar q^\mu$. The system's response to an external source $j$ thus reflects the simplicity of a harmonic system in which each normal mode remains independently excited. On the other hand, the energy-momentum injected through the source $k$, which is coupled to the composite field $\Phi$, gets transmitted through the system in a completely different way. It spreads over all normal modes with the energy-momentum conservation being the only restriction. The dispersion relation of the response is modified in a highly non-trivial manner compared to the previously excited normal modes with $\omega_0(\vec{q})$.

The above statements can also be understood in the following way: after turning on the source $k$, we not only lose the original normal modes but the response of the system begins displaying correlations among those modes. It is important to stress that correlations are usually thought of as the hallmark of genuine interactions. In this language, we can understand the non-interacting nature of the model diagnosed by $j$ as the absence of correlations among the normal modes. In fact, the external source $j(q)\sim\delta(q-p)$, which is local in Fourier space, induces a local response, $\phi(q)\sim\delta(q-p)$. This is no longer the case when the source $k$ is used to diagnose the dynamics since $k(q)\sim\delta(q-p)$ induces a response in the whole Fourier space because the momentum $p^\mu$, injected by the source $k$, is spread over infinitely many plane wave normal modes of the $k\ne0$ model.

It is also instructive to compare the time evolution of $\Phi(x)$ induced by a wave packet in $k(x)$ with the spread of the wave packet in quantum mechanics. In standard quantum mechanics, the spatial Fourier transform of the probability density can be written as
\be\label{sprwpqm}
\rho(t,\vec{p})=\int\frac{d^3q}{(2\pi)^3}e^{\ih[E(\vec{q})-E(\vec{p}+\vec{q})]t}\psi^*(\vec{q})\psi(\vec{p}+\vec{q}) , 
\ee
where $\psi(\vec{q})$ denotes the Fourier transform of the wave function at $t=0$, assumed to be regular and localised in Fourier space. The probability density is also localised at $t=0$ and its $\vec{p}$-dependence displays a convoluted peak. As time passes, the oscillating phase factor makes this peak more and more localised in Fourier space, leading to the spread of the wave packet in position space. The time reversal, being an anti-unitary transformation, leaves the Schr\"odinger equation invariant and the spread of the wave-packet results from the choice of the initial condition rather than the breakdown of time reversal invariance. It is possible to find initial conditions for the wave function with a singular phase in Fourier space so that the wave packet becomes narrower in time.

The field $\Phi(x)$, induced in the field theory \eq{exfth} with $j\ne0$ and $k=0$ is given by
\bea
\Phi(t,\vec{p})&=&-\sum_{ab}\int\frac{d^3q}{(2\pi)^3}e^{i \omega^{(a)}_0(\vec{q}) t - i \omega^{(b)}_0(\vec{p}+\vec{q}) t}   \left[ \mr{Res}\,{\cal L}\left(\omega^{(a)}_0(\vec{q}),\vec{q}\right) \right]^*     \left[ \mr{Res}\,{\cal L}\left(\omega^{(b)}_0(\vec{p}+\vec{q}),\vec{p}+\vec{q}\right)  \right] \nn
&&\times \, j\left(\omega^{(a)}_0(\vec{q}),\vec{q}\right)  j\left(\omega^{(b)}_0(\vec{p}+\vec{q}),\vec{p}+\vec{q}\right),
\eea
where $\mr{Res}\,G^r(\omega^{(a)}_0(\vec{q}),\vec{q})$ is the residue of the retarded Green's function and the summation extends over the poles. The comparison with Eq. \eq{sprwpqm} indicates that if the source $j$ has a regular and localised Fourier transform, then the peak in $\Phi(t,\vec{x})$ spreads as the time passes. The interference among the independent normal modes leads to diffusion-like processes as in the case of Landau damping \cite{clemmow}. As a result, the mixing of infinitely many normal modes may make the effective dynamics of the composite field, $\Phi(x)$, appear to be diffusive.

\subsection{Quantum fields}
To continue with our chain of increasingly complex models, consider a charged scalar quantum field, 
\be
\phi(x)=\sum_c\int\frac{d^3q}{(2\pi)^3}    \left[  \Theta (\omega_0^{(c)} ) \, a_c(\vec{q}) \, e^{-iqx}+\Theta (-\omega_0^{(c)}) \, b^\dagger_c(-\vec{q}) \, e^{-iqx}  \right]_{q^0=\omega_0^{(c)}(\vec{q}) / c},
\ee
where the operators $a_c(\vec{q})$ and $b_c(\vec{q})$ annihilate a particle and an anti-particle, respectively, and the sum is over the different particle modes of the Lagrangian \eq{exfth}. The non-interacting nature of an ideal gas is clearly apparent on the level of the one-particle Green's functions. The factorisation of the higher-order Green's functions of the elementary field, 
\begin{align}
G_{2n}=\left\langle 0 \left|  \CT \left\{ \phi(x_1)\cdots\phi(x_n)\phi^\dagger(y_1)\cdots\phi^\dagger(y_n) \right\}  \right|0 \right\rangle, 
\end{align}
according to Wick's theorem, implies that all connected Green's functions with $n>1$ vanish. In fact, there are $n$ elementary excitations contributing to $G_{2n}$, each of them controlled by a pair of particle or anti-particle operators, $a_c(\vec{q})a^\dagger_c(\vec{q}')$ or $b_c(\vec{q})b^\dagger_c(\vec{q}')$. They can be represented by propagator lines in the Feynman diagrams and $G_{2n}$ is the sum over all possible different combinations of the pairings. Since the elementary excitations are non-interacting, their amplitudes factorise for all the combinations. 

\begin{figure}[h]
\includegraphics[scale=0.8]{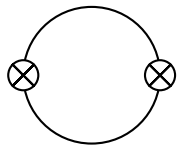}\caption{An example of a scalar two-point Green's function ring diagram for a composite operator.}\label{fig:Ring1}
\end{figure}

\begin{figure}[h]
\includegraphics[scale=0.8]{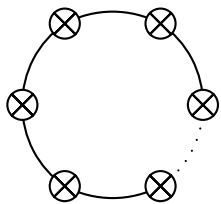}\caption{An example of a scalar $n$-point Green's function ring diagram, which contains $n$ current insertions of $\otimes$ along the ring with propagators connecting the neighbouring pairs of the operator insertions.}\label{fig:Ring4}
\end{figure}

The Green's functions of the composite operator $\Phi(x)$, 
\begin{align}
H_{2n}= \left\langle 0 \left| \CT \left\{ \Phi(x_1)\cdots\Phi(x_n) \right\} \right| 0\right\rangle, 
\end{align}
have a much more involved structure. In fact, each composite operator is made of two elementary excitations and therefore $\Phi(x)$ can be paired with two elementary field lines in a Feynman diagram, as for example in Figures \ref{fig:Ring1} and \ref{fig:Ring4}. In other words, the measurement of $\Phi(x_i)$ generates two elementary excitations, which have to be removed by other operators appearing in $H_{2n}$. The lines drawn between the composite operator insertions can therefore give rise to connected diagrams at arbitrary orders. As a result, the simplicity of the non-interacting, ideal gas becomes completely disguised by the composite operator analysis.

\begin{figure}[h]
\includegraphics[trim=0cm 0.5cm 0cm 0.5cm, scale=0.8]{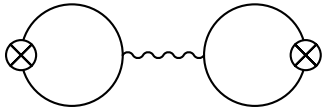}\caption{A simple two-point current-current Green's function diagram in the presence of interactions.}\label{fig:Ring2}
\end{figure}

Even though both the non-interacting and the interacting system share the property that they support non-trivial connected composite operator Green's functions of arbitrarily high orders, there is a marked difference between these Green's functions. For an ideal gas, they are given by a single Feynman loop-integral. In the presence of interactions, on the other hand, there exists an infinite series of distinct Feynman diagrams. For an example of a diagram with added interaction vertices, see Figure \ref{fig:Ring2}. Let us consider a composite operator that is an $N$-th order homogeneous multinomial of an elementary free field. It is easy to see that in the non-interacting case its Green's function with $n$ legs, corresponding to an $n$-point function, is given by a single Feynman diagram with $n(N-2)/2 + 1$ loops. For example, all $n$-point functions for a bilinear operator with $N=2$, as depicted in Figure \ref{fig:Ring4}, are given by one-loop diagrams. The numerical values of the many-body correlations can therefore be as involved in a non-interacting gas as in an interacting system. However, the elementary processes that build up these correlations are far more complicated in the interacting than in the free case. This is consistent with the findings of Section \ref{ints}, where we show that the Coulomb interaction re-summed propagator only modifies the numerical values of the coefficients in the linearised equation of motion for the current, compared to those derived for an ideal gas. We note that our analysis will neglect the vertex corrections as well as the electron self-energy, which would qualitatively change the behaviour of the gas. We will discuss these issues in detail below.

To conclude this section, let us consider a particle described by a complex field $\phi(x)$, which is invariant under the phase transformations $\phi(x)\to e^{i\theta}\phi(x)$ and $\phi^\dagger(x)\to e^{-i\theta}\phi^\dagger(x)$. Its interactions can be described by vertices constructed out of the bi-local field $\Phi(x,y)=\phi(x)\phi^\dagger(y)$. Furthermore, $\Phi(x,y)$ also generates the field's excitations. In a sense, this is a generalisation of the ideas on bosonisation because the excitations are controlled by a bosonic bi-local field, both for bosons and fermions. The dynamics of such excitations is non-local. However, by adopting the idea of the operator product expansion, it can be characterised by a set of infinitely many local fields, 
\be
\Phi^{\mu_1,\ldots,\mu_n}_n(x)=\frac{\partial^n}{\partial_{z^{\mu_1}}\cdots\partial_{z^{\mu_n}}}\phi(x+z)\phi^\dagger(x-z)_{|z=0} .
\ee
One expects that the long wavelength excitations of the system are generated by the $\Phi_n$ fields with small $n$. It is therefore reasonable to look for a simple effective infrared theory in terms of $\Phi_0$. Such effective dynamics may contain dissipative forces because the excitations that are controlled by the fields $\Phi_n$, with higher $n$, represent an environment for the long-distance modes. Note that this environment is well defined even though it has no corresponding direct product structure in the Fock space.

\section{CTP formalism}\label{ctps}

The standard single-time axis formalism of quantum field theory, developed for the calculation of transition amplitudes between pure states, is not sufficient to describe the time evolution of expectation values. To overcome this problem, we will use the Schwinger-Keldysh, Closed-Time-Path (CTP) formalism \cite{schw} in order to derive the equation of motion for the expectation value of the Noether current $j^\mu$. This powerful formalism has been applied to numerous calculations in condensed matter physics as well as in high energy physics, see e.g. \cite{kamenev,calzetta}. To make the paper self-contained, we will devote this section and Section \ref{ctaxs} to the introduction and summary of the relevant CTP techniques employed in this work.

\subsection{Generating functional}

To make our presentation of the CTP techniques as simple as possible, we will work with a specific example. We will consider the dynamics of a neutral scalar field $\phi(x)$ during the time interval $t_i<t<t_f$, defined by the Hermitian Hamiltonian density $H(x)$. In standard QFT, transition amplitudes between {\it pure} initial and final states are calculated by using the generating functional
\be\label{stagfh}
e^{\ih W[j]}=\la\Psi_f|\CT\left\{e^{-\ih\int d^4 x[H(x)-j(x)\phi(x)]}\right\}|\Psi_i\ra.
\ee
The source $j(x)$ is introduced so that we can generate $n$-point Green's functions and facilitate the perturbation expansion. In order to compute {\it expectation values} and allow for an evolution to (and from) {\it mixed} states, one generalises the expression \eqref{stagfh} to the trace of the final density matrix $\rho_f \equiv \rho(t_f)$, written in the Heisenberg representation as
\be\label{ctpgfh}
e^{\ih W[j^+,j^-]}=\Tr\left[ \CT \left\{ e^{-\ih\int d^4 x[H(x)-j^+(x)\phi(x)]} \right\} \rho_i \, \CT^*\left\{e^{\ih\int d^4 x[H(x)+j^-(x)\phi(x)]} \right\} \right],
\ee
where $\rho_i \equiv \rho(t_i) $ stands for the density matrix of the initial state and $\CT^*$ denotes the anti-time ordering. The sources $j^\pm(x)$ generate observables through functional differentiation, $\delta/\delta j^\pm(x)$, and are set to the physical value, $j^+(x)=-j^-(x)=j(x)$, at the end of the calculation. One can distinguish between the two time axes by introducing two time variables $t^+$ and $t^-$ for the time-ordered and the anti-time-ordered products, respectively. Furthermore, it is convenient to define the extended time ordering, $\bar \CT$, in such a manner that it acts as $\CT$ on $t^+$ and $\CT^*$ on $t^-$, while placing each $t^-$ after $t^+$. The result is the condensed expression,
\be
e^{\ih W[j^+,j^-]}=\Tr \left[\bar \CT \left\{e^{\ih\int d^4 x [H(x^-)-H(x^+)+j^+(x^+)\phi(x^+)+j^-(x^-)\phi(x^-)]} \right\} \rho_i\right],
\ee
where $(t,\vec{x})$ goes from $(t^+,\vec{x})$ to $(t^-,\vec{x})$. Moreover, Wick's theorem also becomes explicitly available and can be used in calculations. 

The path integral representation of \eq{ctpgfh} is
\be\label{ctpwelem}
e^{\ih W[j^+,j^-]}= \!\!\!\!\!\!\!\!\!\!\!\!   \int \limits_{\phi^+(t_f,\vec{x})=\phi^-(t_f,\vec{x})} \!\!\!\!\!\!\!\!\!\!\!\!  D[\phi^+]D[\phi^-]\rho_i[\phi^+(t_i,\vec{x}),\phi^-(t_i,\vec{x})]e^{\ih(S[\phi^+]+j^+\phi^+-S^*[\phi^-]+j^-\phi^-)},
\ee
where $S[\phi]$ is the action, which includes Feynman's $i\epsilon$ prescription. In this paper, we will make use of the notation
\be\label{ctpgfpi}
e^{\ih W[\hj]}=\int D[\hphi] \, e^{\ih S[\hat\phi]+\ih \hj\hat\phi},
\ee
where the CTP doublets are $\hat\phi=(\phi^+,\phi^-)$ and $\hj=(j^+,j^-)$, and the action is
\be\label{ctpeact}
S[\hphi]=S[\phi^+]-S^*[\phi^-].
\ee
$S^*$ here denotes the complex conjugate of $S$. The CTP symmetry,
\be\label{ctpasym}
S[\phi^-,\phi^+]=-S^*[\phi^+,\phi^-],
\ee
plays an important role in restricting the structure of the Green's functions and effective actions. 

The unitarity of the time evolution, on the level of the whole system, is expressed by the preservation of the total probability. This is reflected in the fact that the trace of the density matrix, \eq{ctpgfh}, calculated for a physical source, $j^+=-j^-=j$, equals to $1$, or equivalently,
\be\label{unit}
W[j,-j]=0.
\ee

Is the CTP scheme simply a trivial generalisation of the usual, single time axis formalism? The generating functional \eq{ctpgfh} factorises into the product of the conventional transition amplitude \eq{stagfh} and its complex conjugate, with $W[j,-j]=W[j]-W^*[j]$, if the initial state, $|\Psi_i\rangle$, evolves under the influence of the given external source into $|\Psi_f\rangle$ at the final time. Such a trivial relation between the single and the double time axes cases is no longer valid when either the initial state is mixed or we use ``non-physical" sources, $j^+\ne-j^-$. In such cases, the CTP scheme covers new, non-trivial physical phenomena, which could not be described by \eqref{stagfh}. The essential point is that in calculating the expectation values of operators, we always encounter non-physical sources, such as
\be
\Tr[\phi(x)\rho_i]=\fd{W[\hj]}{j^+(x)}_{| j^+=j^- \to 0}=\fd{W[\hj]}{j^-(x)}_{|j^+=j^-\to 0},
\ee
which explains the need for the CTP formalism, as for instance in the theory of linear response. We will make use of this feature in the construction of the effective dynamics for the expectation value of the electric current, described in Section \ref{cdegs}. As mentioned above, another circumstance that makes the CTP formalism a necessity is the presence of a mixed initial state, which happens when the system under consideration is open due to its coupling to an environment. In Section \ref{ctaxs}, we will explain that this extension is in fact essential for effective theories.

\subsection{Propagator}

The CTP formalism with its doubling of the degrees of freedom allows us to set up perturbation expansion for retarded Green's functions. They are completely encoded by the CTP propagator,
\be\label{ctpprop}
iD^{\sigma\sigma'}(x,y)=\Tr \left[\bar \CT \left\{  \phi^\sigma(x)\phi^{\sigma'}(y) \right\}\rho_i\right],
\ee
defined by the generating functional \eq{ctpgfpi}. We will use $\sigma$ to denote the CTP indices $+$ and $-$. The generalised time ordering, $\bar \CT$, agrees with the usual one, $\CT$, on the positive time axis. Therefore, $D^{++}$ is the Feynman propagator,
\be
\Tr \left[\CT\left\{\phi^+(x)\phi^+(y)\right\}|0\ra\la0|\right]
=\la0|\CT\{\phi(x)\phi(y)\}|0\ra.
\ee
The action of $\bar \CT$ is trivial if the two operators belong to different time axes, giving us
\bea\label{wfcctpp}
\Tr \left[\bar \CT \left\{\phi^-(x)\phi^+(y)\right\}|0\ra\la0|\right]
&=&\Tr \left[\phi(x)\phi(y)|0\ra\la0| \right] = \la0|\phi(x)\phi(y)|0\ra.
\eea
Hence, the off-diagonal components of the propagator are the Wightman functions without time ordering. The remaining components of $D^{\sigma\sigma'}(x,y)$ can be found by complex conjugation, leading to the block matrix form
\be\label{bosctppr}
i\hD(x,y)=\begin{pmatrix}\left\langle \CT\left[\phi(x)\phi(y)\right]\right\rangle &\left\langle \phi(y)\phi(x)\right\rangle \cr\left\langle\phi(x)\phi(y)\right\rangle&\left\langle \CT \left[\phi(y)\phi(x)\right]\right\rangle^*\end{pmatrix}.
\ee
The CTP propagators for free bosons and fermions are given in Appendix \ref{propctap}. The CTP identity,
\be
\CT \left\{ A(t_A)B(t_B) \right\} +\CT^*\left\{A(t_A)B(t_B)\right\}=A(t_A)B(t_B)+B(t_B)A(t_A),
\ee
valid for bosonic operators, restricts the propagator to the standard CTP form,
\be\label{stctpform}
\hD=\begin{pmatrix}D^n+iD^i&-D^f+iD^i\cr D^f+iD^i&-D^n+iD^i\end{pmatrix},
\ee
where the functions $D^n$, $D^f$ and $D^i$ appearing in the matrix elements are real. In the bosonic case, the exchange symmetry $(\sigma,x)\leftrightarrow(\sigma',y)$ imposes $D^n(x,y)=D^n(y,x)$, $D^f(x,y)=-D^f(y,x)$, and $D^i(x,y)=D^i(y,x)$. 

The Fourier transform of the Wightman function,
\be\label{WightAsSpect}
iD^{-+}(p)=\Theta(p^0)S(p),
\ee
is the spectral function of excitations, which are generated by $\phi(p)$ in a system with translational invariance and $S(p)\ge0$. The relation \eqref{WightAsSpect} allows us to express both $D^f$ and $D^i$ in terms of the spectral function, which leads to the spectral condition 
\be\label{spectrc}
D^f(p)=\mr{sgn}(p^0)iD^i(p),
\ee
and the CTP propagator can thus be specified by only two real functions,
\be\label{propnf}
\hD(p)=\begin{pmatrix}D^n(p)+\mr{sgn}(p^0)D^f(p)&-2\Theta(-p^0)D^f(p)\cr2\Theta(p^0)D^f(p)&-D^n(p)+\mr{sgn}(p^0)D^f(p)\end{pmatrix},
\ee
where the positive definiteness of the norm imposes the bound
\be\label{posdefn}
i\Theta(p^0)D^f(p)>0.
\ee

The inverse of the propagator \eq{stctpform} is given by
\be\label{istctpform}
\hD^{-1}=\hs\begin{pmatrix}\Delta^n+i\Delta^i&-\Delta^f+i\Delta^i\cr\Delta^f+i\Delta^i&-\Delta^n+i\Delta^i\end{pmatrix}\hs,
\ee
where $\hs$ is a diagonal ``metric tensor" of the form $\hs = \text{diag}\left(1,-1\right)$. Furthermore, 
\begin{align}
&\Delta^{r,a}=1 / D^{r,a}  ,  \label{retadvinv} \\
&\Delta^i=-\Delta^rD^i\Delta^a, \label{didi}
\end{align}
where $\Delta^n(x,y)=\Delta^n(y,x)$, $\Delta^f(x,y)=-\Delta^f(y,x)$, $\Delta^i(x,y)=\Delta^i(y,x)$, $\Delta^r=\Delta^n + \Delta^f$ and $\Delta^a=\Delta^n - \Delta^f$. We also note that the spectral condition, i.e. Eq. \eq{spectrc}, yields
\be\label{imagipropf}
\Delta^f(p)=\mr{sgn}(p^0)i\Delta^i(p).
\ee

Even though the preceding remarks apply to interacting fields, it is instructive to consider free fields in a harmonic model given by the action
\be\label{harmact}
S_{eff}[\hphi]=\hf\int d^4 x \, (\phi^+,\phi^-)\begin{pmatrix}\Delta^n+i\Delta^i&\Delta^f-i\Delta^i\cr-\Delta^f-i\Delta^i&-\Delta^n+i\Delta^i\end{pmatrix}\begin{pmatrix}\phi^+\cr\phi^-\end{pmatrix}.
\ee
The external source, introduced in the generating functional \eq{ctpgfh} generates a non-trivial expectation value, which can be obtained from either one of the two time axes,
\be\label{harmresp}
\left\langle \phi(x) \right\rangle \begin{pmatrix}1\cr1\end{pmatrix}=-\int d^4 y \,\hD(x,y)\begin{pmatrix}j(y)\cr-j(y)\end{pmatrix},
\ee
showing that $D^r=D^n+D^f$ and $D^a=D^n-D^f$ can be identified as the retarded and the advanced Green's function, respectively. Since these Green's functions are real in position space, $D^n(p)=\Re D^r(p)$ and $D^f(p)=i\Im D^r(p)$.

\section{Coupling to the environment}\label{ctaxs}

In the above section, we mentioned that the CTP scheme is important for the analysis of open systems. In this section, we will address the question of how precisely this formalism is able to describe the coupling between a system and its environment.

\subsection{Effective theories}
Let us suppose that the states of a closed dynamical system correspond to the linear space ${\cal H}={\cal H}_s\otimes{\cal H}_e$, and that we are interested in the expectation values of observables acting only on the system factor space, ${\cal H}_s$, assuming that the initial state is pure, $|\Psi_i\rangle\in{\cal H}$. It is well known that the environment, represented by the factor space ${\cal H}_e$, decouples and that the expectation values can be obtained within ${\cal H}_s$ if the state is factorisable, $|\Psi(t)\rangle=|\phi(t)\rangle \otimes|\psi(t)\rangle$, with $|\phi(t)\rangle \in{\cal H}_s$, and $|\psi(t)\rangle \in{\cal H}_e$. This scenario is certainly applicable when the initial state is factorisable and there is no interaction between the system and its environment. But as soon as the system and the environment begin to interact in some manner, the factorisability of the state vector is lost. The system and the environment must be in a pure, entangled state; hence the expectation values within the system can no longer be reproduced by pure states in ${\cal H}_s$. In this case, the system state becomes mixed and must be represented by the reduced density matrix. 

The traditional framework to discuss effective dynamics is based on the single time axis formalism and the generating functional \eq{stagfh}, which is designed to describe the transition amplitudes between pure states. As a result, it is very difficult to use such a formalism to describe mixed states and open systems. An example of such difficulties is the necessity to perform an involved, partial re-summation of the perturbation series to cancel the collinear divergences in the effective theory of point charges \cite{bn,kin,ln,yennie}. This example will be discussed again below, in Section \ref{conss}.

Another property of effective theory, which is not covered by the single time axis formalism is irreversibility. This is because the variational equations are conservative. In the CTP formalism, however, the doubling of the degrees of freedom allows us the realise irreversibility within variational equations by coupling the two time axes. We will further discuss irreversibility in Section \ref{irrevs}.

\subsection{Coupling of the time axes in effective theories}\label{couplingtas}
The distinguishing feature of the CTP formalism is the coupling of the two time axes, which was implemented on the level of a phenomenological effective action for hydrodynamics in \cite{Grozdanov:2013dba}. Such couplings are always present and the simplicity of the action \eq{ctpeact} is misleading as it does not reflect the fact that the dynamical variables on the two time axes are correlated at $t=t_f$. In fact, the final condition, $\phi^+(t_f,\vec{x})=\phi^-(t_f,\vec{x})$, does not fix the dynamical variables as in the single time axes path integral scheme. Instead, it establishes a correlation among $\phi^+(t_f,\vec{x})$ and $\phi^-(t_f,\vec{x})$, leaving the actual value undetermined. Furthermore, if $\phi^+$ and $\phi^-$ are introduced as two independent variables, then the constraint representing the final-time condition implies a non-trivial coupling between the time axes at $t=t_f$. 

The explicit appearance of the condition $\phi^+(t_f,\vec{x})=\phi^-(t_f,\vec{x})$ in the path integral \eq{ctpwelem} violates translational invariance in time. To recover this symmetry, together with the diagonal structure of the Green's functions in frequency space, one can take the limits of $t_i\to-\infty$ and $t_f\to\infty$. One may expect that the coupling of the time axes, generated at $t=t_f$, disappears in this limit. However, a more careful calculation shows that the time axes remain coupled by a time-translationally invariant, infinitesimally strong coupling \cite{arrow}. Such a transmutation of the boundary conditions is actually the origin of the $\CO(\epsilon)$ terms in the quadratic action of a free particle, given by Eqs. \eq{freeinvp}.

In effective theories, the time axes are correlated by finite couplings. As an example, let us consider the system of two interacting fields, $\phi(x)$ and $\psi(x)$, governed by the action 
\begin{align}
S[\phi,\psi]=S_s[\phi]+S_{se}[\phi,\psi],
\end{align} 
where $S_{se}$ describes the interaction between the system and its environment. The CTP generating functional for the effective theory of $\phi(x)$ is
\be
e^{\ih W[\hj]}=\int D[\hphi]D[\hat\psi]\exp \left\{ \ih S[\phi^+,\psi^+]-\ih S^*[\phi^-,\psi^-]+\ih \hj\hat\phi \right\},
\ee
where $\phi^+(t_f,\vec{x})=\phi^-(t_f,\vec{x})$ and $\psi^+(t_f,\vec{x})=\psi^-(t_f,\vec{x})$. After integrating out the fields $\hat\psi$, the bare effective action can be written in the form of Eq. \eq{ctpgfpi} as 
\be\label{effacte}
S_{eff}[\hphi]=S_s[\phi^+]-S_s^*[\phi^-]+S_{infl}[\hphi],
\ee
where the last term on the right-hand side, the {\it influence functional} \cite{Feynman:1963:TGQ}, contains the effective interactions generated by the environment. It is defined by the expression
\be
e^{\ih S_{infl}[\hphi]}=\int D[\hat\psi] \exp \left\{\ih S_{se}[\phi^+,\psi^+]-\ih S^*_{se}[\phi^-,\psi^-] \right\}.
\ee

The form \eq{effacte} of the action does not separate terms that play different roles in the effective dynamics. To distinguish different contributions to the physical properties of the system, it is more illuminating to write
\be\label{ctpeffact}
S_{infl}[\hphi]=S_1[\phi^+]-S_1^*[\phi^-]+S_2[\hphi],
\ee
where $S_2[\phi^+,0]=S_2[0,\phi^-]=0$. $S_1$ is then the {\it self interaction} and $S_2$ the {\it entanglement} functional. They contain the contributions from the single and the double time axes, respectively. Note that the symmetries that act on both components of the CTP doublet identically and simultaneously are preserved by $S_{eff}$, $S_{infl}$ and $S_2$.

\subsection{Perturbative self-energy and the entanglement functional}\label{psientfs}
The physical origin of the separation of the effective vertices of the influence functional into $S_1$ and $S_2$ is easiest to see in the perturbative construction of the effective dynamics. Let us write the action for $\psi$ as
\be
S_{se}[\phi,\psi]=\hf\psi D^{-1}_0\psi+S_{se,int}[\phi,\psi].
\ee
The perturbative expansion of the influence functional is then defined by
\be\label{perfeffth}
e^ {\ih S_{infl}[\hphi]}=\exp\left\{ \ih \left(S_{se}\left[\phi^+,\frac{1}{i}\fd{}{J^+}\right]- S^*_{se}\left[\phi^-,\frac{1}{i}\fd{}{J^-}\right] \right)\right\}\exp\left\{\ih W_0 [\hJ ] \right\},
\ee
where
\bea
e^{\ih W_0 [\hJ]}&=&\int D[\hat\psi]e^{\frac{i}{2\hbar}\hat\psi\hD^{-1}_0\hat\psi+\ih\hJ\hat\psi} =e^{-\frac{i}{2\hbar}\hJ\hD_0\hJ}
\eea
is the free generating functional for the field $\psi$. 

The couplings between the time axes arise from the free CTP propagator, $\hD_0$, given by Eq. \eq{bosctppr}, which has non-vanishing off-diagonal blocks. The propagator lines in the Feynman diagrams, representing the perturbation series \eq{perfeffth}, that connect elementary vertices on different time axes are the ones encoded in these off-diagonal blocks. To find the physical origin of the off-diagonal propagator components, write Eq. \eq{wfcctpp} as
\bea
D^{-+}(x,y)&=&\sum_n\lal n|\phi(x)\phi(y)|0\rar\lal0|n\rar  =\sum_n\lal 0|\phi(x)|n\rar \lal n|\phi(y)|0\rar,
\eea
where $\{|n\ra\}$ is a complete set of basis vectors. $D^{-+}$ thus receives contributions from one-particle states, which together contribute to the trace in the generating functional \eq{stagfh} and the propagator \eq{ctpprop} at $t=t_f$. The internal lines of the Feynman diagrams that connect the time axes therefore correspond to the excitations of the field $\psi$ at the final time.

The states that contribute to the trace and realise the couplings between the time axes are on the mass-shell because the Wightman function in Eq. \eq{wfcctpp} contains no time ordering. $S_2$ is therefore suppressed in the limit of $t_f\to\infty$ if the excitations of the initial state, involving $\psi$-particles, have a gap.

The single time axis contributions to the effective action, $S_1$, describe the type of system-environment interactions that can be successfully handled by the conventional, single time axis formalism. Hence, the separation \eq{ctpeffact} splits the system-environment interactions into two parts, one that keeps a pure state pure and another that generates entanglement.

\subsection{Conservation laws}\label{conss}
The separation \eq{ctpeffact} of the effective interactions is particularly clear in classical systems where the CTP action has an infinitesimal imaginary part. The full theory usually possesses some continuous symmetries, which are realised in the same manner on each of the two time axes, for instance external space-time symmetries or internal global symmetries. Effective vertices in $S_1$ are called conservative because they correspond to a traditional, single time axis action and thereby preserve the conservation laws for all conserved currents of the full theory, derived from the Noether theorem. The functional $S_2$, which mixes the two time axes, contains the remaining, non-conserving and dissipative system-environment interactions, which make the system open \cite{PhysRevD.90.065010}.

Such a formal separation of the CTP effective vertices, based on conservation laws, is less clear in the quantum case where the effective action has an imaginary part and $\Im S_1$ and $\Im S_2$ also contribute to dissipative phenomena. The system-environment interactions generate two kinds of dissipative phenomena. Firstly, the finite life-time of quasi-particles due to the leakage of the pure quasi-particle states into the environment, called damping, is expressed by $\Im S_1$. Secondly, the system-environment entanglement makes the quasi-particle states in the effective theory mixed and generates decoherence, which will be discussed in Section \ref{decohs}. The latter phenomena are encoded by $\Im S_2$.

In the effective theories of charges in QED, the typical example of the effects described by $S_1$ is vacuum polarisation, which is produced by the {\it virtual} soft photon content of charged particles. Virtual intermediate excitations cannot violate conservation laws in the asymptotic, on-shell states. The finite energy resolution of observations, performed in a finite amount of time always leaves the on-shell, soft photon content of charged particle states unresolved and the realistic charged states, which avoid the collinear divergences of non-degenerate perturbation expansion, are defined with a small but finite energy spread. A typical effective interaction described by $S_2$ is therefore due to {\it real} soft photon contributions to charged particle states. 

The cancellation of the collinear divergences of the non-degenerate perturbation expansion in the effective theory of point charges, mentioned above, requires us to sum up the real and the virtual soft photon contributions. This is a very natural procedure within the CTP formalism, as the former and the latter are automatically encoded by $S_1$ and $S_2$, respectively. The same construction in terms of the single time axis formalism is rather involved but is required to compute finite transition amplitudes in the effective theory.

It is essential to note that the mixing of fields in $S_2$ breaks space-time translational symmetries, thereby causing the physical system on each of the time axes to break energy-momentum conservation. This is the way an effective action that encodes the dynamics of only a subset of all degrees of freedom can describe dissipation, i.e. loss of energy from the relevant effective degrees of freedom to those that were integrated out. This fact was used for example in \cite{Grozdanov:2013dba} to write down a classical dissipative action of a fluid with bulk viscosity.

\subsection{Irreversibility}\label{irrevs}
It is important to distinguish between the loss of time reversal invariance and the irreversibility of an effective theory \cite{effthpch}. By assuming that the microscopic dynamics is time reversal invariant, the only source of time reversal non-invariant effective interaction is the initial condition of the environment. The time inversion odd piece of the action \eq{harmact} is $\Delta^f$, which completely determines $S_2$ according to the spectral condition \eq{imagipropf}. Therefore the harmonic theory \eq{harmact} breaks time reversal invariance if $\Delta^f\ne0$. The form \eq{freeinvp} of the inverse free scalar propagator shows that an infinitesimally strong violation of time reversal invariance is sufficient to produce finite time reversal odd expectation values, e.g. $D^f\ne0$ in Eq. \eq{harmresp}. 

The amplification of infinitesimally small symmetry breaking, on the level of the action, to a finite symmetry breaking by expectation values is the hallmark of spontaneous symmetry breaking. Let us again consider a free field, i.e. in the model \eq{harmact} with infinitesimal $\Delta^f$, given by \eq{freeinvp}. Such an amplification takes place for a free field because of the inversion $\hD=\hat\Delta^{-1}$, with its sole role in the dynamics being the representation of the initial conditions. This is no longer the case if $\Delta^f$ assumes a finite value. $S_2$ is then finite and the dynamics, governed by the equation of motion $\hat\Delta\hphi=-\hj$, is open and non-conservative. An environment with a sufficiently large capacity for absorbing energy and a non-trivial spectral weight at vanishing energy then makes the effective dynamics irreversible \cite{irrev}.

The microscopic realisation of irreversibility is dissipation, which is the leakage of the system into the environment. The resulting effect, i.e. damping, is described by $\Im S_1$ and $\Re S_2$, with the former being responsible for the finite life-time of the excitations and the latter contributing to diffusive forces in the equation of motion derived from the effective theory. In a classical effective theory, such as the one considered in \cite{Grozdanov:2013dba}, the only way to introduce dissipation is through $S_2$. This is because the imaginary part of the Lagrangian is infinitesimally small.

The emergence of a complex action in a conventional, single time axis path integral signals non-unitary time evolution. However, the unitarity condition, expressed by Eq. \eq{unit}, is preserved in the effective theories, which therefore always preserve the unitarity of the microscopic theory. Such a global unitarity is maintained by compensating for the leakage of the system state into the environment, described by $\Im S_1$, with the system-environment entanglement expressed by $S_2$.

\subsection{Decoherence}\label{decohs}
Decoherence \cite{zeh,zurek} is the suppression of the off-diagonal density matrix elements and is a basis-dependent phenomenon. In fact, the density matrix, being Hermitian, is always diagonal in a suitable basis. Here, we wish to consider the decoherence of a local operator $\Phi(\vec{x})=F[\phi(\vec{x})]$. More precisely, we are interested in the suppression of the off-diagonal elements of the reduced density matrix for $\Phi(\vec{x})$ in the field-diagonal basis where the basis vectors are the eigenvectors of $\Phi(\vec{x})$. It is convenient to use the Fourier transform, $\Phi(\vec{k})$, rather than the coordinate-dependent field because the eigenstates of this operator contribute independently to the suppression in our truncated effective theory.

For the purposes of analysing decoherence, it is useful to generalise the CTP formalism, which is aimed at computing the trace of the density matrix, to the Open-Time-Path (OTP) scheme, which is able to reproduce the full density matrix. Instead of Eq. \eq{ctpgfh}, one defines the OTP generating functional as a matrix element of the density matrix, identified by the spatial configurations $\phi^\pm_f(\vec{x})$ in the presence of external sources,
\begin{align}
e^{\ih W[j^+,j^-;\phi_f^+,\phi_f^-]} &= \rho [\phi^+,\phi^-;t_f]  \nn
&= \left\langle \phi_f^+ \left| \CT\left\{ e^{-\ih\int d^4 x[H(x)-j^+(x)\phi(x)]} \right\} \rho(t_i) \CT^* \left\{ e^{\ih\int d^4 x[H(x)+j^-(x)\phi(x)]} \right\} \right|\phi_f^-\right\rangle,
\end{align}
with finite $t_f$. The path integral expression for this functional is
\be
e^{\ih W[\hj,\hphi_f]}=\!\!\!\!\!\!\!\!\!   \int \limits_{\hat\phi_f (\vec{x})=\hat\phi(t_f,\vec{x})} \!\!\!\!\!\!\!\!\!           D[\hphi] \, e^{\ih S[\hphi,\hphi_f]+\ih\hj\hphi},
\ee
where the action satisfies the OTP symmetry,
\be
S[\phi^+,\phi^-,\phi^+_f,\phi^-_f]=-S^*[\phi^-,\phi^+,\phi^-_f,\phi^+_f].
\ee

The generating functional for the reduced density matrix of a composite operator $F[\phi(\vec{x})]$, the OTP analog of the effective theory, is then
\be\label{effdmnonPathInt}
e^{\ih W[j^+,j^-;\Phi_f^+,\Phi_f^-]}=\left\langle \Phi_f^+ \left| \CT \left\{e^{-\ih\int d^4 x \left[H(x)-j^+(x)F(\phi(x)) \right]} \right\} \rho(t_i) \CT^* \left\{ e^{\ih\int d^4 x \left[H(x)+j^-(x)F(\phi(x)) \right] } \right\} \right|\Phi_f^- \right\rangle,
\ee
where the state $|\Phi(\vec{x})\ra$ is an eigenstate of $F[\phi(\vec{x})]$ with an eigenvalue $\Phi(\vec{x})$. In the path integral language, the expression \eqref{effdmnonPathInt} can be written as 
\be\label{effdm}
e^{\ih W[\hj,\hat\Phi_f]}=\int D[\hphi] \, e^{\ih S[\hat\phi]+\ih \hj F(\hphi)}\prod_\vec{x}\delta \left(\Phi_f^\pm(\vec{x})-F\left[\phi^\pm(t_f,\vec{x})\right]\right),
\ee
or in terms of an effective action,
\be\label{effdmkk}
e^{\ih W[\hj,\hat\Phi_f]}= \!\!\!\!\!\!\!\!\!  \int \limits_{\hat\Phi(t_f,\vec{x})=\hat\Phi_f(\vec{x})}   \!\!\!\!\!\!\!\!\!   D[\hat\Phi]  \, e^{\ih S_{eff}[\hat\Phi]+\ih\hj F[\hphi]}.
\ee
The effective action in Eq. \eqref{effdmkk} is given by 
\be\label{effctpact}
S_{eff}[\hat\Phi]=-i\hbar\log\left[\int D[\hphi] \, e^{\ih S[\hat\phi]}\prod_\vec{x}\delta\left(\Phi^\pm(x)-F\left[\phi^\pm(x) \right] \right)\right],
\ee
which has the structure of Eq. \eq{ctpeffact}. If $F[\phi(\vec{x})]$ does not determine $\phi(\vec{x})$ uniquely or if the initial state, described by $\rho(t_i)$, is mixed, then the path integral on the right-hand side of Eq. \eq{effctpact} becomes non-trivial and $S_{infl}\neq 0$.

Decoherence of $F[\phi(\vec{x})]$, i.e. the suppression of the off-diagonal reduced density matrix elements for the composite operator, is governed in the OTP formalism by $\Im S_2[\Phi^+,\Phi^-]$. It is a property of the state at each given instant, which gradually builds up through temporal evolution. Its dynamical origin is consistency \cite{grif}. This is the suppression of the contributions to the density matrix that correspond to well-separated OTP doublet trajectories, where $\Phi^+(x)-\Phi^-(x)$ assumes significant values in large regions of space-time. Such a build-up of the suppression can be detected without having access to the full density matrix. In fact, decoherence must also appear as the suppression of the CTP doublet trajectories that are well separated for a significant amount of time. In our calculation, we will give up the details provided by the access to the off-diagonal density matrix elements of the OTP formalism. Instead, we will use the simpler CTP effective theory and only compute the trace of the density matrix to analyse decoherence in the electron gas.

\subsection{Generalised fluctuation-dissipation theorem}\label{fdts}

Quasi-particles are dressed by the conservative interactions and have a finite life-time if $\Im S_1$ makes the effective theory irreversible, as explained in Section \ref{irrevs}. Non-conservative interactions make the final state mixed even if the initial state is pure and $\Im S_2$ generates decoherence, cf. Section \ref{decohs}. A common element of these phenomena is that both irreversibility and decoherence result from the quasi-degeneracy of a continuous spectrum. Such a common origin is actually rooted in a more fundamental identity of the CTP two-point functions, which can be considered as a generalisation of the {\it fluctuation-dissipation theorem}.

The fluctuation-dissipation relation is usually discussed in the context of a linear equation of motion. Hence, we consider a translationally invariant, harmonic effective action given by Eq. \eq{harmact} where the quasi-particles are defined by the single time axis action
\be\label{soh}
S_0[\phi]= \hf \int d^4 x \, \phi\left(\Delta^n+i\Delta^i\right)\phi.
\ee
The entanglement with the environment can described by the correlation functional
\be\label{scorrh}
S_{corr}[\hphi]=\int d^4 x \, \phi^+\left(\Delta^f-i\Delta^i\right)\phi^- .
\ee
We will use the Keldysh basis \cite{keldysh}, $\phi^\pm=\phi\pm\phi^d/2$, in which
\be
S_{eff}[\hphi]=\hf \int d^4x \left[\phi\Delta^a\phi^d+\phi^d\Delta^r\phi+i \phi^d\Delta^i\phi^d \right] .
\ee
We also introduce the parametrisation $j^\pm=j/2\pm j^p$ of the external sources, where $j^p(x)$ is the physical source. This represents the experimental setup in which the system is driven adiabatically from the perturbative vacuum at $t_i=-\infty$ to a desired initial state at $t=0$. The source term is thus
\be
\hj\hphi=j\phi+j^p\phi^d,
\ee
indicating that the expectation value $\left\langle \phi(x)\right\rangle$ is generated by the book-keeping source $j(x)$, which should be set to zero after the functional derivatives had been taken.

The translationally invariant operators commute and each CTP block is diagonal in Fourier space, a property which simplifies the discussion. The common origin of irreversibility and decoherence in this simple model is the fact that the eigenvalue $\Delta^i(q)$ plays two different roles in the effective dynamics. On the one hand, it is the imaginary part of the inverse Feynman propagator of the action \eq{soh} and as such it describes the inverse life-time of the plane wave $q$,
\be
\Delta^i(q)=1/ \tau(q).
\ee
It serves as a measure of strength of the breakdown of time reversal invariance. Note that $\Delta^i(q)\ge0$ as a result of the positive definiteness of the norm in the effective theory, cf. Eq. \eq{posdefn}. On the other hand, $\Delta^i(q)$ enters into the imaginary part of the double time axis action and
\be
\left|e^{\ih S_2[\hphi]}\right|=\exp \left\{-\hf\int\frac{d^4q}{(2\pi)^4}\Delta^i(q)|\phi^d(q)|^2 \right\},
\ee
indicating that the off-diagonal matrix elements of the density matrix are exponentially suppressed in $\phi^d=\phi^+-\phi^-$, and that the width of the corresponding Gaussian is $1/\Delta^i(q)$. In other words, $\tau(q)$ is a measure of the contribution of the plane waves to decoherence. It is worthwhile noting that a lesson of Eq. \eq{imagipropf} is that both decoherence and time reversal invariance arise when the time-translationally invariant couplings of the two time axes have a finite strength.

The coupling between the time axes, $\Delta^f$, may be local in time. However, the decoherence of the state builds up in time in a non-local manner. In fact, $\mr{sgn}(\omega)$, which appears in Eq. \eq{imagipropf}, contains slow time dependence,
\be
\int\frac{d\omega}{2\pi}e^{-i\omega t}\mr{sgn}(\omega)=-P\frac{i}{\pi t},
\ee
if the integration is carried out over $t$. We use $P$ to denote the {\it principal value} of the integral.

What happens in an interacting system? The form \eq{propnf} of the two-point function, which is also valid for interacting, composite operators, shows that the imaginary parts of the diagonal and off-diagonal blocks are given by the same real function, i.e. the far field component. The structure of the action \eq{harmact} of our simple harmonic model therefore also remains present in the quadratic part of any effective action for interacting theories. As a result, irreversibility and decoherence have a common origin as far as small fluctuations around a stable state are concerned.

The double role of $\Delta^i$, discussed above, can be considered as an extension of the fluctuation-dissipation theorem. On the one hand, $\Delta^i(p)$ is the imaginary part of the inverse Feynman propagator of the action \eq{soh}. Furthermore, it also provides the imaginary part to the inverse retarded Green's function, $\Delta^f(p)=\mr{sgn}(p^0)\Delta^i(p)$, and thus controls the ``leakage'' of the quasi-particle states towards the environment, which is an elementary dissipative process. On the other hand, $\Delta^{-+}(p)=2\Theta(p^0)\Delta^i(p)$ is the quadratic form of $S_2$ and therefore represents the interactions with the environment, which are the origin of fluctuations. Such a relation between fluctuations and dissipation holds even in the absence of thermodynamical reservoirs so long as the linearised equation of motion is reliable, i.e. the system is in a state that is resilient against microscopic fluctuations.

\section{The electric current in an electron gas}\label{cdegs}

We are now ready to begin analysing the dynamical properties of the electric current in an electron gas by using the full CTP machinery introduced above. The relevant theory of the microscopic physics of electrons and photons is quantum electrodynamics (QED). Our goal will be the derivation of the effective action for the Noether current $j^\mu$ in the quadratic and one-loop approximations of QED at finite density and temperature.

\subsection{Effective theory for the current}

We begin by introducing the microscopic theory of QED, which we will use to compute the Green's functions for the current, sourced by $a^\mu$, as well as the electromagnetic field, sourced by $j^\mu$. The relevant generating functional is
\be\label{genfqed}
e^{\ih W[\ha,\hj]}=\int D[\hsi]D[\hsib]D[\hA] \exp \left\{\ih \int d^4x\left[ \hsib \left(\hF^{-1}+\ha\bre-\frac{e}c\hs\hA\bre \right) \hsi+\frac{1}{2}\hA\hD^{-1}_0\hA+ \hj\hA \right] \right\}.
\ee
The free photon propagator is
\be
\hD^{\mu\nu}_0(p)=-\hD(p;0)\,T^{\mu\nu},
\ee
with $T^{\mu\nu}=g^{\mu\nu}-\partial^\mu\partial^\nu/\Box$, and the free electron propagator is
\be\label{freeelpr}
\hF(p)=\left(p\br+\frac{mc}\hbar\right)\left[\hD(p;m)+2\pi i\delta\left(p^2-\frac{m^2c^2}{\hbar^2}\right)n(p)\begin{pmatrix}1&-1\cr-1&1\end{pmatrix}\right],
\ee
where
\be
\hD(p;m^2)=\begin{pmatrix}\frac1{p^2-\frac{m^2c^2}{\hbar^2}+i\epsilon}&-2\pi i\delta\left(p^2-\frac{m^2c^2}{\hbar^2}\right)\Theta(-p^0)\cr-2\pi i\delta\left(p^2-\frac{m^2c^2}{\hbar^2}\right)\Theta(p^0)&-\frac1{p^2-\frac{m^2c^2}{\hbar^2}-i\epsilon}\end{pmatrix}.
\ee
The occupation number density $n(p)$ is given by the expression
\be
n(p)=\frac{\Theta(p^0)}{e^{\beta(\epsilon_\vec{p}-\mu)}+1}+\frac{\Theta(-p^0)}{e^{\beta(\epsilon_\vec{p}+\mu)}+1},
\ee
in terms of the single particle energy, $\epsilon_\vec{p}=c\sqrt{m^2c^2+\hbar^2\vec{p}^2}$, and the chemical potential $\mu$. The UV divergences are the same as in the single time axis case, but special care is required to ensure that the composite operator Green's functions are finite. The similarity between the current-current two-point function in the case of QED with non-interacting electrons and the one-loop photon polarisation tensor in the standard interacting QED implies that we require an $\CO(e^0)$ counter-term even for the ideal gas \cite{ed}. 

The functional $W[\ha,\hj]$ can be used to construct the effective action through a functional Legendre transform,
\be\label{legtr}
W[\ha,\hj]=\Gamma[\hJ,\hA]+\ha\hJ+\hj\hA,
\ee
where the new variables
\begin{align}\label{legtrv}
\hJ=\fd{W[\ha,\hj]}{\ha}, && \hA=\fd{W[\ha,\hj]}{\hj},
\end{align}
are the expectation values, 
\begin{align}
J^\mu(x)\equiv \left\la j^\mu(x) \right\ra =\Tr[\rho\psib(x)\gamma^\mu\psi(x)], &&A_\mu(x)=\Tr[\rho A_\mu(x)],
\end{align} 
for $\ha=\hj=0$.

In what is to follow, we will make use of the parametrisation 
\be\label{para}
a^\pm=\frac{\bar a}2\pm a,
\ee
for the sources. The function $a$ represents a source that can be adjusted by a suitably chosen physical environment and leads to unitary time evolution, cf. Eq. \eq{unit}. The other component, $\bar a$, is only used as a formal book-keeping device. The Legendre transform of a real, convex function can be defined either geometrically or algebraically. We follow the latter route and use Eqs. \eq{legtr} and \eq{legtrv} to define the effective action, a complex functional, in an algebraic manner. The inverse Legendre transform is based on the variables
\begin{align}\label{eqmja}
\ha = - \fd{\Gamma[\hJ,\hA]}{\hJ},&&\hj = - \fd{\Gamma[\hJ,\hA]}{\hA}.
\end{align}
Hence, the variational equations of the effective action are satisfied by the expectation values, which are obtained by setting the external sources to zero. 

We are interested in the effective current dynamics and the corresponding effective action, $\Gamma[\hJ]$, which can be obtained by eliminating $\hA$ from $\Gamma[\hJ,\hA]$ by making use of the second equation in \eq{eqmja}. The effective action is real in the physical case, $j^+=-j^-$, $a^+=-a^-$; therefore it is sufficient to retain only the real part, $\Re\Gamma$, of the equation of motion.

We write the external source coupling as
\be
\ha\hJ=\bar aJ+aJ^d,
\ee
which yields
\begin{align}\label{phvar}
J=\fd{W}{\bar a}=\hf(J^++J^-),&&J^d=\fd{W}{a}=J^+-J^-,
\end{align}
showing that $J$ is the current expectation value and $J^d=0$ in the physical case with $\bar a=0$.

\begin{figure}
\includegraphics[trim=0cm 0cm 0cm 1cm, scale=0.8]{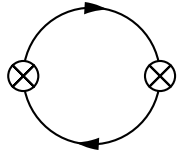}\caption{Current-current two-point Green's function diagram in QED.}\label{fig:Ring1QED}
\end{figure}

To find the linearised equation of motion, we need the quadratic parts of $W$ and $\Gamma$, which will be calculated by expanding in $\hbar$ and keeping the first two orders. We start by integrating out the electron field,
\be
e^{\ih W[\ha,\hj]}=\int D[\hA] \exp \left\{\int d^4 x\left[ \Tr[\hF^{-1}+\ha\bre-\frac{e}c\hs\hA\bre]+\frac{i}{2i\hbar}\hA\hD^{-1}_0\hA+\ih\hj\hA \right] \right\},
\ee
and keeping the $\ord{\hbar^0\ha^2}$ terms, giving us
\be\label{paa}
e^{\ih W[\ha,\hj]}=\int D[\hA] \exp \left\{\int d^4 x \left[ -\frac{i}{2\hbar}\ha\hG\ha+\ih\hat k\hA+\frac{i}{2i\hbar}\hA\hD^{-1}\hA \right] \right\}.
\ee
The inverse of $\hat D$,
\be\label{photpr}
\hD^{-1}=\hD_0^{-1}-\frac{e^2}{c^2}\hs\hG\sigma,
\ee
and the source,
\be\label{sourcek}
\hat k=\hj+\frac{e}c\hs\hG\ha,
\ee
are given in terms of the one-loop current-current Green's function, represented diagrammatically in Figure \ref{fig:Ring1QED},
\be\label{curprop}
G^{\sigma_1\sigma_2}_{\mu_1\mu_2}(x_1,x_2)=-i\hbar \, \tr[\hF^{\sigma_1\sigma_2}(x_1,x_2)\gamma_{\mu_2}\hF^{\sigma_2\sigma_1}(x_2,x_1)\gamma_{\mu_1}].
\ee
The presence of a neutralising, homogeneous background charge was assumed in Eq. \eq{sourcek} to avoid divergent tadpole contributions. The Gaussian integral \eq{paa} yields
\be
W[\ha,\hj]=-\hf\ha\hG\ha-\frac{e}c  \hj\hD_0\sigma\hG\ha-\hf\hj\left(\hD_0+\frac{e^2}{c^2}\hD_0\hs\hG\hs\hD_0\right)\hj,
\ee
at the quadratic order in the sources.

Although the orders of the expansion of $W[\ha,\hj]$ in $\hbar$ and in the number of loops correspond to each other in standard QFT calculations, this is no longer the case when the effective action is considered. The reason is that the variables of the effective action have different orders in $\hbar$, $\hJ=\ord{\hbar}$ and $\hA=\ord{\hbar^0}$, as opposed to the variables of $W$, $\ha=\ord{\hbar^0}$, $\hj=\ord{\hbar^0}$. The Legendre transform \eq{legtr} therefore gives
\be
\Gamma[\hJ,\hA]=\hf\hJ\hG^{-1}\hJ+\hf\hA\hD^{-1}_0\hA-e\hA\hs\hJ
\ee
for the next-to-leading order, $\ord{\hbar}$ effective action.

\begin{figure}
\includegraphics[trim=0cm 1.5cm 0cm 1.5cm, scale=0.8]{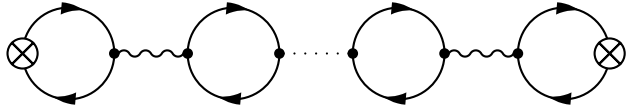}\caption{Current-current two-point Green's function with the Schwinger-Dyson re-summed propagator.}\label{fig:SDresum}
\end{figure}

The Maxwell's equation,
\be
\hA=e\hD_0\sigma\hJ,
\ee
can be used to eliminate the photon field so that we obtain an effective action,
\be\label{currctpact}
\Gamma[\hJ]=\hf\hJ\hat{\cal L}\hJ,
\ee
with the Schwinger-Dyson re-summed propagator, represented in Figure \ref{fig:SDresum}, giving us
\be\label{klineom}
\hat{\cal L}=\hG^{-1}-\frac{e^2}{c^2}\hs\hD_0\hs.
\ee
The calculation of $\hG$ is summarised in Appendix \ref{grfnctap}.

\subsection{Equation of motion}
All bosonic two-point functions, including $\hG$, have the same CTP matrix structure as Eq. \eq{stctpform}. This allows us to define the retarded and advanced parts of the linearised equation of motion for the current,
\be\label{invret}
{\cal L}^{r,a}=(G^{r,a})^{-1}-\frac{e^2}{c^2}D_0^{r,a},
\ee
where $D_0^{r,a}$ are given by Eqs. \eq{retccgf} and \eq{retgfph}, respectively. The effective action, expressed in terms of the near and the far components of ${\cal L}$, ${\cal L}^n=({\cal L}^r+{\cal L}^a)$ and ${\cal L}^f=({\cal L}^r-{\cal L}^a)$, is
\be\label{leffactpm}
\Gamma[J^+,J^-]=\hf(J^+,J^-)\begin{pmatrix}-i({\cal L}^n-{\cal L}^f)^2+{\cal L}^n&i({\cal L}^n-{\cal L}^f)^2+{\cal L}^f\cr i({\cal L}^n-{\cal L}^f)^2-{\cal L}^n&-i({\cal L}^n-{\cal L}^f)^2-{\cal L}^n\end{pmatrix}\begin{pmatrix}J^+\cr J^-\end{pmatrix},
\ee
cf. Eq. \eq{istctpform}, which results in 
\be\label{leffactd}
\Gamma[J,J^d]=-\frac{i}2J^d{\cal L}^r{\cal L}^aJ^d+\hf J^d{\cal L}^rJ+\hf J{\cal L}^aJ^d,
\ee
when expressed in terms of the Keldysh basis, i.e. Eq. \eq{phvar}. The form of \eq{leffactpm} shows that $\Re S_2$ and $\Im S_2$ are given by the kernels ${\cal L}^f$ and $({\cal L}^n-{\cal L}^f)^2$, respectively. The form \eq{leffactd} of the effective action then reveals that $\Im\Gamma$ governs decoherence. On the other hand, the equation of motion for the physical expectation value $J$, for which $\bar a=0$, i.e. the variational equation for $J^d$ at $J^d=0$, 
\be\label{eomgam}
a=-{\cal L}^rJ,
\ee
arises from $\Re\Gamma$. We note that an effective action with the structure of this type, i.e. mixing $J^+$ and $J^-$, was used in \cite{Grozdanov:2013dba} to describe a classical dissipative fluid.

{The Legendre transform of the effective action \eq{currctpact},
\be
W[a]=-\hf\hat a\hat{\cal L}\hat a,
\ee
is the generator functional of the connected Green's functions for the current after the resummation of the ring diagrams with photon line insertions. The corresponding, improved Kubo formula for the induced current,
\be
J=-({\cal L}^r)^{-1}a,
\ee
cf. Eq. \eq{kfw}, corresponds to the inversion of the equation of motion \eq{eomgam}. In other words, the equation of motion, used in this paper, could have been obtained from the traditional Kubo formulae for the electric current after the resummation of the ring diagrams and an inversion, which expresses the external source as the function of the induced response. For a review of the linear response formalism and its connection to the CTP formalism, see Appendix \ref{linresps}.

The non-relativistic limit, $c\to\infty$, has been well understood in QED, where the minimal coupling is given by the term $-ej^\mu A_\mu/c$ in the Lagrangian. The source term that generates the electric current, $j^\mu a_\mu$, is not suppressed by $c$. The higher-than-quadratic order terms in $W$, in the source $a^\mu$, therefore represent correlations, which come from relativistic effects in QED. In other words, the non-relativistic mechanical equation of motion for the electric current of the non-interacting Dirac sea can contain relativistic terms in the presence of an external electromagnetic field. At the leading order, i.e. the quadratic level of the generating functional, studied in this work, there are no such unusual mixing terms.

We can now use the non-relativistic retarded Green's functions \eq{retccgf} and the three-dimensional parametrisations $a^\mu=(\phi/c,\vec{a})$ and $J^\mu=(\rho c,\vec{j})$ to find the equation of motion in Fourier space for the mode $q^\mu=(\omega/c,\vec{q})$,
\be
\begin{pmatrix} \phi/c \cr\vec{a}\end{pmatrix}=-\frac1c\left[\frac1c \, {\cal L}_\ell\begin{pmatrix}1&\vec{n}\xi\cr\vec{n}\xi&\xi^2\vec{L}\end{pmatrix}+c \, {\cal L}_t\begin{pmatrix}0&0\cr0&\vec{T}\end{pmatrix}\right]\begin{pmatrix}c\rho\cr-\vec{j}\end{pmatrix}   ,
\ee
where the overall factor of $1/c$ on the right-hand side is due to the definition \eq{scprod} of the scalar product in Fourier space and $\xi=\omega/c|\vec{q}|$. We are also using $\vec{n}=\vec{q}/|\vec{q}|$, $\vec{L}=\vec{n}\otimes\vec{n}$ and $\vec{T}=\openone-\vec{L}$, or in terms of the index notation, $L^{ij} = n^i n^j$. We decomposed the above expression in terms of the longitudinal,
\be\label{leomop}
{\cal L}_\ell=\frac1{G^r_\ell}-\frac{e^2}cD_0^r,
\ee
and the transverse part,
\be\label{teomop}
{\cal L}_t=\frac1{G^t}\vec{T}.
\ee
Finally, we are in position to derive the desired non-relativistic equations of motion,
\begin{align}
\phi=-{\cal L}_\ell \, \rho, && \vec{a}={\cal L}_t\,\vec{j}.
\end{align}
It is worth noting that the second equation indicates that the external source in a non-relativistic gas should be in the Coulomb gauge, $\vec{\nabla}\cdot\vec{a}=0$.

\section{Hydrodynamical limit}\label{hydrlims}

The question we wish to address in this section is whether local equilibrium can be formed in the presence of weak, slowly changing external sources, in space-time, leading to hydrodynamic behaviour of the electron gas at non-zero density, even in the absence of interactions. Somewhat counterintuitively, we wish to argue that this is indeed possible. While local equilibrium is assumed in the usual phenomenological approach to hydrodynamics, it is our goal to understand its emergence through a detailed derivation of an effective theory. We hope that such calculations may reveal new microscopic insights into hydrodynamics. To define our terminology more precisely, we consider the system to be in local equilibrium when the following two conditions are satisfied: 

\vbox{%
\begin{enumerate}[label=(\Roman{*}),ref=(\roman{*})]
	\item{The equations of motion are local in time and contain a finitely ranged smearing in space. While the equations of motion of an interacting system may be local in space-time, such a smearing is necessary in an ideal gas due to the absence of elementary relaxation processes.}
	\item{The retarded Green's functions, describing the solution of the linearised equation of motion, display damped time dependence, leading to relaxation.}  
	\end{enumerate}
}

In order to satisfy the condition (I), we have to restrict our attention to phenomena with sufficiently slow dependence on the space-time coordinates. To ensure local response, the external perturbations should therefore remain slow compared to the characteristic velocity. However, determining the precise temporal resolution for the ``hydrodynamical" processes to be observable can be rather complicated in an ideal gas with no interactions. In the traditional phenomenological approach, developed for interacting systems, one assumes the analyticity of the dynamics in the infrared region, i.e. in a small vicinity of the point $\omega=0$, $\vec{q}=0$ in Fourier space. This renders the determination of the allowed space-time resolution in a hydrodynamic regime straightforward. Although the conservation laws tend to generate slow, long-range modes, the interactions generate a finite life-time, $\tau_0$, for quasi-particles. This life-time acts as an infrared cutoff for the frequency and makes the dynamics local for $\omega\ll1/\tau_0$ and $|\vec{q}|\ll1/r_{mfp}$, where $v_F=\hbar k_F/m$ and $r_{mfp}=v_F\tau_0$ denote the Fermi velocity and the mean free path. The absence of elementary (microscopic) relaxation processes in an ideal gas, i.e. as $\tau_0\to\infty$, forces us to look for other sources of smearing mechanisms to establish local dynamics. 

It is shown in Appendix \ref{grfnctap} that the response to the external field can be conveniently expressed in our case in terms of two dimensionless variables, 
\begin{align}\label{xyvars}
x=\frac{m\omega}{\hbar|\vec{q}|k_F } &&\text{and} &&y=\frac{|\vec{q}|}{k_F}.
\end{align}
The singularities of the linearised, one-loop dynamics of the ideal gas then arise at the threshold of the particle-hole excitations, i.e. at the border of the domain,
\be\label{domain}
{\cal D}_{hydr}=\left\{(y,x)\, \big| \, \left| |x|-y/2 \right|<1 \right\},
\ee
on the $(y,x)$ plane. We claim that in order for the current to display hydrodynamic-like behaviour, $\omega$ and $\vec{q}$ must be restricted to ${\cal D}_{hydr}$, where the functions \eq{lfuncdef} and \eq{ilfuncdef} are analytic. More precisely, the dynamics will be most closely analogous to the usual hydrodynamics when $y \ll 1$ and $|x| < 1$, having used Eq. \eqref{domain}. The resulting Laurent series for $\la j^\mu \ra$ in the magnitude of the wave vector, $|\vec{q}|$, starts with a non-zero, negative power of $|\vec{q}|$. The situation here is therefore significantly different compared to the traditional phenomenological expansion, where $\la j^\mu \ra$ is given by a Taylor series in $\omega$ and $\vec{q}$. This explains the appearance of the factor $1/|\vec{q}|$ in the equations of motion, announced in the Introduction.

Such a hydrodynamical regime is more restricted than that of the interacting systems, described by phenomenological hydrodynamics, where the analyticity of $\omega$ and $\vec{q}$ allows us to consider the two limits $\omega\to0$ and $\vec{q}\to0$, independently. In the absence of interactions, the IR limit depends on the ratio of $\omega$ and $|\vec{q}|$, as the point $\omega=\vec{q}=0$ is approached on the $(y,x)$ plane. To understand the physical meaning of the restriction imposed by ${\cal D}_{hydr}$, note that $x=v_{ph}/v_F$, with $v_{ph}=\omega/|\vec{q}|$, is the ratio of the phase velocity of the external probe to the Fermi velocity. As argued above, the external perturbations should remain slow compared to the characteristic velocity, thus giving us a restriction on the size of $x$. As soon as interactions are turned on, the quasi-particles acquire finite life-time, which then acts as an IR cutoff and begins screening the $1/|\vec{q}|$ singularities. More precisely, it is the imaginary part of the self-energy at the Fermi surface, correcting the loop integrals \eq{ctpblfsp}, that makes the Green's functions analytic in $q^\mu$. The traditional phenomenological approaches for interacting systems, which rely on the analyticity in $q^\mu$, therefore correctly assume that the hydrodynamical regime extends to arbitrarily large values of $|x|$.

We will see that the above considerations also make condition (II) satisfied for the discussed ranges of $x$ and $y$. Details will be presented in the sections below.

\subsection{Spectral function at zero temperature}\label{sepctrfs}
The simplest insight into the response of the dynamics to an external perturbation is provided by the spectral weight of the excitations, given in our case by
\be\label{spfnct}
iG^{-+}_{\mu\nu}(q)=\left\langle 0 \left| j_\mu(-q)j_\nu(q) \right|0 \right\rangle.
\ee
This correlator can be recovered from the imaginary part of the retarded Green's function in momentum space, i.e. the far field component. A more detailed view of the response is provided by the identification of the collective modes. The dispersion relation of the collective modes, $\omega (\vec{q})$, of a harmonic system is easiest to define by the use of the residue theorem, if the inverse retarded Green's function is an analytic function on the complex frequency plane. The dispersion relations are then defined by the zeros of the inverse retarded Green's function, $[G^r(\omega,\vec{q})]^{-1}=0$, and the real and the imaginary parts of $\omega_0(\vec{q})$ give the frequency and the inverse life-time, respectively. A simplification of this prescription occurs when the frequency dependence of $[G^r(\omega,\vec{q})]^{-1}$ is linear and $|\Im\omega (\vec{q})|\ll |\Re\omega (\vec{q})|$. The approximate normal mode dispersion relation is then given by the vanishing of $\Re[ G^r(\omega,\vec{q})]^{-1} = 0$ and the inverse life-time can be approximated by evaluating $\Im[G^r(\omega,\vec{q})]^{-1}$ on the dispersion relation.

It is important to note that our spectral function is {\it not} analytic on the entire complex frequency plane at $T=0$, cf. Eq. \eq{tsndiag}, hence the residue theorem-based arguments, mentioned above, do not apply. One can presumably recover analyticity at non-zero temperature, however, the non-analytic part of the Green's function is still present in the low-temperature expansion. The picture becomes clearer if we restrict our attention to a sufficiently small region near $x=y=0$, where we do indeed recover analiticity and are able to find well-defined collective modes. This feature explains the universal importance of collective modes for the hydrodynamical description, as they can be introduced in the IR even if they become ill-defined at shorter time and length scales.

The abundance of particle-hole states can be estimated by the spectral weight \eq{spfnct}, given by the $+-$ components of the propagators in Eq. \eq{parproj} and presented in a more detailed way in Eq. \eqref{tsndiag}. We depict the longitudinal and transverse spectral functions, $-\Im \CT^{+-}$ and $-\Im {\cal{S}^{+-}}$, respectively, in Figs. \ref{tspmf} and \ref{tspmf2}. The numerical results are presented at metal density with $k_F=10^8\,\text{cm}^{-1}$, in units of $\lambda_c=\hbar/mc=1$. The spectral weight at $y\sim0$ is proportional to $x$ and drops rapidly to zero for $|x|=1-\ord{y}$. For $y\gg1$, the Fermi sphere is negligible and the holes carry negligible energy-momentum. Hence, the particle-hole excitations, both in the longitudinal and in the transverse sectors, obey approximately the same dispersion relation of a free particle and spread over the interval $2|x|-1<y<2|x|+1$, for $|x|\gg1$.

However, the spectral weights in the longitudinal and the transverse sectors are markedly different. The majority of the longitudinal modes are present at long wavelengths and their number diminishes at shorter wavelengths. On the other hand, the transverse excitations are more common at short wavelengths. Of course, these are relative statements, as the longitudinal spectral weight is several orders of magnitude larger than the transverse one in the kinematical regime of interest.

\begin{figure}
\includegraphics[scale=0.56]{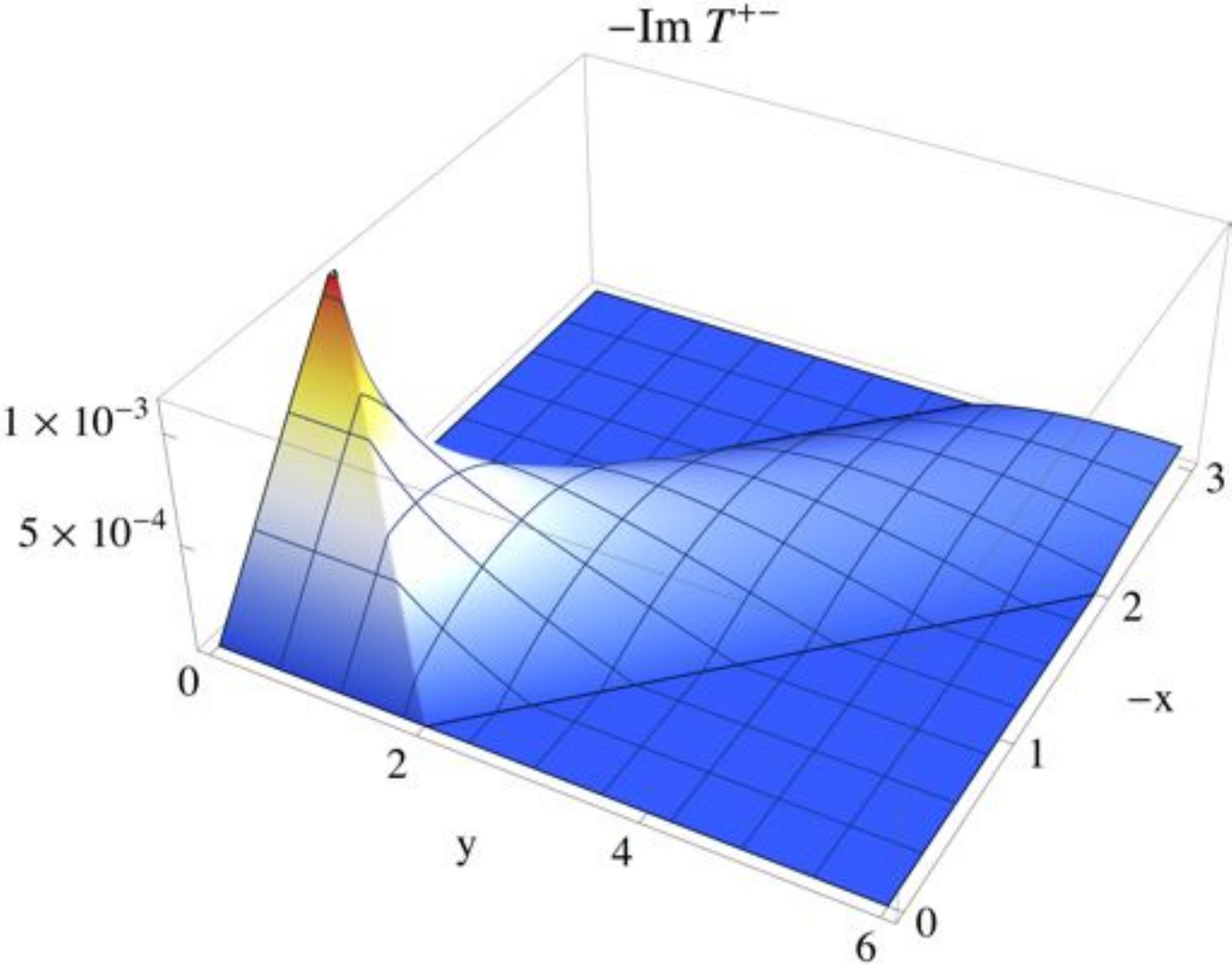}\hskip0.65cm\includegraphics[scale=0.53]{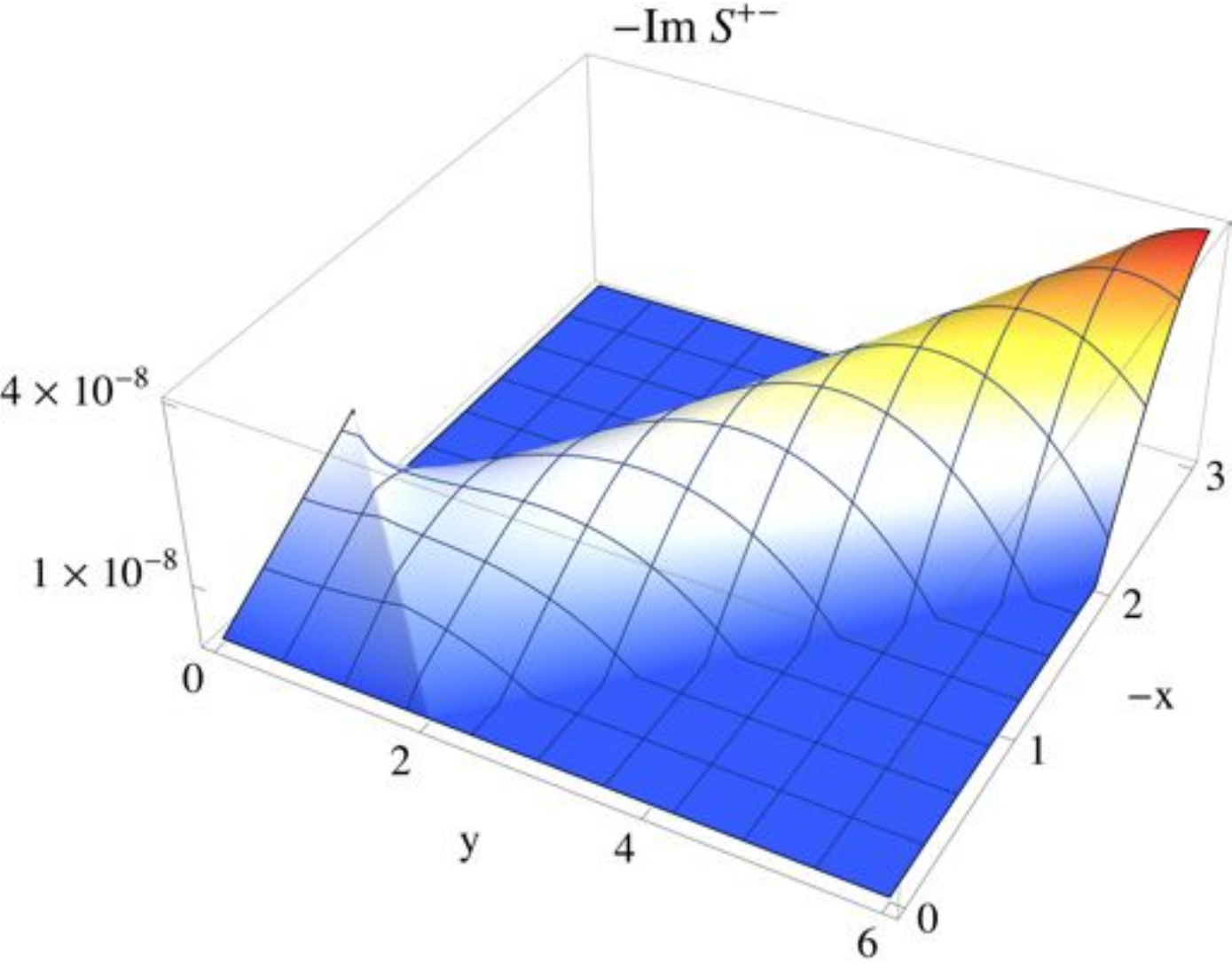}
\centerline{(a)\hskip10cm(b)}
\caption{Plots of the longitudinal and transverse spectral functions, shown over the $(y,-x)$ plane, in units of $m=1$. (a): The longitudinal component, $-\Im \CT^{+-}(x,y)$, with values between $0$ and $10^{-3}$. (b): The transverse component, $-\Im {\cal{S}^{+-}} (x,y)$, with values between $0$ and $4\times 10^{-8}$.}\label{tspmf}
\end{figure}

\begin{figure}
\includegraphics[scale=0.56]{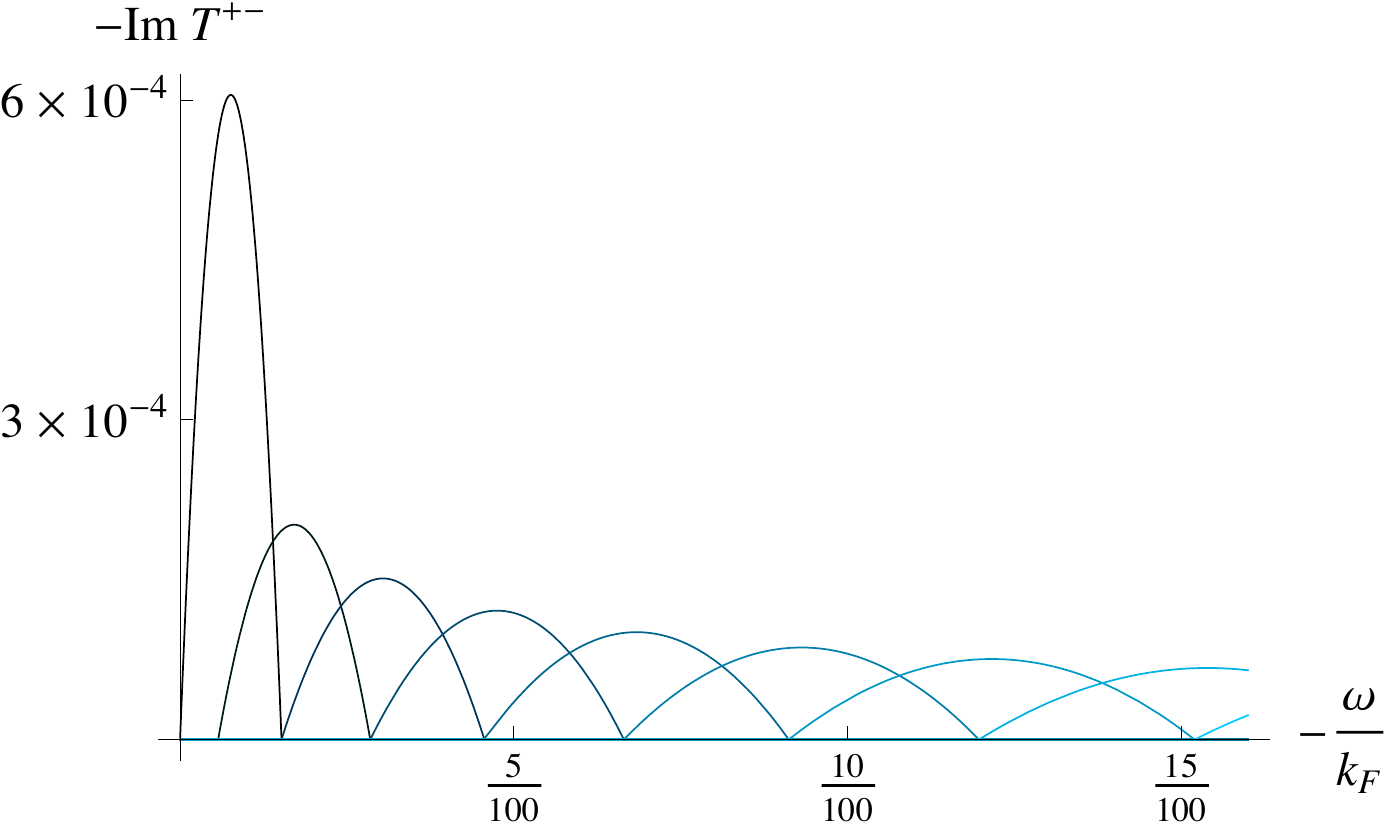}\hskip0.5cm\includegraphics[scale=0.56]{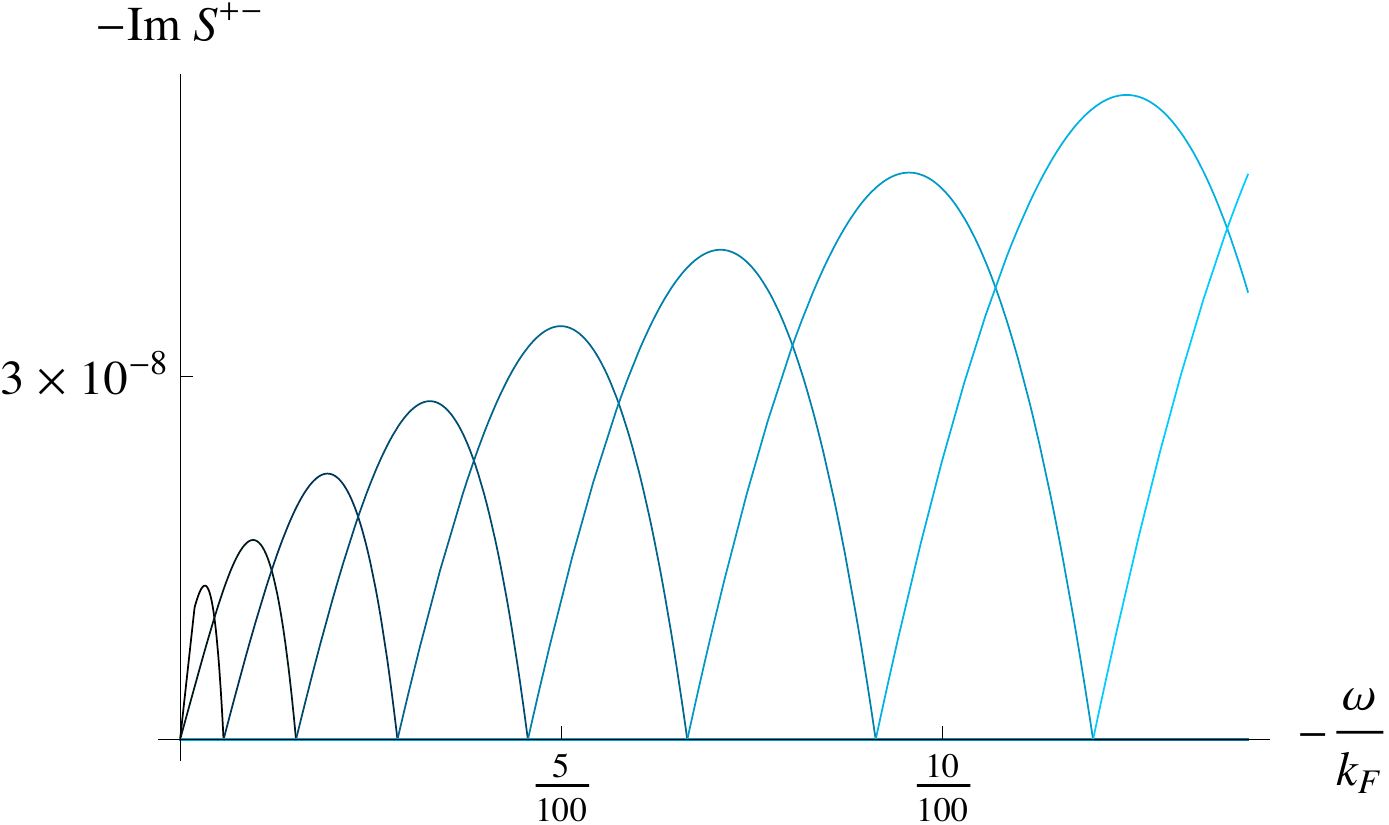}
\centerline{(a)\hskip8cm(b)}
\caption{Plots of the longitudinal and transverse spectral functions, shown as functions of  $-\omega/k_F$, in units of $m=1$. The curves with different colours represent spectral functions at different dimensionless momenta $y = |\vec{q}| / k_F$. In both plots the colours range from black ($y=1$) to progressively lighter blue colours (up to $y=9$) in steps of $1$. (a): The longitudinal component, $-\Im \CT^{+-}(\omega,|\vec{q}|)$. (b): The transverse component, $-\Im {\cal{S}^{+-}} (\omega,|\vec{q}|)$.}\label{tspmf2}
\end{figure}

\subsection{Excitations at zero temperature}
We begin a more detailed analysis of the current dynamics by first considering an ideal gas with $e=0$ at zero temperature, $T=0$. The retarded Green's function and its inverse can be presented in the hydrodynamical limit, discussed above, by the Laurent expansion in Fourier space. The {\it longitudinal} retarded Green's function,
\be
G_\ell^r=\frac{k_Fm}{\pi^2\hbar^2}\tilde G_\ell,
\ee
is given by the dimensionless function $\tilde G_\ell^r$, which assumes the following form in the IR limit,
\be\label{lgrfig}
\tilde G_\ell=\sum_{j,k=0}^\infty a_{\ell,j,k,0}(ix)^jy^{2k}.
\ee
The first few coefficients of the series are given in Table \ref{cghra}, with the actual order of the truncation of the series to be justified later, cf. the discussion after Eq. \eq{stflj}. The inverse Green's function can be written as
\be\label{glinv}
[G_\ell^r]^{-1}=\frac{\pi^2\hbar^2}{k_Fm}\tilde H_\ell,
\ee
and contains the dimensionless function
\be
\tilde H_\ell=\sum_{j,k=0}^\infty b_{\ell,j,k,0}(ix)^jy^{2k}.
\ee
The expansion coefficients $b_{\ell,j,k,0}$ are listed in Table \ref{cghrb}.

As discussed in Section \ref{sepctrfs}, the real part of $\omega(\vec{q})$ at which the inverse Green's function vanishes, gives the dispersion relation of the collective mode. We therefore need to look for $\tilde H_\ell = 0$. What we find is a strongly damped sound wave, propagating with velocity
\be
v^{(0)}_s=\lim_{\vec{q}\to0}\frac{\omega_\ell(\vec{q})}{|\vec{q}|}
=v_F\sqrt{\frac{b_{\ell,0,0,0}}{b_{\ell,2,0,0}}} \approx 0.83 \, v_F.
\ee

Fig. \ref{lqpigf} shows the exact longitudinal dispersion relation, $x_\ell(\vec{q})=m\omega_\ell(\vec{q})/|\vec{q}|k_F$, which was found numerically by plotting the line $\Re_x[G^r_\ell]^{-1}=0$ for the full retarded Green's function, given by Eqs. \eq{tsdiag}, \eq{tspm} and \eq{retccgf}. The analytical form of Eqs. \eq{lgrfig} reproduces the inverse of the full Green's function very accurately, given by Eqs. \eq{tsdiag} and \eq{tsndiag}, along with the sound wave dispersion relation for $y\ll1$. Note that the sound wave is within the hydrodynamical regime due to the fact that $b_{\ell,0,0,0}/b_{\ell,2,0,0}<1$.

\begin{figure}
\includegraphics[scale=0.5]{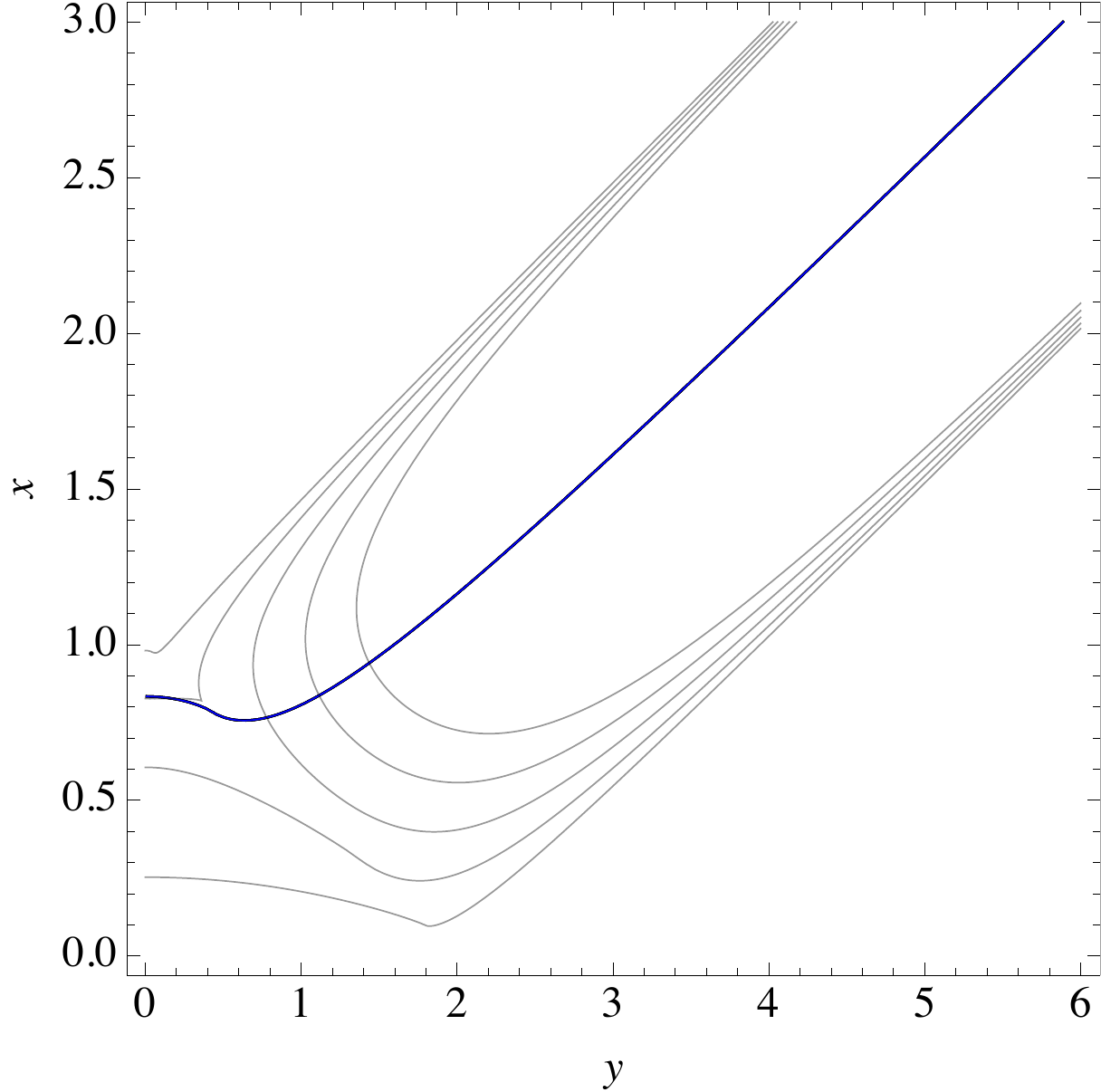}\caption{The blue solid line shows the full, numerically computed longitudinal collective mode dispersion relation, $x_\ell(\vec{q})$, for an ideal gas at metal density, as the function of $|\vec{q}|/k_F$. The black contour lines correspond to $\Im[G^r_\ell]^{-1}$, which reaches its minimum around $|\vec{q}|\sim0.7k_F$. The plot uses the dimensionless $(y,x)$ variables for the axes.}\label{lqpigf}
\end{figure}

\begin{figure}
\includegraphics[scale=0.5]{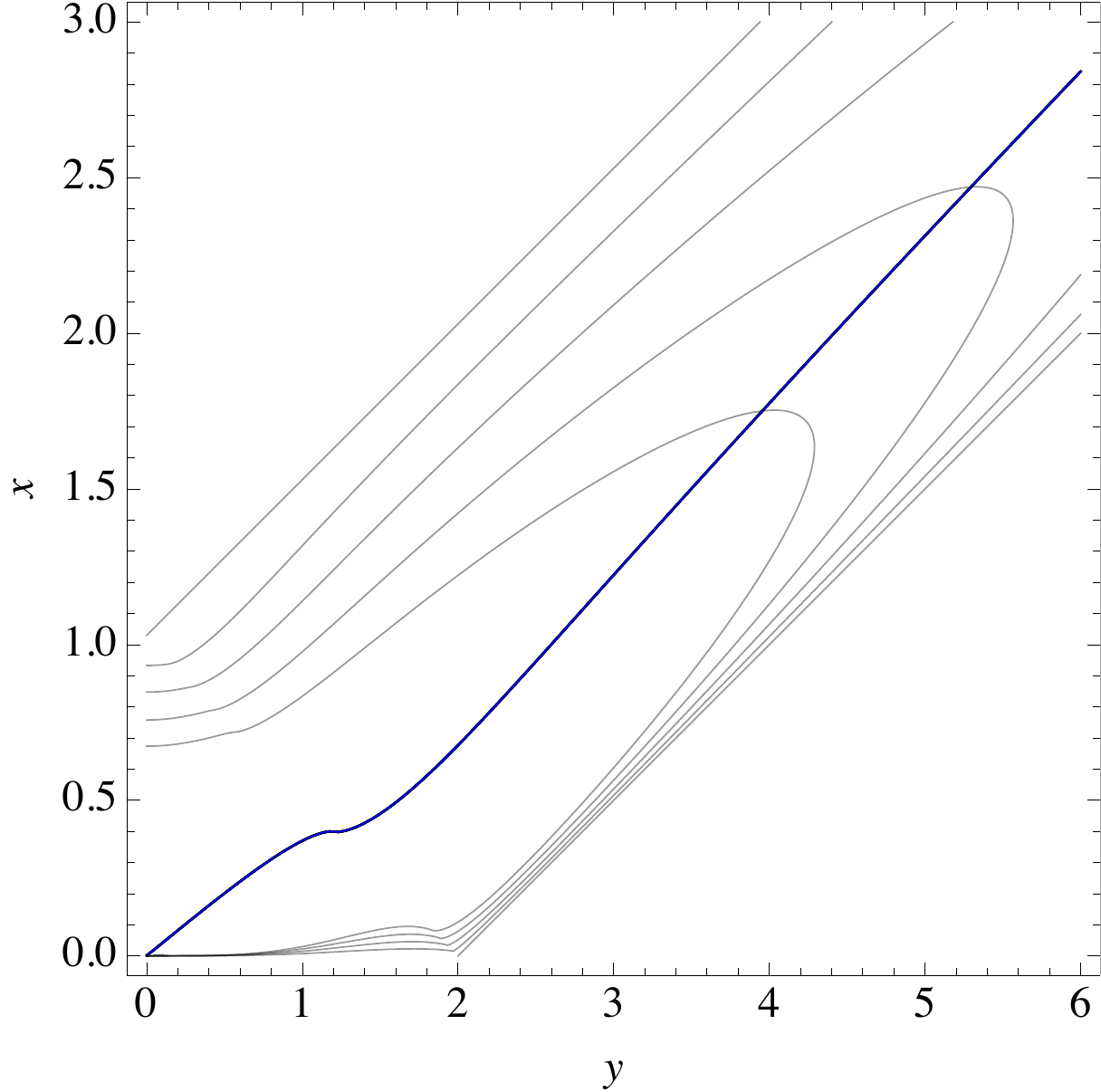}\caption{The plot depicts the same information as Fig. \ref{lqpigf}, except for the transverse collective modes. The imaginary part of the equation of motion, $\Im[G_t^\ell]^{-1}$ approaches 0 from below as $|\vec{q}|$ increases.}\label{tqpigf}
\end{figure}

One could na\"{i}vely expect that the transverse sector should display no collective modes since the solutions of the Schr\"odinger's equation only describe longitudinal waves. However, the collective mode spectrum, the curve $\Re_x[G^r_t]^{-1}=0$, obtained numerically from Eqs. \eq{tsdiag}, \eq{tspm} and \eq{retccgf} and plotted in Fig. \ref{tqpigf}, suggests the presence of quasi-particles with an approximatively free dispersion relation, in agreement with the well known transverse waves in dissipative fluids \cite{landau2013fluid}. This phenomenon may take place in an ideal gas by combining particle and hole plane wave modes, which are one-by-one longitudinal, but have a transverse component with respect to their total momentum. For the transverse Green's function, we find
\be
G_t^r=\frac{k_Fmv_F^2}{\pi^2\hbar^2}\tilde G_t,
\ee
with
\be
\tilde G_t=\sum_{j,k=0}^\infty a_{t,j,k,0}(ix)^jy^{2k},
\ee
in the infrared regime, cf. Table \ref{cghra}. The absence of the term $\ord{x^0y^0}$ signals that there are transverse collective modes in the infrared regime, but that their equation of motion is non-local. The dispersion relations in the longitudinal and the transverse sectors are similar for $y\gg1$, which is in agreement with the remarks made in Section \ref{sepctrfs}.

To conclude this sub-section, we note that the plan to derive hydrodynamical equations has surprisingly already been partially realised in the case of a zero-temperature electron gas at finite density. Indeed, the longitudinal sector possesses a hydrodynamical limit, it satisfies a local equation of motion in the IR and displays a collective mode for $|\vec{q}| \to 0$ with a linear dispersion relation, $\omega\sim|\vec{q}|$. The collective excitation is therefore a sound wave. However, because of the fact that it is composed out of non-interacting constituents, the mode is neither zero nor first sound. Instead, its existence stems from the composite nature of the operator generating it, i.e. the Noether current. This collective mode can thus be thought of as a {\it composite sound} mode consisting of infinitely many simple plane waves. To establish its existence, no additional thermodynamical considerations were needed as the assumption of local equilibrium is replaced by the infrared conditions needed to derive the local equations of motion. The transverse sector of the current has collective modes that can be approximated by the free dispersion relation, but their equation of motion is completely non-local.

\subsection{Excitations at small, non-zero temperature}
In order to recover local equation of motion for the transverse modes we bring the ideal gas into contact with a heat bath. Because the loop-integrals in Eqs. \eq{ctpblfsp} cannot be computed in closed form for arbitrarily large temperatures, we will restrict our attention to the low-temperature case with $T\ll\epsilon_F=\hbar^2k_F^2/2m$. Within this approximation, the discontinuous jump of the occupation number at the Fermi surface is slightly softened at low temperatures. Hence, the low-temperature expansion yields a power series in the dimensionless small parameter $z=\pi T/2\epsilon_F$, with expansion coefficients containing the derivatives of the zero-temperature result with respect to $k_F$. Non-zero temperature introduces a new length scale, the thermal wavelength $\lambda_T=\hbar\sqrt{2\pi/mT}$. The small parameter of the low-temperature expansion can then be written as $z=2\pi^2/(k_F\lambda_T)^2$. In terms of $\lambda_T$, we will be working in the regime of $\lambda_Tk_F\gg1$.

The low-temperature expansion of the Green's function, given by a one-loop Feynman diagram, produces artificial divergences where the spectral weight \eq{spfnct} is non-analytical. Even though it is reasonable to assume that these singularities are not present when the full temperature dependence is taken into account, we continue here with the inspection of the Green's function away from these singularities, only in the IR. The longitudinal and the transverse current-current Green's functions take the following expanded form at low temperature,
\be
\tilde G_o=\sum_{j,k,m=0}^\infty a_{o,j,k,m}(ix)^jy^{2k}z^{2m},
\ee
where the index $o$ may either take the value $\ell$ or $t$, indicating whether the Green's function is longitudinal or transverse. The constants, read off from Eqs. \eq{ltctpbl}, are again listed in Table \ref{cghra}. The different signs of $a_{\ell000}$ and $a_{\ell001}$, shown in the Table, reflect the fact that thermal fluctuations tend to weaken the polarisation of the Fermi sphere and that the low-temperature expansion breaks down when the $\ord{x^0y^0}$ term changes its sign. The inverse of the longitudinal Green's function can be written in the form of Eq. \eq{glinv}, while the transverse Green's function now takes the form
\be\label{gtinv}
[G_t^r]^{-1}=\frac{k_F\hbar^2}{mT^2} \tilde H_t,
\ee
with
\be
\tilde H_o=\sum_{j,k=0}^\infty b_{o,j,k}(ix)^jy^{2k},
\ee
where
\begin{align}
b_{\ell ,j,k}=\sum_{m=0}^\infty b_{\ell,j,k,m}z^{2m},&&b_{t,j,k}=\sum_{m=0}^\infty b_{t,j,k,m}z^{-2m}.
\end{align}
The coefficients at low orders of the expansion are collected in Table \ref{cghrb}. The zero-temperature longitudinal Green's function is invertible in the IR limit, enabling us to express its temperature dependence as a power series in $z$. Contrary to this situation, the presence of $T\neq 0$ temperature is essential for the inversion of the transverse Green's function, which can therefore be written as a power series in $1/z$.

\begin{table}
\caption{Coefficients of the retarded Green's functions up to terms $\ord{x^2}$, $\ord{y^4}$ and $\ord{T^2}$ in an ideal gas.}\label{cghra}
\begin{ruledtabular}
\begin{tabular}{lcccr}
$jk$&$a_{\ell,j,k,0}$&$a_{\ell,j,k,1}$&$a_{t,j,k,0}$&$a_{t,j,k,1}$\\
\hline
00&$-1$&$\frac16$&$0$&$\frac18$\\
10&$-\frac\pi2$&0&$\frac\pi4$&0\\
20&$-1$ &$-\hf$&$1$&$-\frac{13}{24}$\\
01&$\frac1{12}$&$\frac1{24}$&$\frac16$&$\frac1{288}$\\
11&0&0&$\frac\pi{16}$&0\\
21&$-\frac16$&$-\frac5{12}$&$\frac1{12}$&$\frac1{24}$\\
02&$\frac1{240}$&$\frac1{96}$&$-\frac1{60}$&$-\frac1{120}$\\
12&0&0&0&0\\
22&$-\frac3{80}$&$-\frac7{32}$&$\frac7{240}$&$\frac7{96}$
\end{tabular}
\end{ruledtabular}
\end{table}

\begin{table}
\caption{The same as Table \ref{cghra} for the inverse retarded Green's function.}\label{cghrb}
\begin{ruledtabular}
\begin{tabular}{lcccccr}
$jk$&$b_{\ell,j,k,0}$&$b_{\ell,j,k,1}$&$b_{t,j,k,0}$&$b_{t,j,k,1}$&$b_{t,j,k,2}$&$b_{t,j,k,3}$\\
\hline
00&$-1$&$-\frac16$&8&0&0&0
\\
10&$\frac\pi2$&$\frac\pi6$&0&$-16\pi$&0&0
\\
20&$1-\frac{\pi^2}4$&$\frac56-\frac{\pi^2}8$&$\frac{104}3$&-64&$32\pi^2$&0
\\
01&$-\frac1{12}$&$-\frac5{72}$&$-\frac29$&$-\frac{32}3$&0&0
\\
11&$\frac\pi{12}$&$\frac\pi{12}$&0&$-\frac{28\pi}9$&$\frac{128\pi}3$&0
\\
21&$\frac13-\frac{\pi^2}{16}$&$\frac{13}{18}-\frac{7\pi^2}{96}$&$-\frac{124}{27}$&$-\frac{848}9$&$\frac{512+40\pi^2}3$&$128\pi^2$
\\
02&$-\frac1{90}$&$-\frac1{45}$&$\frac{437}{810}$&$\frac{224}{135}$&$\frac{128}9$&0
\\
12&$\frac{7\pi}{480}$&$\frac{43\pi}{1440}$&0&$-\frac{263\pi}{135}$&$\frac{128\pi}{45}$&$-\frac{256\pi}3$
\\
22&$\frac{17}{180}-\frac{13\pi^2}{960}$&$\frac{29}{72}-\frac{167\pi^2}{5760}$&$\frac{149}{810}$&$\frac{1864}{135}$&$\frac{7552}{45}+\frac{974\pi^2}{135}$&$\frac{1024}3+\frac{1664\pi^2}{45}$
\end{tabular}
\end{ruledtabular}
\end{table}

We have now all of the necessary ingredients to return to the problem mentioned in Section \ref{sepctrfs}, namely, the understanding of the analytical properties of the spectral weight, needed to define simple collective modes. The point is that the spectral functions of an ideal gas are indeed analytic in the sufficiently small vicinity of $x=y=0$. If the external source $a^\mu(p)$ is an analytic function and is negligible beyond the region of analyticity of the spectral weights, then the approximation of the Green's functions, which is based on the analytical, small $x$ and $y$ behaviour, is justified. 

The poles of the truncated Green's functions clearly depend on the level of truncation. The truncation dependence is stronger in the transverse sector where the inversion of the Green's function is possible only at finite temperature. The normal mode that arises from the $\ord{x}$ truncation is
\be\label{fordtrm}
x=-\frac{i}{2\pi}z^2+\ord{y^2},
\ee
which is to be contrasted with the $\ord{x^2}$ result,
\be\label{sordtrm}
x=\frac{3i\pi\pm\sqrt{27\pi^2-72+39z^4}}{24-12\pi^2-13z^4}z^2+\ord{y^2}.
\ee
In other words, the transverse normal mode is fully damped at the $\ord{x}$ truncation as can be seen from \eq{fordtrm}. On the other hand, the $\ord{x^2}$ truncation gives rise to a damped, but also oscillatory normal mode. As for the truncation of the series with respect to $y$, we note that the qualitative behaviour of the normal modes is already stable at the $\ord{y^2}$ truncation. The $\ord{x^2}$ dispersion relation qualitatively follows the numerical curve and is quantitatively wrong by a factor of approximately 2.5 for $0<y<1$. The truncations at the level of $\ord{x^2}$ and $\ord{y^4}$, used for simplicity in all of the Tables and Figures, thus already provide a qualitatively correct description of the dispersion relations. Since $\left|\Re [x]\right|<1$ at low temperature, the transverse sound wave is within the hydrodynamical regime. If we had used higher orders of $x$, the functions \eq{lfuncdef} and \eq{ilfuncdef} could be approximated with even better (quantitative) precision and the plots of the truncated Laurent series would lie closer to those of the exact linearised equations of motion.

The real and imaginary parts of the normal mode frequencies, $\omega_\pm(\vec{q})=\pm\omega_r(\vec{q})+i\omega_i(\vec{q})$, are plotted in Figs. \ref{polex} for $z=0.87$. The real parts were found to follow a quadratic dependence in the wave vector, $\omega^{(\ell)}_r(\vec{q})\sim\omega_{0\ell}-\hbar\vec{q}^2/2m_\ell^*$ and $\omega^{(t)}_r(\vec{q})\sim\omega_{0t}+\hbar\vec{q}^2/2m_t^*$, respectively, with $m_\ell^*,m_t^*>0$. An upper bound on the radius of convergence for the analytic form of Eqs. \eq{glinv} and \eq{gtinv} is given by $|\omega_+(\vec{q})|$. The collective mode \eq{sordtrm} is therefore a strongly damped, collective sound wave with both the speed of propagation and the inverse lifetime of the order $\ord{T^2}$. The damping of the normal modes makes condition (II), mentioned at the beginning of this Section, satisfied in the effective theory.

\begin{figure}
\includegraphics[scale=0.5]{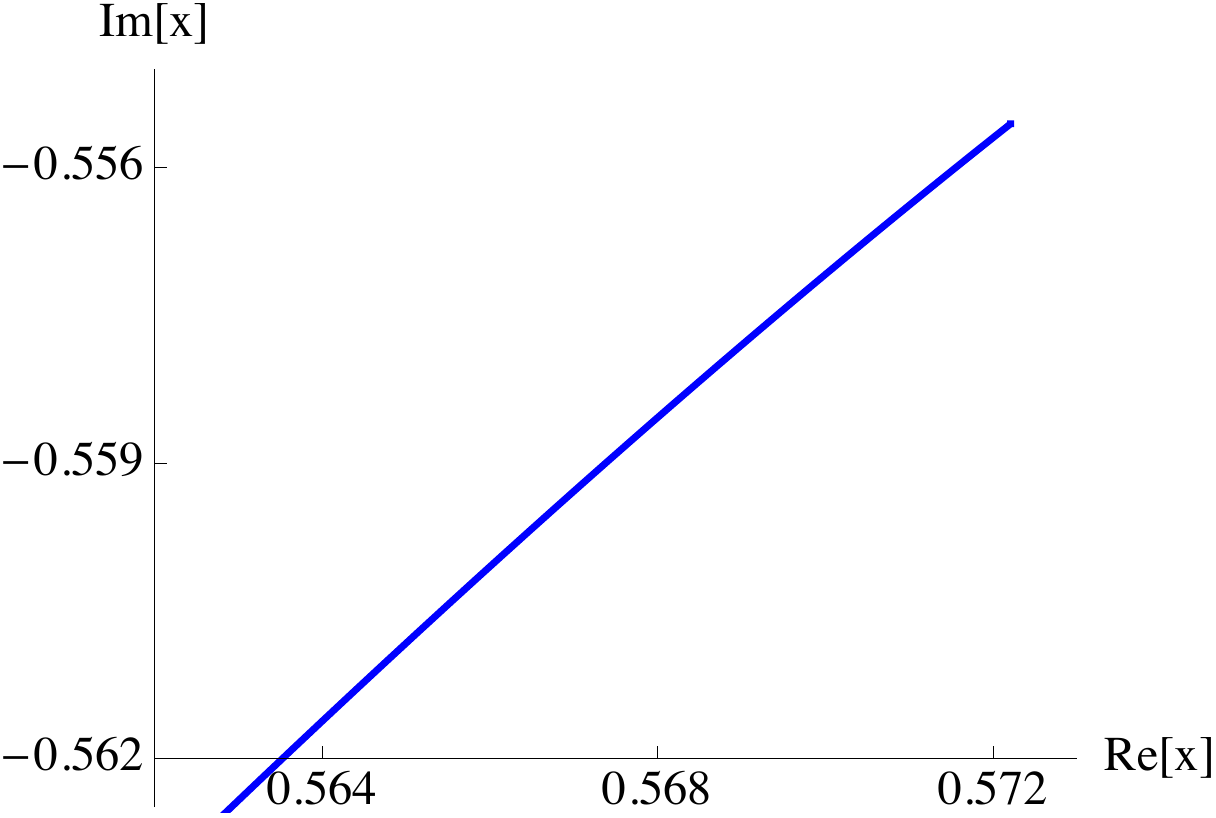}\hskip2cm
\includegraphics[scale=0.5]{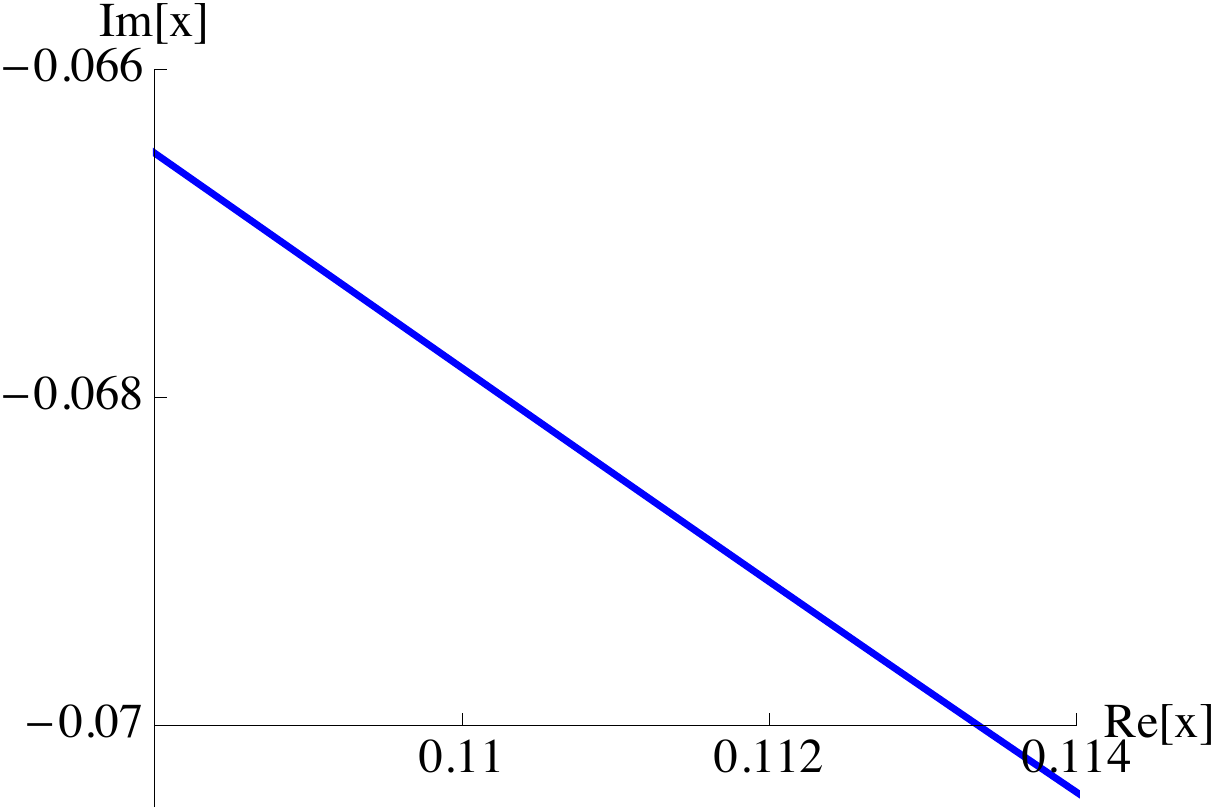}
\centerline{(a)\hskip8cm(b)}
\caption{The dispersion relation of the normal modes on the complex $x=\hbar m\omega_+(\vec{q})/k_F|\vec{q}|$ plane. (a): The longitudinal mode for $0<y<0.4$. (b): The transverse mode for $0<y<0.2$.}\label{polex}
\end{figure}

\subsection{Excitations in the presence of interactions}\label{ints}

In this section, we will include the photon-fermion one-loop electromagnetic interactions of the type depicted in Fig. \ref{fig:Ring2} into the calculation to show that not all interactions are able to change the qualitative behaviour of the non-interacting gas. This can be achieved by modifying the linear operator $\cal L$ from Eq. \eq{klineom} to contain the Coulomb interactions in its longitudinal component, while the contributions of the interactions to the transverse sector are negligible in the non-relativistic limit. We can further improve the precision of the one-loop calculation by making use of the full one-loop re-summed photon propagator, given by Eq. \eq{photpr} and represented diagrammatically in Fig. \ref{fig:SDresum}. In this work, we will not include the vertex corrections nor the electron self-energy to the current-current two-point function, which would qualitatively change the behaviour of the electron gas and bring its dynamics closer to the usually discussed regime. 

The Schwinger-Dyson re-summation results in the replacement of $[G^r_\ell]^{-1}$ by
\be\label{longeomi}
{\cal L}_\ell=[G^r_\ell]^{-1}-\frac{e^2}{\vec{q}^2-e^2G^r_\ell}.
\ee
The Laurent expansion now yields
\be
\CL_\ell=\frac{\pi^2\hbar^2}{k_Fm}\sum_{jkmn=0}^\infty b_{\ell,j,k,m,n}(ix)^jy^{2k}z^{2m}(\pi^2a_0k_F)^n,
\ee
showing the emergence of a new length scale related to the Coulomb interactions. The coefficients $b_{\ell,j,k,m,n}$ are listed in Table \ref{clin}. We note that the partial Schwinger-Dyson re-summation of the photon propagator produces a non-perturbative result at finite temperature and density. Furthermore, the longitudinal equation of motion changes at the $\CO(\vec{q}^0)$-order due to the expression \eq{longeomi}. It changes the equation by taking ${\cal L}_\ell\to2{\cal L}_\ell$. Intriguingly, the Coulomb interactions only introduce minor quantitative changes to the equations of motion and the sound wave remains the only collective mode at long wavelengths. The speed of sound is higher than in the ideal gas while the damping is only weakly influenced by the instantaneous Coulomb interactions, cf. Fig. \ref{poleemlt}. 

The inclusion of other one-loop corrections, the self-energy of electrons and the vertex corrections, would qualitatively change the hydrodynamical regime. They would generate a finite life-time and render the infrared limit, $\omega \to 0$, $|\vec{q}|\to0$, of the loop integral \eqref{ctpblfsp} well defined. The loop integral would include an improved electron propagator whose complex self-energy would regulate the $1/|\vec{q}|$ singularities, thereby restricting the ideal gas hydrodynamical regime to
\be
{\cal D}_{hydr}=\left\{(y,x)\, \big| \, \left| |x|-y/2 \right|<1 \wedge y>1/k_Fr_{mfp} \right\},
\ee
instead of the domain presented in Eq. \eqref{domain}. Here, $r_{mfp}$ denotes the mean free path. These types of interactions open the way to the traditional phenomenological approach, which is based on the hydrodynamical regime with $y<1/k_Fr_{mfp}$ and an arbitrary $x$.

\begin{table}
\caption{Coefficients of the longitudinal equation of motion of the Coulomb gas up to terms $\ord{x^2}$, $\ord{y^4}$ and $\ord{T^2}$.}\label{clin}
\begin{ruledtabular}
\begin{tabular}{lcccccr}
$jk$&$b_{\ell,j,k,0,0}$&$b_{\ell,j,k,1,0}$&$b_{\ell,j,k,0,1}$&$b_{\ell,j,k,1,1}$&$b_{\ell,j,k,0,2}$&$b_{\ell,j,k,1,2}$\\
\hline
00&$-2$&$-\frac13$&0&0&0&0
\\
10&$\frac\pi2$&$\frac\pi6$&0&0&0&0
\\
20&$1-\frac{\pi^2}4$&$\frac56-\frac{\pi^2}8$&0&0&0&0
\\
01&$-\frac1{12}$&$-\frac5{72}$&1&$\frac13$&0&0
\\
11&$\frac\pi{12}$&$\frac\pi{12}$&0&0&0&0
\\
21&$\frac13-\frac{\pi^2}{16}$&$\frac{13}{18}-\frac{7\pi^2}{96}$&0&0&0&0
\\
02&$-\frac1{90}$&$-\frac1{45}$&0&0&-1&-$\hf$
\\
12&$\frac{7\pi}{480}$&$\frac{43\pi}{1440}$&0&0&0&0
\\
22&$\frac{17}{180}-\frac{13\pi^2}{960}$&$\frac{29}{72}-\frac{167\pi^2}{5760}$&0&0&0&0
\end{tabular}
\end{ruledtabular}
\end{table}

\begin{figure}
\vskip0.5cm
\includegraphics[scale=0.6]{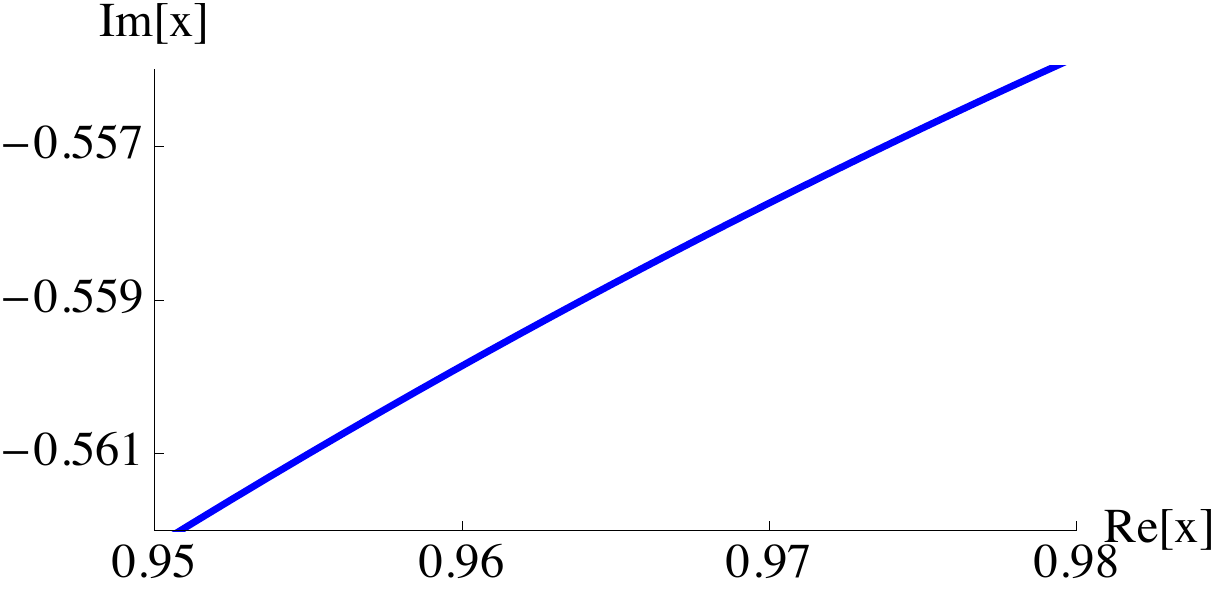}
\caption{The dispersion relation for the longitudinal mode in an interacting electron gas, plotted on the complex $x=\hbar m\omega_+(\vec{q})/k_F|\vec{q}|$ plane. The range of $y$ is the same as in Fig. \ref{polex}, $0<y<0.4$.}\label{poleemlt}
\end{figure}

\subsection{Equation of motion in momentum space}\label{ref:EoMinFS}

To write down the equation of motion, we use the deviation of the density from the homogeneous value, $n(q)=\left[j^0(q)-j^0(0) \right]/c$, split the current into longitudinal and transverse parts, $\vec{j}=\vec{j}_\ell+\vec{j}_t$, where $\vec{\nabla}\cdot\vec{j}_t=0$, giving us
\be\label{leomf}
-\frac{v_F}{\pi^2\hbar}\phi=\left[\sum_{j,k=0}^\infty b_{\ell,j,k}\left(\frac{i\omega}{v_F|\vec{q}|}\right)^j\left(\frac{\vec{q}^2}{k_F^2}\right)^k\right]n,
\ee
and
\be\label{teomf}
\frac{T^2}{v_F\hbar}\vec{a}=\left[\sum_{j,k=0}^\infty b_{t,j,k}\left(\frac{i\omega}{v_F|\vec{q}|}\right)^j\left(\frac{\vec{q}^2}{k_F^2}\right)^k\right]\vec{j}_t.
\ee
The temperature and the coupling constant dependence are included in the coefficients $b_{\ell jk}$ and $b_{tjk}$. As a reminder, we are using the parametrisations $a^\mu=(\phi/c,\vec{a})$ and $J^\mu=(\rho c,\vec{j})$ of the external source $a^\mu$ and the vacuum expectation value of the Noether current. 

The continuity equation, $\partial_tn+\vec{\nabla}\cdot\vec{j}=0$, can now be used to rewrite the equation of motion for the full current in the form of
\be
\frac{T^2}{v_F\hbar}\vec{a}-\frac{b_{t,2,0}}{b_{\ell,2,0}}\frac{v_F}{\pi^2\hbar}\frac{\omega\vec{q}}{\vec{q}^2}\phi=\left[\sum_{j,k=0}^\infty b_{t,j,k}\left(\frac{i\omega}{v_F|\vec{q}|}\right)^j\left(\frac{\vec{q}^2}{k_F^2}\right)^k+\frac{\vec{q}\otimes\vec{q}}{\vec{q}^2}\sum_{j,k=0}^\infty b_{j,k}\left(\frac{i\omega}{v_F|\vec{q}|}\right)^j\left(\frac{\vec{q}^2}{k_F^2}\right)^k\right]\vec{j},
\ee
where the coefficients $b_{j,k}$ are given by
\be
b_{j,k}=\frac{b_{t,2,0}}{b_{\ell,2,0}}b_{\ell,j,k}-b_{t,j,k}.
\ee
The $b_{j,k}$ have been chosen so as to have a simple $\ord{\omega^2}$ piece. As before, the $\vec{q}\otimes\vec{q}$ notation indicates the Kronecker product, giving us a matrix acting on $\vec{j}$. 

It is important to note that the linearised equation of motion breaks time reversal invariance. The pieces with even and odd powers of $i\omega$ arise from the $\Re S_1$ and $\Re S_2$, respectively, in terms of the classification introduced in Section \ref{couplingtas}, cf. Eq. \eq{leffactpm}. The finite strength of ${\cal L}^f$ in the linearised equation of motion generates true irreversibility because ${\cal L}^f$ is odd under time reversal. Its presence leads to damping and relaxation to equilibrium, which are reflected in the non-conservation of the energy-momentum tensor, associated with the gas \cite{Grozdanov:2013dba}.

It is instructive to look for the stationary transverse flow with $\omega \propto x = 0$ and $j^\mu=(0,j(q_y),0,0)$, which is driven by $a^\mu=(0,a(q_y),0,0)$. It satisfies the equation of motion
\be\label{statem}
\frac{T^2}{v_F\hbar}a(q)=\left(b_{t,0,0}-|b_{t,0,1}|\frac{q^2}{k_F^2}+b_{t,0,2}\frac{q^4}{k_F^4}\right)j(q),
\ee
with the solution
\be\label{stflj}
j(q)=-\frac{T^2}{v_F\hbar}\frac{a(q)}{(q^2-q^2_+)(q^2-q^2_-)},
\ee
where
\be\label{qpm}
q^2_\pm=k_F^2\frac{|b_{t,0,1}|}{2b_{t,0,2}}\left(1\pm\sqrt{1-4\frac{b_{t,0,0}}{|b_{t,0,1}|}}\right).
\ee
The argument of the square root is negative and therefore the stationary flow is a Gaussian wave packet in coordinate space with the width $\Delta x=1/|\Im q_\pm|$. Note that the role of the driving force of the flow, the pressure in hydrodynamics, is played by the external source $a^\mu$ in our scheme. The real and the imaginary parts of the wave vector are shown in units of $k_F$ as functions of the temperature in Fig. \ref{poleq}. They characterise the characteristic length scale of the flow pattern and the dissipation of the flow, respectively. The lesson here is that the inverses of both length scales increase linearly with the temperature when the thermal energy is well below the Fermi energy.

\begin{figure}
\includegraphics[scale=0.6]{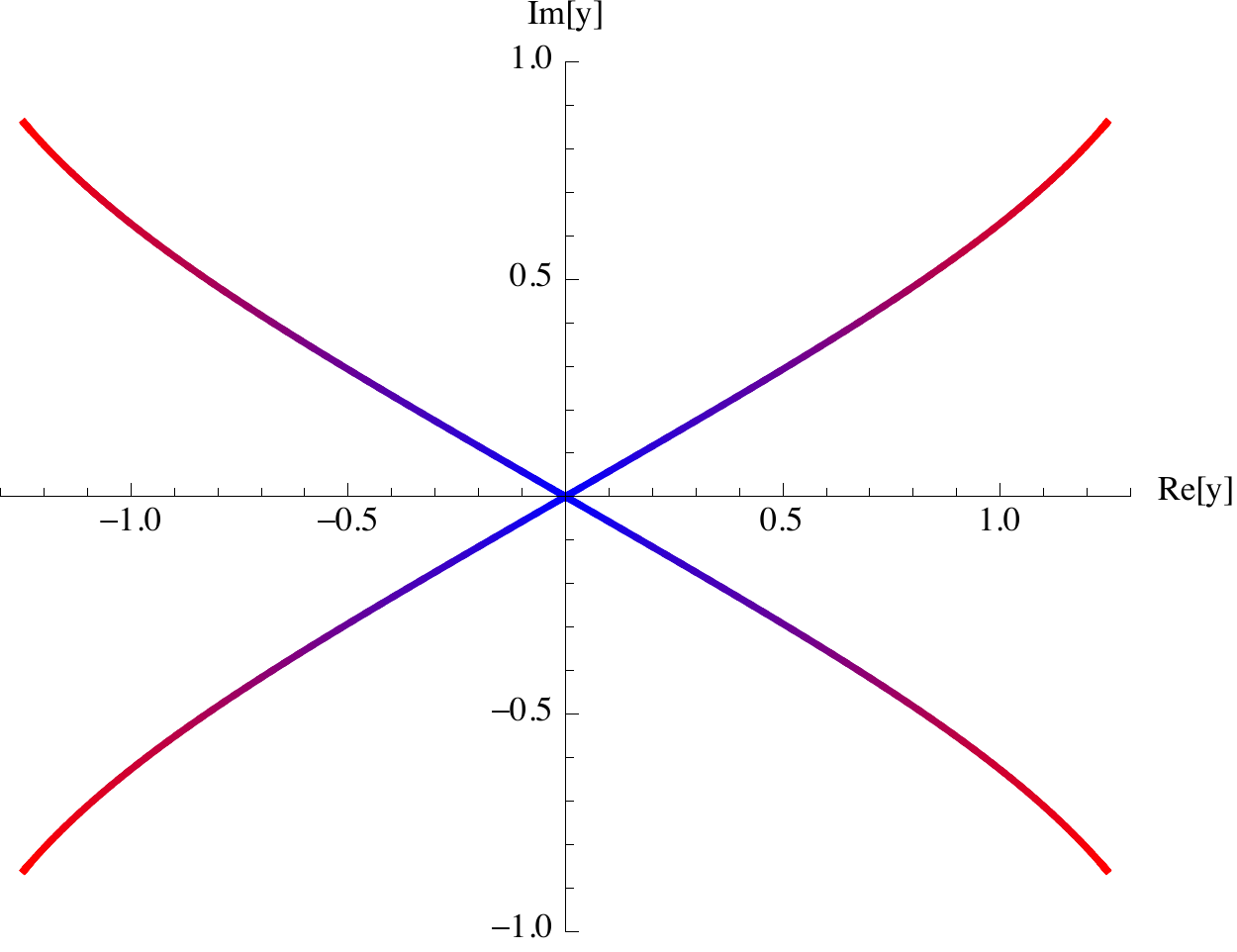}
\caption{The location of the poles of the transverse current in Eq. \eq{stflj} on the complex $y$ plane, at $x=0$, plotted as the functions of the temperature $T$, in units of $m=1$ and $k_F = 3.8 \times 10^{-3}$. The colouring indicates the values of the temperature that grows from blue to red colour. At the origin of the $y$ plane, $T=0$ and we use the blue colour. The endpoint red colour is at $T=10^{-5}$.}\label{poleq}
\end{figure}

\subsection{Equation of motion in space-time}\label{Sec:EoMSpaceTime}
It is easy to find the equations of motion in space-time. All we need is to work out the operators that correspond to the multiplicative factors $Q^{(n)}=|\vec{q}|^n$, acting on the coordinate-dependent functions. Straightforward integration yields
\begin{align}
&Q^{(1)}(\vec{x},\vec{y}) = -P\frac1{\pi^2(\vec{x}-\vec{y})^4}, \label{Qker1}\\
&Q^{(-1)}(\vec{x},\vec{y})=P\frac1{2\pi^2(\vec{x}-\vec{y})^2}, \label{Qker2}\\
&Q^{(-2)}(\vec{x},\vec{y})=P\frac1{4\pi|\vec{x}-\vec{y}|}. \label{Qker3}
\end{align}
The short distance singularities of the Fourier transformation can be regulated by adding an infinitesimal imaginary piece to the radius, which amounts to the use of the principal value prescription in carrying out the integrals.

The equations of motion, written as
\begin{align}\label{lineom}
-\frac{v_F}{\pi^2\hbar} \, \phi={\cal L}_\ell \, n,&&  \tilde{\vec{a}}={\cal L} \, \vec{j},
\end{align}
cf. Eqs. \eq{leomop} and \eq{teomop}, have
\bea\label{eomspt}
{\cal L}_\ell&=&b_{\ell,0,0}-\frac{b_{\ell,1,0}}{v_F}Q^{(-1)}\partial_t+\frac{b_{\ell,2,0}}{v_F^2}Q^{(-2)}\partial_t-\frac{b_{\ell,0,1}}{k_F^2}\Delta-\frac{b_{\ell,1,1}}{v_Fk_F^2}Q^{(1)}\partial_t\nn
&&+\frac{b_{\ell,2,1}}{v_F^2k_F^2}\partial_t^2+\frac{b_{\ell,0,2}}{k_F^4}\Delta^2+\frac{b_{\ell,1,2}}{v_Fk_F^4}Q^{(1)}\Delta\partial_t-\frac{b_{\ell,2,2}}{v_F^2k_F^4}\Delta\partial_t^2+ \cdots,
\eea
\bea\label{eomspt2}
{\cal L}^{ij}&=&\delta^{ij}\biggl[b_{t,0,0}-\frac{b_{t,1,0}}{v_F}Q^{(-1)}\partial_t+\frac{b_{t,2,0}}{v_F^2}Q^{(-2)}\partial_t-\frac{b_{t,0,1}}{k_F^2}\Delta-\frac{b_{t,1,1}}{v_Fk_F^2}Q^{(1)}\partial_t+\frac{b_{t,2,1}}{v_F^2k_F^2}\partial_t^2+\frac{b_{t,0,2}}{k_F^4}\Delta^2  \nn
&&+\frac{b_{t,1,2}}{v_Fk_F^4}Q^{(1)}\Delta\partial_t-\frac{b_{t,2,2}}{v_F^2k_F^4}\Delta\partial_t^2\biggr]\nn
&&+\nabla^i\nabla^jQ^{(-2)}\biggl(-\frac{b_{1,0}}{v_F}Q^{(-1)}\partial_t+\frac{b_{2,0}}{v_F^2}Q^{(-2)}\partial_t-\frac{b_{0,1}}{k_F^2}\Delta-\frac{b_{1,1}}{v_Fk_F^2}Q^{(1)}\partial_t\nn
&&+\frac{b_{2,1}}{v_F^2k_F^2}\partial_t^2+\frac{b_{0,2}}{k_F^4}\Delta^2+\frac{b_{1,2}}{v_Fk_F^4}Q^{(1)}\Delta\partial_t-\frac{b_{2,2}}{v_F^2k_F^4}\Delta\partial_t^2\biggr) + \cdots,
\eea
and
\be
\tilde{\vec{a}}=\frac{T^2}{v_F\hbar}\vec{a}-\frac{b_{t,2,0}}{b_{\ell,2,0}}\frac{v_F}{\pi^2\hbar}Q^{(-2)}\vec{\nabla}\partial_t\phi.
\ee

The linear response formulae for the components of the current, generated by an external electromagnetic field, which normally lead to the Kubo formulae, give $J^\mu$ in terms of $a^\mu$. Our linearised equations of motion, i.e. Eq. \eq{lineom}, are similar to the Kubo-type equations, except that we now have $a^\mu$ expressed in terms of $J^\mu$. The inversion of the kernel of the linear response formulae provides an additional piece of information, beyond the Kubo formulae, namely the linearised equation of motion for $J^\mu$, which is needed to make contact with the hydrodynamical evolution equations. As long as this equation can be considered local in time, it characterises the dynamics of the electron gas alone for $t>0$, that is after the external source, $a^\mu$, has been switched off, as explained in Section \ref{fdts}.

The kernels of the linearised equations of motion are given up to $\ord{x^2}$ in Eqs. \eqref{eomspt} and \eqref{eomspt2}. When higher orders are retained, these kernels contain non-local terms. In fact, the $\ord{x^n}$ contributions to the kernels are given in terms of the Fourier transforms of $|\vec{q}|^{n'}$, with $n'>-n$. This results in the appearance of the smearing kernels $Q^{-n'}$, which are of the order $\ord{|\vec{x}-\vec{y}|^{-n'-3}}$, for $n'\ne-3$, and proportional to $\ln|\vec{x}-\vec{y}|$, for $n'=-3$, cf. Eqs. \eqref{Qker1}-\eqref{Qker3}. The matrix elements of the operators $Q^{-n'}$, with $n'<-2$, grow with spatial separation at a fixed frequency. However, such a non-locality is suppressed by the frequency in ${\cal D}_{hydr}$, see Eq. \eqref{domain}, where the matrix elements remain bounded. Systematical and convergent improvement of the equation of motion, within the hydrodynamical regime, can thus be achieved by including higher powers of $x$.

As an aside, we note that the long-range correlations, experienced by the composite sound modes, are reminiscent of the gluon dynamics in the confining vacuum of Yang-Mills theories where the single gluon modes develop strong correlations beyond a characteristic distance scale and glueballs are formed. Note that $Q^{-4}(\vec{x},\vec{y})$ grows linearly with $|\vec{x}-\vec{y}|$ so that the composite sound is localised with a linear potential at this order of the Laurent series. Higher orders generate correlations that increase even faster with spatial separation.

Returning to the dynamics of the electron gas, the boost invariance requires special care because the truncation of the expansion of the effective action in the fields breaks the Galilean boost symmetry. A minimal extension that can be understood as a partial re-summation of the functional Taylor expansion to reinforce the Galilean boost-invariance, is provided by the replacement of the time derivative by a ``covariant'', convective derivative $\partial_t\to\partial_t+\vec{v} \cdot \vec{\nabla}$. It results in the replacement
\be\label{MinExtNonLinerDer}
\partial_t\to\partial_t+\frac{\vec{j}\cdot\vec{\nabla}}{\rho_0+n}  ,
\ee
in our case with $\rho_0=k_F^3/3\pi^2$, and makes the equations of motion non-linear. The novel feature of these equations, compared to phenomenological hydrodynamics, is the presence of the $Q^{(n)}$ factors and the two constants $b_{\ell,0,0}$ and $b_{t,0,0}$.

It is interesting to again, as in Section \ref{ref:EoMinFS}, consider a {\it stationary} flow, this time in position space. The linearised equation of motion operator for this type of a flow takes the form
\be
{\cal L}^{ij}_{stat}=\delta^{ij}\left(b_{t,0,0}-\frac{b_{t,0,1}}{k_F^2}\Delta+\frac{b_{t,0,2}}{k_F^4}\Delta^2\right)+\nabla^i\nabla^j  Q^{(-2)}\left(b_{0,0}-\frac{b_{0,1}}{k_F^2}\Delta+\frac{b_{t,0,2}}{k_F^4}\Delta^2\right).
\ee
The appearance of the corresponding operator of the Navier-Stokes equation,
\be
{\cal L}^{ij}_{NS,stat}=-\eta\delta^{ij}\Delta-\left(\zeta + \frac\eta3\right)\nabla^i\nabla^j,
\ee
shows the presence of shear viscosity and bulk viscosity terms, which are smeared in space. This is of course not surprising from the point of view of an effective field theory.

Finally, we can look for the flow corresponding to a homogeneous external field, $\tilde{\vec{a}}'=(a,0,0)$, $\phi=0$, assuming the form of $\vec{j}=(j(y),0,0)$ and $n=0$. The solution of the equation of motion is
\begin{align}\label{jSolAnyy4Trunc}
j=\frac{a}{b_{t,0,0}}+\sum_{\sigma=\pm}[j^{(+)}_\sigma(e^{\sigma q_+y}+e^{\sigma q_-y})+ij^{(-)}_\sigma(e^{\sigma q_+y}-e^{\sigma q_-y})].
\end{align}
It contains two undetermined real parameters $j^{(\pm)}_\sigma$ and is valid at any order of the truncation in $y$. The constants $q_\pm$ were defined in Eq. \eqref{qpm}. The necessity to retain at least the $\ord{y^4}$ terms in the above equation of motion can be seen by noting that the $\ord{y^2}$ equation is solved by
\begin{align}\label{jSolQuadTrunc}
j=\frac{a}{b_{t,0,0}}+j_+\cos \left[\sqrt{\frac{b_{t00}}{b_{t01}}}k_Fy \right] +j_-\sin \left[\sqrt{\frac{b_{t00}}{b_{t01}}}k_Fy \right].
\end{align}
The current in Eq. \eqref{jSolQuadTrunc} is purely oscillating. The level of the truncation thus clearly influences the position of the poles of the Green's function and changes both qualitative and quantitative features of the solution. While quadratic truncation leads to a reversible oscillating current in \eqref{jSolQuadTrunc}, the quartic truncation introduces an imaginary part into the poles (cf. Fig. \ref{poleq}) leading to damping.

\begin{figure}
\includegraphics[scale=0.53]{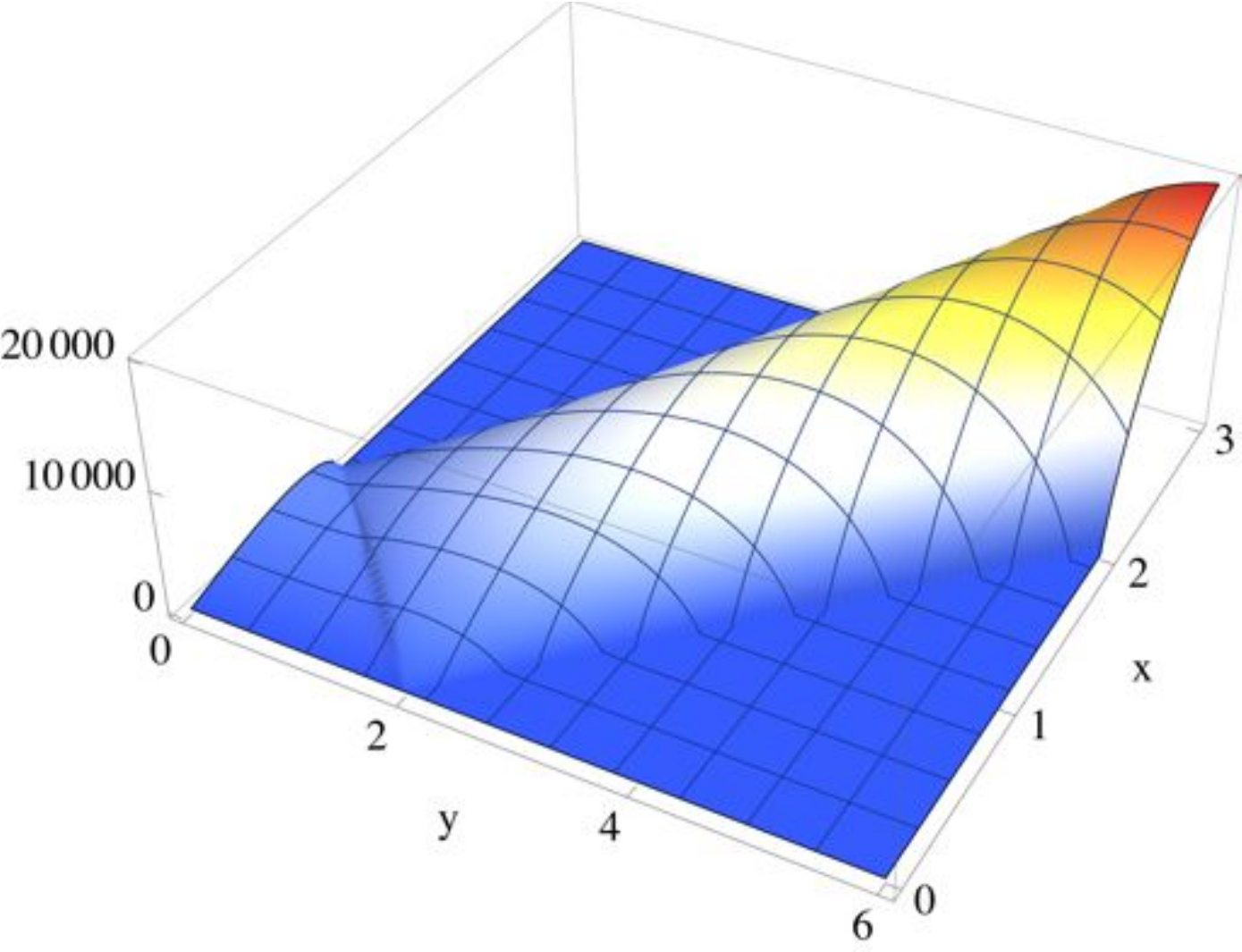}\hskip1.2cm
\includegraphics[scale=0.53]{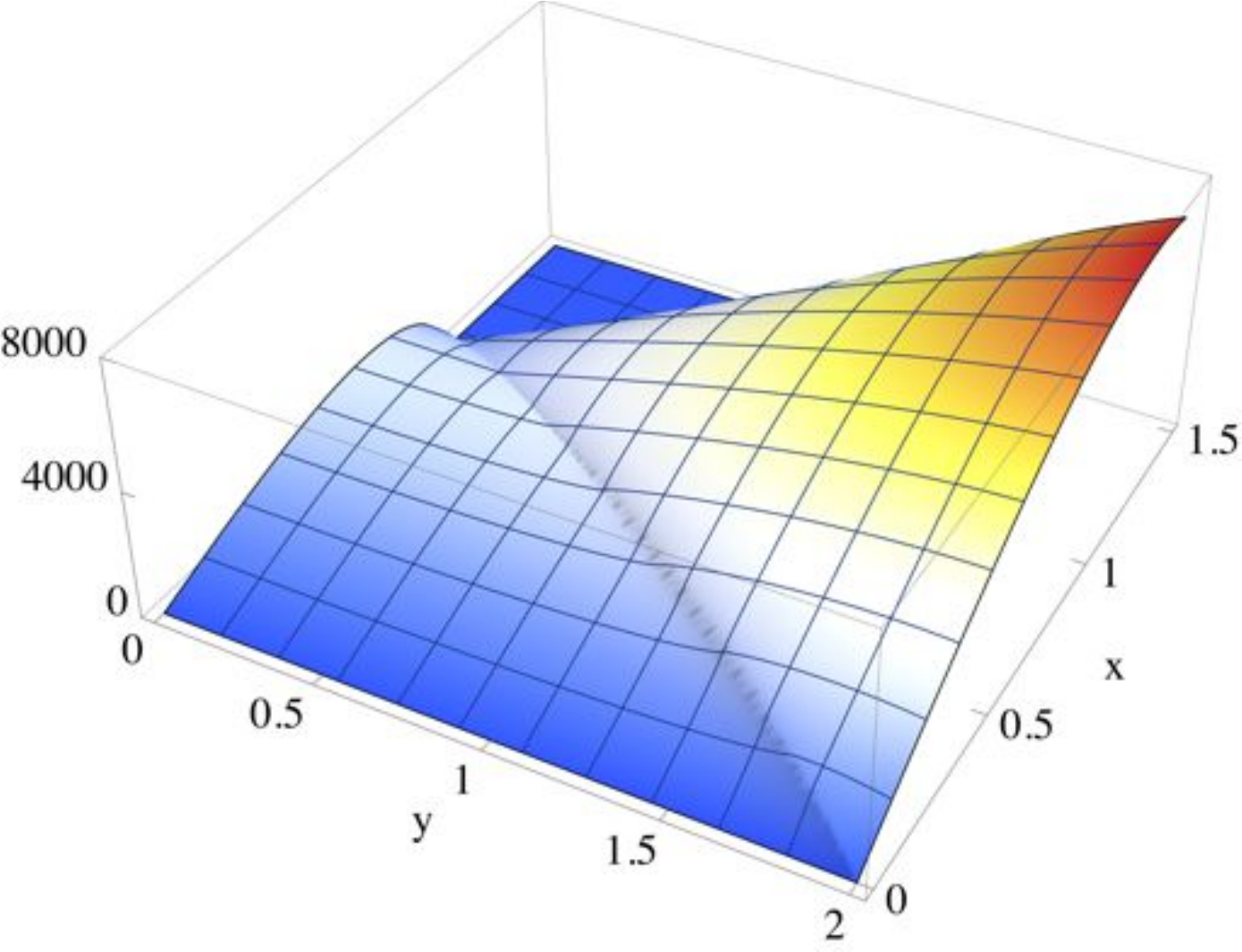}
\centerline{(a)\hskip9cm(b)}
\vskip0.6cm
\includegraphics[scale=0.53]{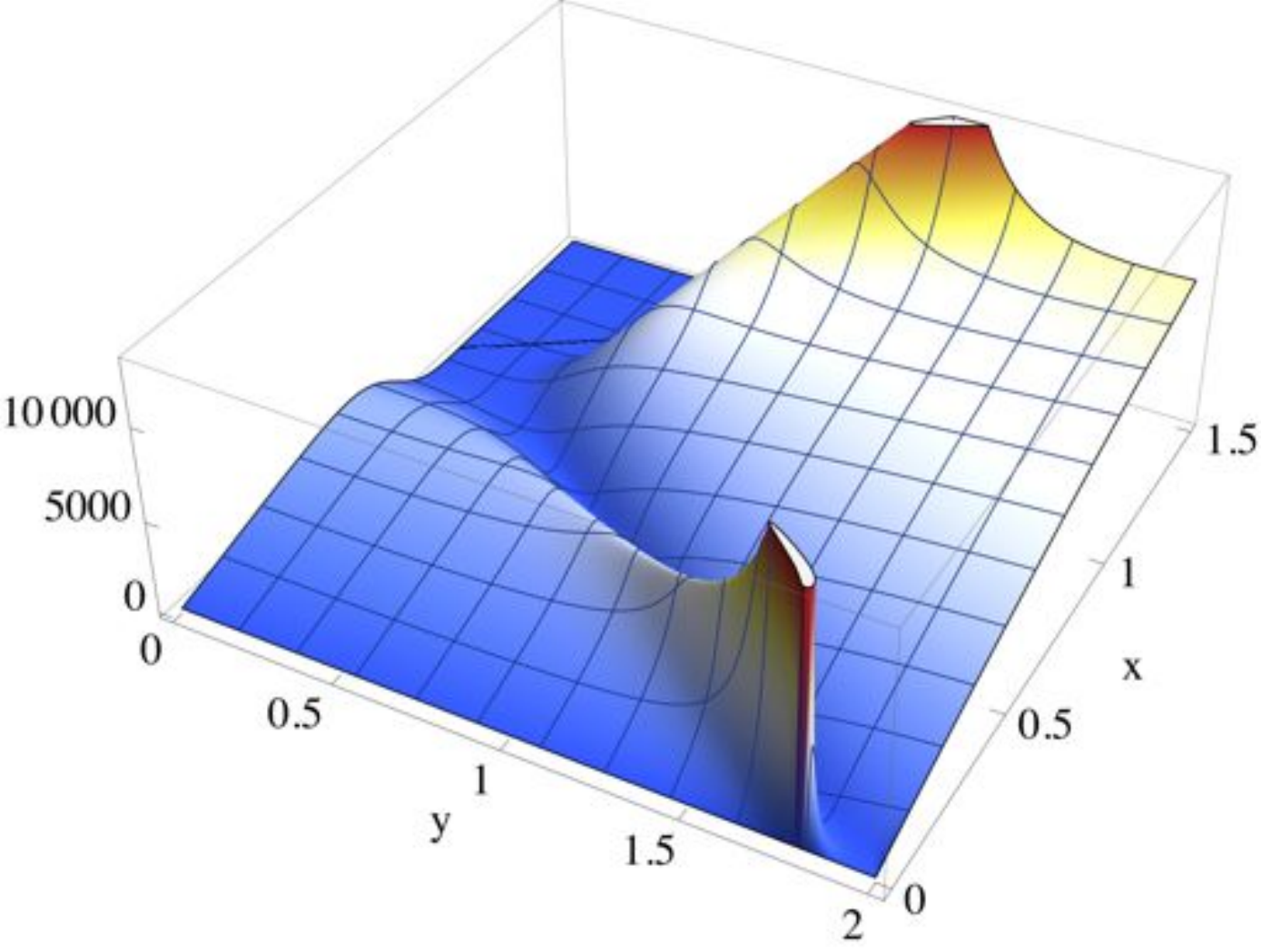}\hskip1.2cm
\includegraphics[scale=0.53]{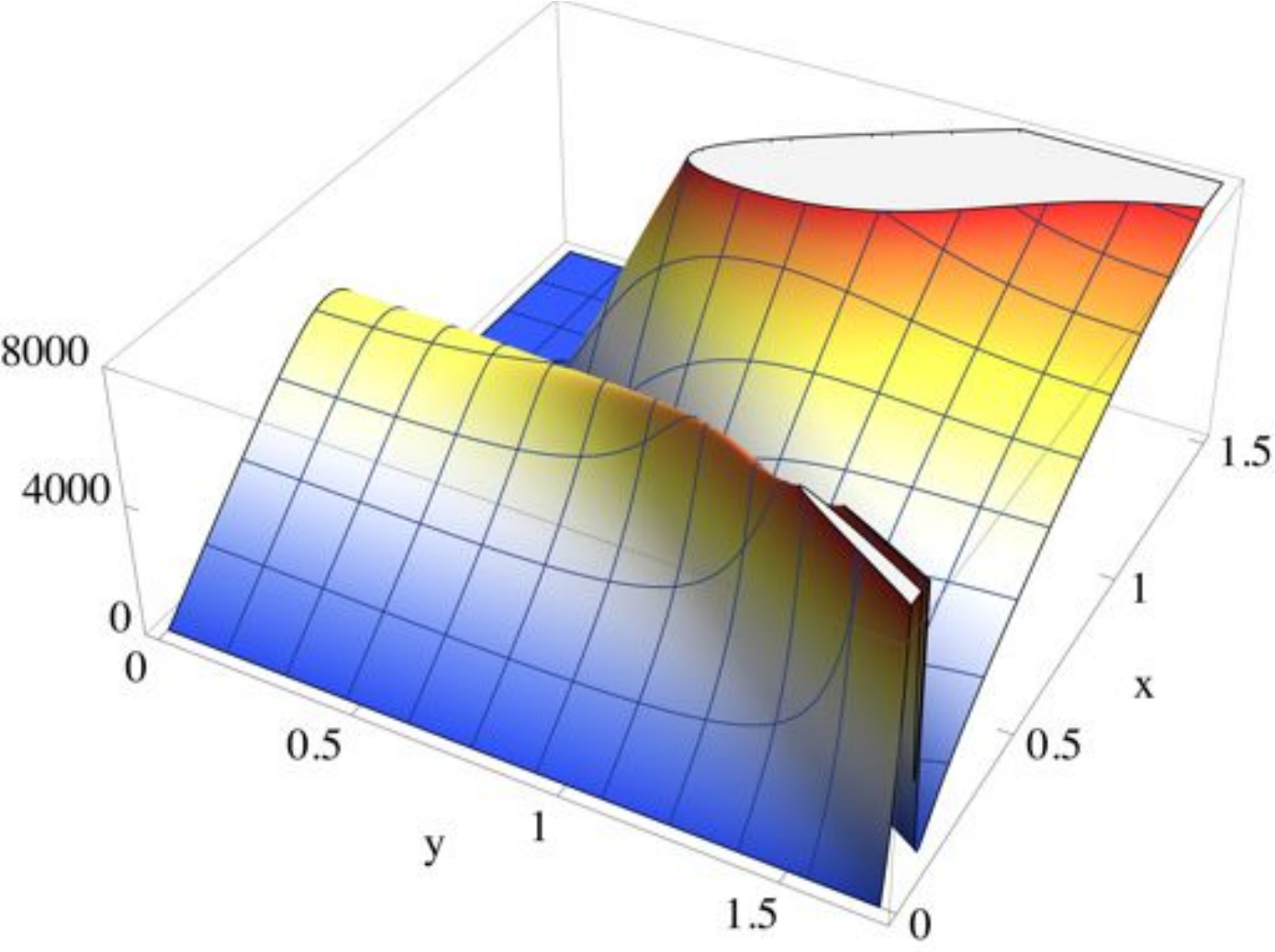}
\centerline{(c)\hskip9cm(d)}
\caption{The longitudinal decoherence strengths, $i\,\mr{sgn}(\omega)\Im{\cal L}_\ell(x,y)$, as functions of $x$ and $y$ in the following cases: (a): $T=0$, $e=0$, (b): zoom into (a), (c): $T>0$, $e=0$, (d): $T>0$, $e>0$. Note that the singular lines in (c) and (d) are caused by the derivatives appearing in the low-temperature expansion in Eq. \eqref{ITexp}.}\label{decl}
\end{figure}

\subsection{Decoherence and irreversibility}
The build-up of decoherence in space-time is governed by the imaginary parts of the spectra of the quadratic effective action, $i\,\mr{sgn}(\omega)\Im{\cal L}_\ell(\omega,\vec{q})$ and $-i\,\mr{sgn}(\omega)\Im{\cal L}_t(\omega,\vec{q})$. The imaginary parts of the spectra of Eqs. \eq{retccgf} and \eq{longeomi}, plotted in Figs. \ref{decl} and \ref{dect}, vanish where the particle-hole spectral weights vanish, as demanded by Eq. \eq{didi}. The longitudinal and the transverse decoherence of the density of the ideal gas at vanishing temperature, depicted in Figs. \ref{decl} (a) and \ref{dect} (a), show that the longitudinal density and the transverse current modes are classical at short and long distances, respectively. It is remarkable that the effect of temperature is qualitatively different in the longitudinal and the transverse sectors. In fact, the thermal fluctuations make the decoherence of the longitudinal modes stronger as one can see by comparing Figs. \ref{decl} (b) and \ref{decl} (c). The deep valley on the latter plot is due to the artificial singularity of the low-temperature expansion at $|r|=1$ and should be ignored. The decoherence of the transverse modes becomes stronger with the temperature at higher frequencies, but it is strongly weakened in the infrared, according to Figs. \ref{dect} (b) and \ref{dect} (c). Note that the composite sound experiences strong decoherence both in the longitudinal and the transverse sectors.

\begin{figure}[h]
\includegraphics[scale=0.53]{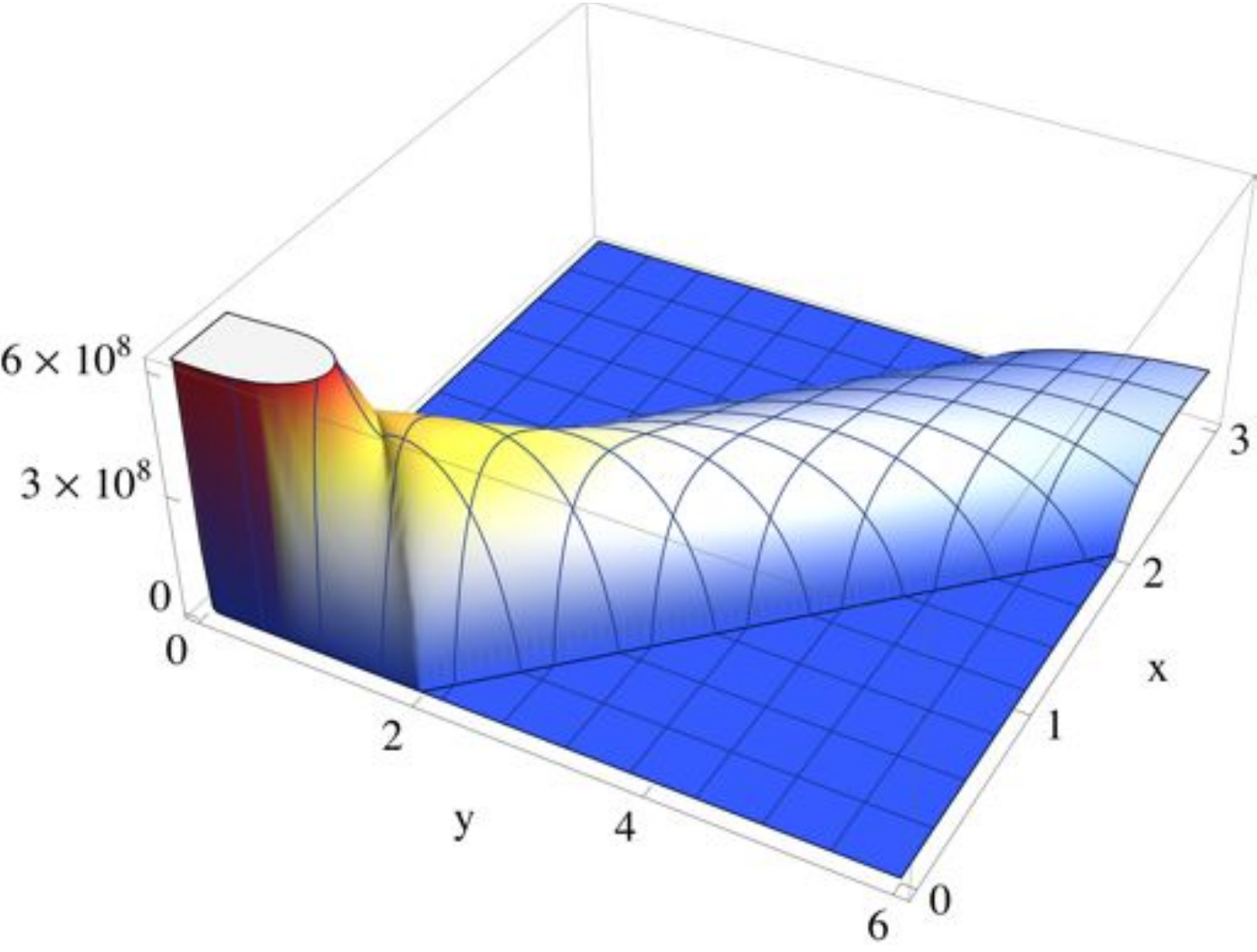}\hskip1.2cm
\includegraphics[scale=0.53]{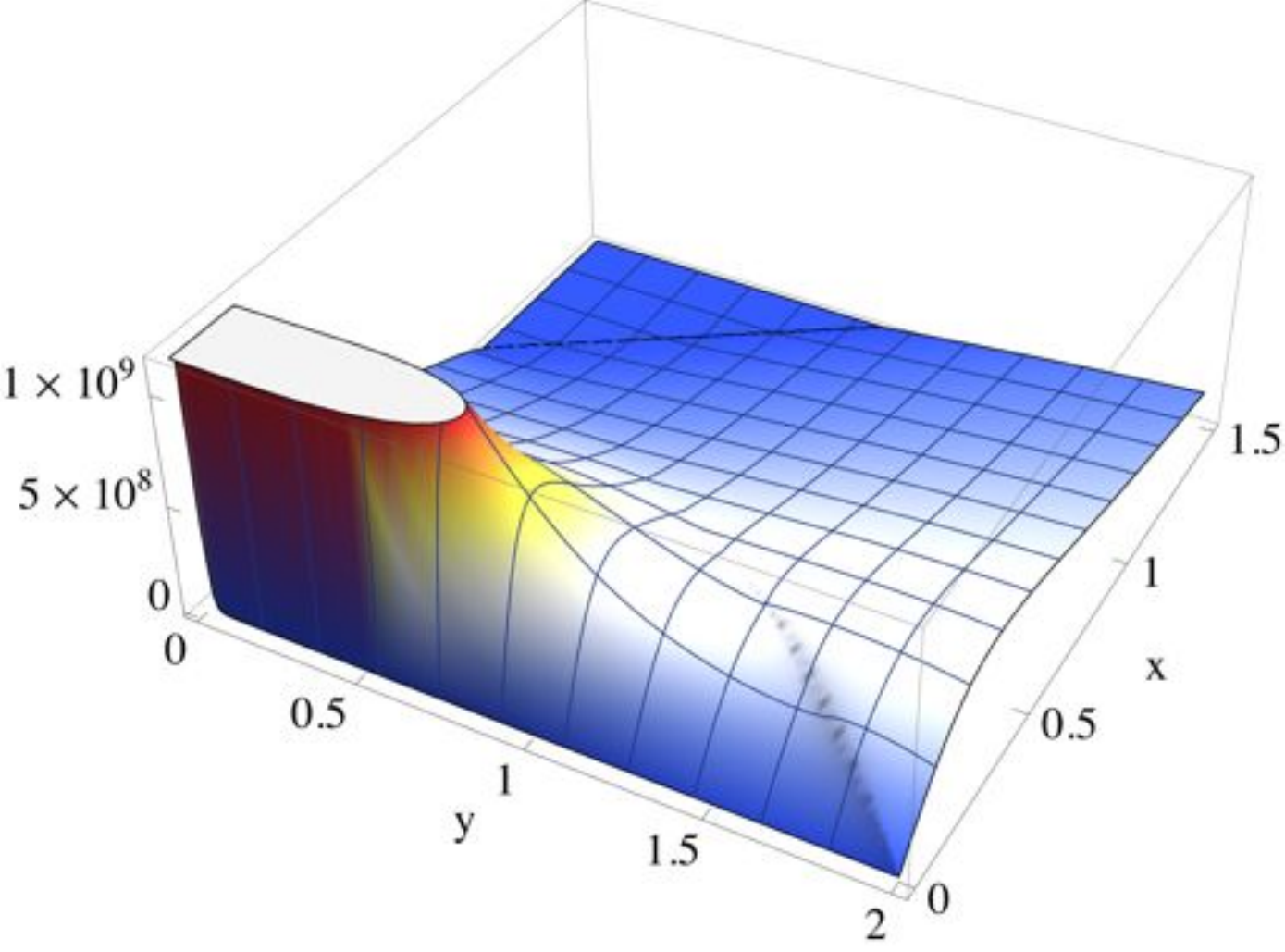}
\centerline{(a)\hskip9cm(b)}
\vskip0.6cm
\includegraphics[scale=0.53]{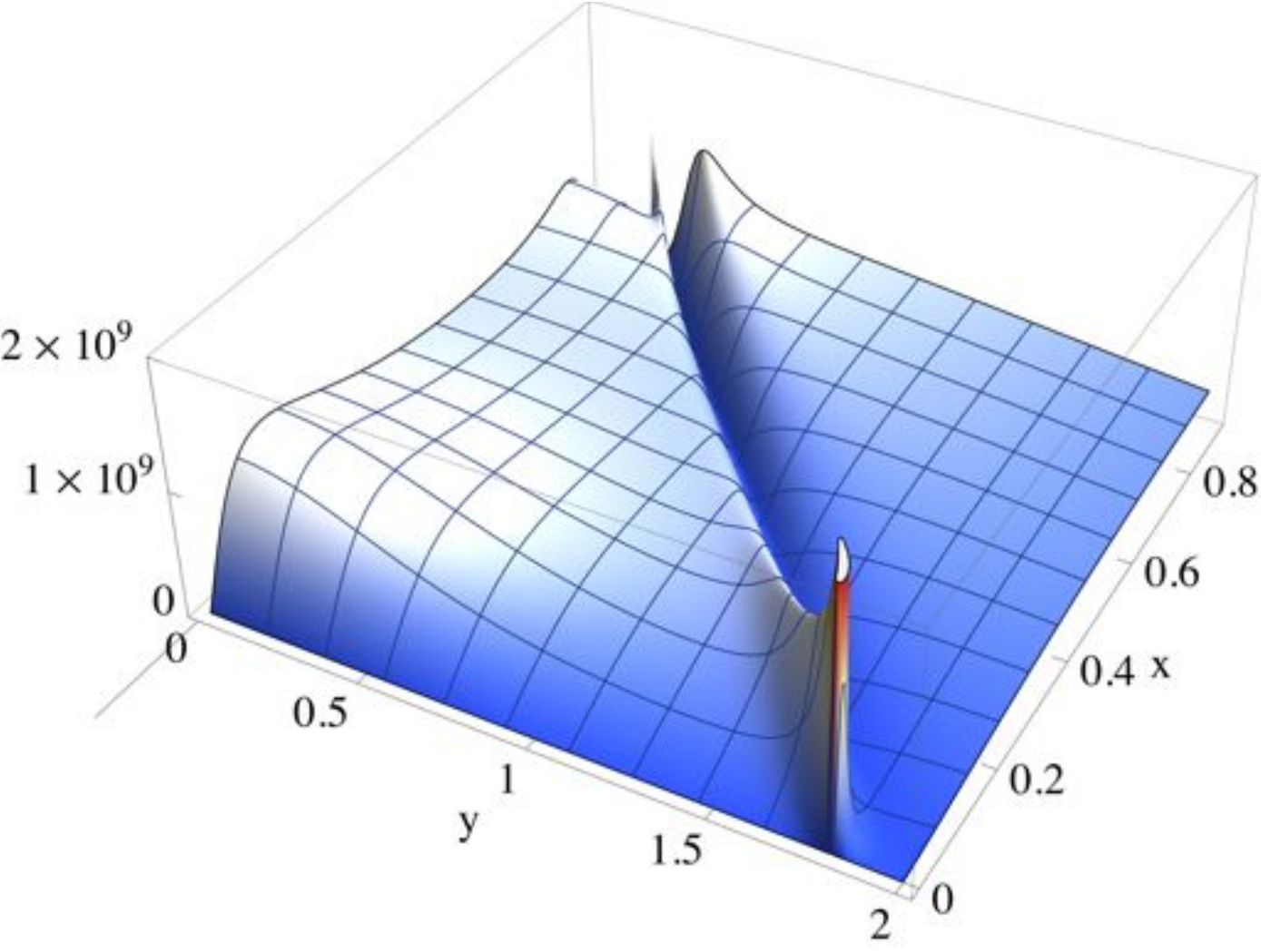}
\centerline{(c)}
\caption{The plots of the transverse decoherence strength, $i\,\mr{sgn}(\omega)\Im{\cal L}_t(x,y)$, as functions of $x$ and $y$ in the following cases: (a): $T=0$, (b): zoom into (a), (c): $T>0$. As in Fig. \ref{decl} (c) and (d), the singular line at non-zero $T$ is again caused by the derivatives appearing in the low-temperature expansion in Eq. \eqref{ITexp}.}\label{dect}
\end{figure}

Decoherence and the breakdown of time reversal invariance share the same dynamical origin in a quadratic theory, as was pointed out in Section \ref{fdts}. Hence, the effective dynamics of the current experiences decoherence and breaks time reversal invariance even in an ideal gas. The latter symmetry breaking manifests itself as the damping of the normal modes. When the system is placed into a finite space-time volume, then its spectrum becomes discrete and the imaginary part of the current-current Green's function \eq{cgfctp} becomes infinitesimal, $\ord{\epsilon}$. As a result, the far field component of the quadratic part of the effective action for the current vanishes according to Eq. \eq{imagipropf} and the dynamics of the current becomes coherent and reversible. The emergence of irreversibility in the thermodynamical limit is another piece of circumstantial evidence of the spontaneously broken time reversal invariance.

%%%%%%%%

\section{Discussion and conclusion}\label{concls}

In this work, we studied the equation of motion for the electric current in the low-energy, non-relativistic regime of an electron gas at non-zero density. Contrary to expectations, we found that the dynamical properties of the current are highly non-trivial even in the non-interacting, ideal gas limit. The simplest way to understand this behaviour was presented at the beginning of the paper where we argued that non-linear transformations of coordinates generate inertial forces, thus making the dynamics of non-linear coordinates (or fields) complicated and far from obvious.

We showed that the electric current, which is bi-linear in the fundamental free electron field, obeys a highly non-linear equation of motion even in the absence of electromagnetic interactions. The linearised equation of motion for the electric current was first derived in Fourier space. We found it to include dissipative terms and to display a non-analytic structure at vanishing frequency and momentum. The collective modes were sound waves, both in the longitudinal and the transverse sectors. It is important to stress that these modes are neither zero nor first sound modes of the usual Fermi gas. This {\it composite sound} mode, which was to our knowledge thus far unknown, is a result of the non-trivial redistribution of the energy-momentum, injected into the system through an external perturbation of the relevant composite operator, i.e. the Noether current. Our linearised equation of motion, an inhomogeneous linear equation for the current, is the inverse of the Kubo linear response formula for the current in the presence of an external vector potential. Our results are therefore equivalent to those derived from the linear response formalism. There is however an obvious difference between the two ways the dynamics is presented in the two schemes, namely, the equation of motion displays the collective modes in a more explicit manner than the linear response formulae. This fact may explain why the modes identified in the paper had not been found previously within the usual linear response formalism. Decoherence, which goes beyond the linear response formalism was also analysed in this work. We found it to be to be very strong for the composite sound modes, thus making them important also in the classical limit.

The equation of motion for the longitudinal component of the current was found to be local in time for an ideal gas at vanishing temperature, so long as the motion was slower than the Fermi velocity. We could recover a similar local equation for the transverse current, but only at non-zero temperature. However, both the longitudinal and the transverse equations of motion remained non-local in space. 

After obtaining the linearised equation of motion for the current, we studied its non-linear extension by replacing the time derivatives with convective derivatives. This minimal extension enabled us to restore Galilean invariance. Although the equation of motion for a stationary flow showed resemblance with the Navier-Stokes equation, important differences were present in comparison with phenomenological hydrodynamics. In particular, our system of ``mechanical" equations was closed without evoking any thermodynamical considerations. Furthermore, spatial non-locality arose due to the factor of $1/|\vec{q}|$. This non-locality followed from the singularity of the current-current two-point Lindhard function at $\omega=\vec{q}=0$, generated by the gapless particle-hole excitations.  

The effects of the microscopic QED interactions on the dynamics of the current can be most easily assessed in the framework of the skeleton expansion. Intriguingly, what we found was that the re-summation of the Coulomb interaction into the photon self-energy only changed the linearised equation of motion quantitatively, while preserving all of the qualitative features of the behaviour of a non-interacting electron gas. We therefore determined that qualitatively different behaviour would have been generated by vertex corrections and the electron self-energy insertions. Thus, such corrections would be required to bridge the gap between the electron gas regime studied in this work and the interacting regime that is usually discussed in physics.  

At the end of the paper, we turned our attention to the discussion of decoherence and irreversibility. We argued that it is the non-vanishing far field component of the current-current Green's function that connects irreversibility with decoherence. This relation could be interpreted as a generalisation of the fluctuation-dissipation theorem. As a result, the longitudinal and the transverse modes displayed strong irreversibility and decoherence at high and low frequencies, respectively, even in the absence of interactions.

We believe that beyond our presentation of the intricacies and complexities that govern the dynamics of non-interacting, ideal gases, the results presented in this work have methodological value and may lead to the development of a systematic way to derive hydrodynamical equations for more involved models. Furthermore, our calculation gave us the exact linearised equation of motion with irreversibility and decoherence; a result that would not be invalidated by calculations at higher orders. 

This work opens several questions: What is the signature of the free, non-interacting nature of a dynamical system when expressed in terms of non-linear functions of the coordinates? The structure of the connected Green's functions completely disguises the underlying harmonic system. Is there perhaps a simpler way to recognise the existence of the uncorrelated degrees of freedom than by studying the properties of its Noether current? How precisely do the interactions, in particular the vertex corrections and the electron self-energy change the picture? Does our approach allow us to recover the full analyticity of the equation of motion for a Fermi liquid in Fourier space, assumed in phenomenological approaches? 

Beyond these issues, what we find to be most important at present is to assess the relevance of the phenomena discussed in this work for the actual experimental observations. The central question is whether the correlations, established by the ideal gas equation of motion, remain experimentally accessible in an interacting system. To address this issue, one should distinguish between three independent length scales. The first is the average inter-particle separation, $1/k_F$. The interactions introduce the second length scale, the mean free path, $r_{mfp}$. The experimental observations are then carried out at the third length scale, $r_o$. The three physical scales normally obey the relation $1/k_F<r_{mfp}<r_o$. The elementary particle regime is restricted to the length scales $\ell<1/k_F$. In this regime, the energy-momentum is mainly carried by the particles. This is because the holes, which have $|\vec{q}|<k_F$, carry a relatively small fraction of the total energy-momentum. Therefore, no collective modes can exist for $\ell<1/k_F$. The ideal gas collective modes appear at $\ell>1/k_F$, but they remain restricted to the {\it composite} length scale regime, $1/k_F<\ell<r_{mfp}$, because the particle collisions generate a finite life-time for the dressed quasi-particle modes for $\ell>r_{mfp}$. The usual experimental observations at $r_o$ allow one to reconstruct the physics of the quasi-particle regime, $r_{mfp}<\ell<r_o$, which is dominated by the ``true" quasi-particles of the Fermi liquid. Their finite life-time screenes the $1/|\vec{q}|$ singularity of the ideal gas equation of motion, thus justifying the usual phenomenological treatment. The composite sound waves are therefore the collective modes that play a central role in the composite particle regime by forming the internal structure of the quasi-particles. However, they are suppressed on scales beyond the mean free path, i.e. in the quasi-particle regime.\footnote{We find it interesting that there exists a similarity between the composite sound and the dynamics of gluons in QCD. This is because both the single gluon excitations inside the glueballs, i.e. the quasi-particles of QCD, and the composite sound modes are localised into a finite region in space by strong correlations that increase with the distance. Outside the quasi-particles, the modes are screened. In QCD, this is due to confinement.} To uncover the physics at the composite particle scale, we would have to resolve the structure of the quasi-particles and the effective couplings. The main question is therefore whether there is an experimental method to zoom into the short distance features of the dynamics in order to identify the collective modes, down to the microscopic edge of the collective phenomena, at $1/k_F$. We leave the resolution to this question open in this work.

We further believe that it would be interesting to extend the calculation to the relativistic domain where dynamical degrees of freedom of the electromagnetic field appear and one could systematically derive equations, which are analogous to magnetohydrodynamics. We expect that the radiation reaction would enhance the dissipative forces and generate new collective modes.

Finally, we may also wonder if turbulence can occur in our system when the equations of motion include the minimal non-linear extension, cf. Eq. \eqref{MinExtNonLinerDer}. As far as dimensional analysis is concerned, the characteristic scales of the non-interacting electron gas at zero temperature are the mass $m$ and the average particle separation, $1/k_F$. We can thus construct a Reynolds number, 
\begin{align}
R_F=\frac{u k_F\ell}{v_F} =\frac{ u \hbar \ell}{m}, 
\end{align}
where $\ell$ and $u$ denote the typical length and velocity scales of the flow. One finds that $R_F\sim u\ell$ in CGS units. By assuming the flow velocity to be of the order of the magnitude of the metallic Fermi velocity, $u=v_F\sim10^8$cm/s, we find that $R_F\sim10^8\ell$. In viscous fluids, turbulence sets in at around $R_F=4000$. If such a na\"{i}ve dimensional analysis can be carried over to our case then one expects that a homogeneous flow of the size $\ell\sim400$nm indeed reaches $R_F\sim4000$ and thereby experiences turbulence. The presence of a heat bath introduces further length scales, the thermal wavelength $\lambda_T$ and the Coulomb interaction brings in the Bohr radius, $a_0$, rendering the dimensional analysis more involved. In fact, these gases have two further Reynolds numbers, $R_T=R_F/(\lambda_Tk_F)^2$ and $R_C=R_F/(a_0k_F)^2$, where $R_T$ characterises the ideal gas at finite temperature and $R_C$ the Coulomb gas. However, $R_T\ll R_F$ in the low-temperature limit, $\lambda_T\gg1/k_F$, and $R_C<R_F$ for a weakly coupled Coulomb gas, where $a_0\gg1/k_F$. Hence, $R_F$ remains the relevant Reynolds number for determining the onset of turbulence.

We end this paper by noting that it would be fascinating to connect our results with the low-energy analytic structure of a field theory dual to a higher-spin Vasiliev theory. The reason for this possible analogy is the fact that the presence of higher-spin conserved currents constrains the field theory correlators to be those of a free theory of bosons or fermions \cite{Maldacena:2011jn}. We hope to explore some of the research directions and open questions raised in this work in the future, particularly in models with direct relevance for experimental observation.

\acknowledgments It is a pleasure to thank J\'anos Hajd\'u for numerous encouraging and enlightening discussions, which were essential to express the ideas of this paper. We are also very grateful to Aleksey Cherman for his comments on the manuscript. Furthermore, we would like to thank Richard Davison, Nikolaos Kaplis, David Kralji\'{c} and Andrei Starinets for helpful discussions. SG is supported in part by a VICI grant of the Netherlands Organization for Scientific Research (NWO), and by the Netherlands Organization for Scientific Research/Ministry of Science and Education (NWO/OCW).

\appendix
\section{Linear response and hydrodynamics}\label{linresps}
The linear response formalism \cite{kubo} is a widely used technique for calculating the expectation value of an observable, $O_S$, in the Schr\"odinger representation, provided that the Hamiltonian is of the form $H=H_0+H_1(t)$. The expectation value,
\be
\la\la O(t)\ra\ra=\mr{Tr}\rho(t)O(t),
\ee
can be written in the {\em interaction picture} representation, using $H_0$ and $H_i(t)$ as the non-perturbed Hamiltonian and its perturbation, respectively. In this representation, an operator $A$ is then given by the expression 
\be
A_i(t)=e^{\frac{i}{\hbar}H_0}A_Se^{-\frac{i}{\hbar}H_0},
\ee
and
\be
\rho_i(t)=U(t,t_i)\rho_0U^\dagger(t,t_i)
\ee
is the density matrix, where the time evolution operator, $U(t,t_i)$, satisfies the equation of motion
\be
i\hbar\partial_tU(t,t_i)=H_{1i}(t)U(t,t_i).
\ee
The equation of motion for the density matrix,
\be
i\hbar\partial_t\rho_i(t)=[H_{1i}(t),\rho_i(t)],
\ee
can be solved iteratively,
\be
\rho_i(t)=\rho_0-\frac{i}{\hbar}\int_{t_i}^tdt'[H_{1i}(t'),\rho_0] +\ord{H_{1i}^2}.
\ee
The first, linear approximation in $H_1$ gives
\bea
\delta\la\la O(t)\ra\ra&=&\la\la O(t)\ra\ra-\mr{Tr}\rho_0O(t)\nn
&=&\frac{i}{\hbar}\int_0^tdt'\mr{Tr}\rho_0[H_{1i}(t'),O_i(t)] .
\eea
One usually assumes the separability of the time dependence in the perturbation, $H_1(t)=a(t)B_i$, and writes
\be
\delta\la\la O(t)\ra\ra=-\int_{t_0}^{t_f} dt' G^R_{O,B}(t,t')a(t') ,
\ee
where the retarded Green's function is defined by
\be
i\hbar G^R_{A,B}(t,t')=\Theta(t-t')\la\la[A_i(t),B_i(t')]\ra\ra.
\ee

To relate this result to the CTP formalism, let us suppose that we want to find the expectation value of a local composite operator, $F(\phi(x))$, in the scalar field theory, defined by the action
\be
S_h[\phi]=S[\phi]+\int d^4x\,a(x)G(\phi(x)).
\ee
The perturbation series of the expectation value of our composite operator in $\hbar$ can be calculated by means of the generator functional 
\be
e^{\ih W[\hat a,\hj]}=\int D[\hphi] \, e^{\ih S[\phi^+]-\ih S^*[\phi^-]+\ih\int d^4xa^\sigma(x)G(\phi^\sigma(x))+\ih\int d^4xj^\sigma(x)F(\phi^\sigma(x))},
\ee
as
\bea
\la\la F(\phi(x)\ra\ra&=&\frac\hbar{i}\fd{}{j^\sigma(x)}
\sum_{n_+,n_-=0}^\infty\frac{(-1)^{n_-+n_+}}{\hbar^{n_-+n_+}n_+!n_-!}
\left(\sum_{\sigma^+=0}^\infty\int d^4x^+a^+(x^+)\fd{}{a^+(x^+)}\right)^{n_+}\nn
&&\times\left(\sum_{\sigma^-=0}^\infty\int d^4x^-a^-(x^-)\fd{}{a^-(x^-)}\right)^{n_-}
{e^{iW[\hat a,\hj]}}_{|a^+=-a^-=a,\hj=0},
\eea
for either $\sigma=+$ or $\sigma=-$. The linear response formula is the $\ord{\hbar}$ result,
\be
\delta\la\la F(\phi(x)\ra\ra=\int d^4y\left(\fdd{}{j^\sigma(x)}{a^+(y)}-\fdd{}{j^\sigma(x)}{a^-(y)}\right){e^{iW[\hat a,\hj]}}_{|a^+=-a^-=a,\hj=0},
\ee
which can be written as
\be\label{kfw}
\delta\la\la F(\phi(x)\ra\ra=\int d^4yD^r(x,y)a(y),
\ee
where
\be
D^r(x,y)= - \sum_{\sigma'}\sigma'\fdd{W[\hat a,\hj]}{j^\sigma(x)}{a^{\sigma'}(y)}_{|a^+=-a^-=a,\hj=0}.
\ee 

Hydrodynamics addresses the inverse problem. There, we are interested in the equations (of motion), satisfied by the expectation values, in which the external sources appear linearly. In case of the linear response theory, it is easy to find the equation in question, 
\be
a(x)= - \int d^4 y \left[D^r (x,y) \right]^{-1}\delta\la\la F(\phi(y))\ra\ra.
\ee
The only subtlety is the potential necessity to exclude the null-space from the domain of the inverse Green's function.

The generalisation of such an inverse linear response formula beyond $\ord{a}$ is provided by the functional Legendre transform of $W[\hat a,\hj]$, the effective action,
\be\label{fltr}
\Gamma[\hat F]=W[\hat a,\hj]-\hat a\hat F,
\ee
where
\be
\hat F=\fd{W[\hat a,\hj]}{\hat F}.
\ee
The inverse Legendre transform is then also given by Eq. \eq{fltr}, with
\be
\hat a=-\fd{\Gamma[\hat F]}{\hat F}.
\ee
This equation plays the role of the equation of motion and produces a non-linear extension of the hydrodynamical equations. The inverse Legendre transform generates the non-linearity, which is necessary to close the equations, without an introduction of auxiliary variables, such as thermodynamical functions.

\section{Free propagators}\label{propctap}
The detailed calculation of the free propagator is easiest to carry out in the conventional operator formalism, see e.g. \cite{kamenev,calzetta}. We summarise the results below. The boson propagator, given by Eq. \eq{bosctppr}, yields
\be
\hD(k)=\begin{pmatrix}\frac1{k^2-\frac{m^2c^2}{\hbar^2}+i\epsilon}&-2\pi i\delta(k^2-\frac{m^2c^2}{\hbar^2})\Theta(-k^0)\cr
-2\pi i\delta(k^2-\frac{m^2c^2}{\hbar^2})\Theta(k^0)&-\frac1{k^2-\frac{m^2c^2}{\hbar^2}-i\epsilon}\end{pmatrix}-i2\pi\delta\left(k^2-\frac{m^2c^2}{\hbar^2}\right)n_B(k)\begin{pmatrix}1&1\cr1&1\end{pmatrix},
\ee
in Fourier space where the Bose-Einstein distribution function is 
\be
n_B(k)=\frac{\Theta(-k^0)}{e^{\beta(\epsilon_\vec{k}+\mu)}-1}
+\frac{\Theta(k^0)}{e^{\beta(\epsilon_\vec{k}-\mu)}-1}.
\ee
The inversion, cf. Eqs. \eq{retadvinv} and \eq{didi}, gives the following expressions for $\hat\Delta=\hD^{-1}$,
\begin{align}\label{freeinvp}
\Delta^n(k)=k^2-m^2,&&\Delta^i(k)=\epsilon,&&\Delta^f(k)=i\,\mr{sgn}(k^0)\epsilon.
\end{align}

The free fermionic propagator, defined by the generating functional
\be
e^{\ih W[\hj,\bar{\hj}]}=\int D[\hsi]D[\hsib]e^{\ih\hsib\hG^{-1}\hsi+\ih\bar{\hj}\hsi+\ih\hsib\hj},
\ee
can be written as
\be
\hG^{\alpha\beta}(x,y)=\begin{pmatrix}\la0|T[\psi^\alpha(x)\psib^\beta(y)]|0\ra&-\la0|\psib^\beta(y)\psi^\alpha(x)|0\ra\cr\la0|\psi^\alpha(x)\psib^\beta(y)|0\ra&\la0|T[(\gamma^0\psi(y))^\beta(\psib(x)\gamma^0)^\alpha]|0\ra^*\end{pmatrix},
\ee
and the detailed expression written in terms of the scalar propagator in Fourier space is
\be
\hG(k)=\left(k\br+\frac{mc}\hbar\right)\hD_k.
\ee
In case of finite temperature and density, one uses the occupation number density,
\be
n_F(k)=\frac{\Theta(k^0)}{e^{\beta(\epsilon_\vec{k}-\mu)}+1}+\frac{\Theta(-k^0)}{e^{\beta(\epsilon_\vec{k}+\mu)}+1},
\ee
and the full propagator becomes
\be
\hG_k=(k\br+m)\left[\hD_k+2\pi i\,\delta(k^2-m^2)n_F(k)\begin{pmatrix}1&1\cr1&1\end{pmatrix}\right].
\ee

\section{Current-current two-point function at finite density in the non-relativistic limit}\label{grfnctap}
The result of the calculation of the current-current Green's function, given by Eq. \eq{curprop},
\be\label{cgfctp}
G^{\mu\nu}_{\sigma\tau}(q)=-i\int\frac{d^4p}{(2\pi)^4}\tr \left[ \gamma^\mu F_{\sigma\tau}(q+p)\gamma^\nu F_{\tau\sigma}(p) \right],
\ee
at finite density and vanishing temperature in the non-relativistic limit, $c\to\infty$, is briefly summarised in this Appendix. The real and imaginary parts if the Green's function are always defined in position space; hence,
\begin{align}
\Re_xG(q)=\hf \left[G(q)+G^*(-q) \right], && i\Im_xG(q)=\hf \left[G(q)-G^*(-q) \right].
\end{align}

\subsection{Lorentz structure}
The two-point function is symmetric, $G_{(\sigma\mu)(\sigma'\nu)}(p)=G_{(\sigma'\nu)(\sigma\mu)}(-p)$, and transverse, $p^\mu G_{\mu\nu}(p)=0$. Furthermore, it is covariant and depends on two four-vectors, $p^\mu$ and $\beta^\mu$, defined as $\beta^\mu=(1,\vec{0})$ in the inertial frame where the electron gas is at rest. This gives two independent kinematical, scalar combinations, $\vec{q}^2=-[q-u(uq)]^2$ and $\xi=uq/|\vec{q}|=\omega/c|\vec{q}|$, where the notation $q^\mu=(\omega/c,\vec{q})$ is used. Such a tensor can be parametrised by two Lorentz scalars as
\be\label{parproj}
G^{\mu\nu}=G_\ell P^{\mu\nu}_\ell+D_tP^{\mu\nu}_t,
\ee
where $P_t$ and $P_\ell$ are projectors onto the three-dimensional transverse and longitudinal subspaces,
\bea
P^{\mu\nu}_t&=&-\begin{pmatrix}0&0\cr0&\vec{T}\end{pmatrix},\nn
P^{\mu\nu}_\ell&=&\frac1{1-\xi^2}\begin{pmatrix}1&\vec{n}\xi\cr\vec{n}\xi&\xi^2\vec{L}\end{pmatrix},
\eea
respectively, with $\vec{n}=\vec{k}/|\vec{k}|$, $\vec{L}=\vec{n}\otimes\vec{n}$ and $\vec{T}=\openone-\vec{L}$. The inverse, defined by $G^\mu_{~\rho}G^{-1\rho\nu}=T^{\mu\nu}$ is given by
\be
G^{-1}=\frac1{G_\ell}P_\ell+\frac1{G_t}P_t.
\ee
For future reference, the retarded and advanced photon propagators are
\begin{align}\label{retgfph}
D_{0,\ell}^{r}(q)=-D_{0,t}^{r}(q)=-\frac1{(q^0 + i\epsilon)^2-\vec{q}^2}, && D_{0,\ell}^{a}(q)=-D_{0,t}^{a}(q)=-\frac1{(q^0 - i\epsilon)^2-\vec{q}^2}.
\end{align}

\subsection{Vacuum contribution}
The two-point function is the sum of the vacuum and the finite density contributions,
\be
\hG^{\mu\nu}=\hG^{\mu\nu}_{vac}+\hG^{\mu\nu}_{gas},
\ee
and both the vacuum and the finite density contributions are of the form \eq{stctpform}. The vacuum contributions, $\hG_{vac}^{\mu\nu}=\hG_{vac}T^{\mu\nu}$, are easy to find. The diagonal CTP block, $G^{++}_{\ell,vac}=G^{++}_{t,vac}=G^{++}_{vac}$, gives the standard result,
\bea
G^{++}_{vac}(q)&=&\frac1{3\pi}q^2\left\{\frac13+2\left(1+\frac{2m^2c^2}{q^2}\right)\left[\sqrt{\frac{4m^2c^2}{q^2}-1}~\mr{arccot}\left( \sqrt{\frac{4m^2c^2}{q^2}-1} \right)-1\right]\right\}\nn
&=&\frac{q^2}{15\pi}\left[\frac{q^2}{m^2c^2}+\ord{\left(\frac{q^2}{m^2c^2}\right)^2}\right].
\eea
The off-diagonal CTP block, calculated by using the free propagator \eq{freeelpr}, is $G^{+-}_{\ell,vac}=G^{+-}_{t,vac}=G^{+-}_{vac}$, with
\be
G^{+-}(q)=\frac{i}3\int\frac{d^4p}{(2\pi)^4}2\pi\delta((p+q)^2-m^2)\Theta(-p^0-q^0)2\pi\delta(p^2-m^2)\Theta(p^0)\tr N(p+q,q),
\ee
where the trace is taken over the Lorentz indices of the trace formula,
\bea
N^{\mu\nu}(p,q)&=&\tr\,\gamma^\mu \left[(p_\alpha\gamma^\alpha+mc)\gamma^\nu(q_\beta\gamma^\beta+mc)\right]\nn
&=&4(m^2c^2-pq)g^{\mu\nu}+4p^\mu q^\nu+4p^\nu q^\mu.  \label{Ntrace}
\eea
Simple steps lead to the expression
\be
G^{+-}_{vac}=\frac{i(c^2m^2+\frac{q^2}2)}{3\pi c|\vec{q}|}\Theta(-q^0-m)\int_0^{\sqrt{q^{02}-m^2c^2}}\frac{dpp}{\omega_p}\Theta\left(2p|\vec{q}|-|q^2+2\omega_pq^0|\right),
\ee
which can be neglected in the non-relativistic limit because the Heaviside function vanishes for non-relativistic frequencies with $|q^0| \ll m$.

\subsection{Fermi sphere contribution}
To find the non-zero density contributions to $\hG_\ell$ and $\hG_t$ in the electron gas we need to compute the CTP blocks corresponding to the time components, $\hat{\cal T}=\hG_{gas}^{00}$, and the spatial trace, $\hat{\cal S}=\hG^{jj}_{gas}$. For this purpose, we need to use the trace factors, cf. Eq. \eqref{Ntrace},
\bea
t&=&N^{00}(p+q,p)_{|p^2=m^2c^2}=t'-2(q^2+2pq),\nn
s&=&N^{jj}(p+q,p)_{|p^2=m^2c^2}=s'+2(q^2+2pq),
\eea
where $t'=8((p^{0})^2+p^0q^0)+2q^2$ and $s'=8(p^0q^0+\vec{p}^2)-2q^2$ in the loop integrals
\bea\label{ctpblfsp}
\left(^{{\cal T}(q)}_{{\cal S}(q)}\right)^{++}&=&i\int_p \left(^t_s\right)\biggl[2\pi\delta((q+p)^2-m^2c^2)n_{q+p}2\pi\delta(p^2-m^2c^2)n_p\nn
&&-i\frac{2\pi\delta(p^2-m^2c^2)n_p}{(p+q)^2-m^2c^2+i\epsilon}-i\frac{2\pi\delta((p+q)^2-m^2c^2)n_{p+q}}{p^2-m^2c^2+i\epsilon}\biggr],\nn
\left(^{{\cal T}(q)}_{{\cal S}(q)}\right)^{+-}&=&-i\int_p \left(^{t'}_{s'}\right)  2\pi\delta((q+p)^2-m^2c^2)2\pi\delta(p^2-m^2c^2)  \nn
&&\times \left[\Theta(-p^0-q^0)n_p+\Theta(p^0)n_{p+q}-n_{q+p}n_p\right].
\eea
At this point, it is advantageous to introduce the integrals
\bea
I_1[q;f]&=&\int\frac{d^4p}{(2\pi)^4}f(p,q)2\pi\delta(p^2-m^2c^2),\nn
I_2[q;f]&=&\int\frac{d^4p}{(2\pi)^4}f(p,q)2\pi\delta(p^2-m^2c^2)2\pi\delta(q^2+2pq),
\eea
and write Eqs. \eq{ctpblfsp} as
\bea
\left(^{\cal T}_{\cal S}\right)^{++}(q)&=&I^{(1)}\left[q;\frac{\left(^t_s\right)n_p}{q^2+2pq+i\epsilon}\right]+\frac{i}2I^{(2)} \left[q;\left(^{t'}_{s'}\right)n_pn_{q+p} \right]+(q\to-q),\nn
\left(^{\cal T}_{\cal S}\right)^{+-}(q)&=&-iI^{(2)} \left[q;\left(^{t'}_{s'}\right)\left(\Theta(-p^0-q^0)n_p+\Theta(p^0)n_{p+q}-n_{q+p}n_p\right) \right].
\eea
These integrals are then evaluated in the non-relativistic limit, $|\vec{p}|\ll mc$, by assuming that the integrands are spherically symmetric and non-vanishing for $p^0>0$. We find
\bea
I^{(1)}[q;f(p^0,\vec{p})]&=&\frac1{4\pi^2mc}\int dpp^2f(mc^2,\vec{p}),\nn
I^{(1)}\left[q;\frac{g(p^0,\vec{p})}{q^2+2pq+i\epsilon}\right]&=&\frac1{16\pi^2|\vec{q}|mc}\int_0^\infty dppg(mc^2,\vec{p})\log\frac{k+p+i\epsilon}{k-p+i\epsilon},\nn
I^{(2)}\left[q;h\left((p^0,\vec{p}),(\frac\omega{c},\vec{q})\right)\right]&=&\frac1{16\pi^2|\vec{q}|mc}\int d^3ph\left((mc^2,\vec{p}),(\frac\omega{c},\vec{q})\right)\delta(rk_F-p_z),
\eea
where the following dimensionless parameter has been introduced,
\be
r=\frac{q^2+2m\omega}{2|\vec{q}|k_F}\approx\frac{2m\omega-\vec{q}^2}{2|\vec{q}|k_F}.
\ee

We now need to find the Fourier transform of the real part of the CTP diagonal block at the leading order in $1/c$,
\bea
\!\!\!\!\!\!\!\!\!\!\!\!\!\!\!\!   \Re_x\left(^{{\cal T}(q)}_{{\cal S}(q)}\right)^{++} \! &=&\Re I_1\left[\frac{\left(^t_s\right)n_p}{q^2+2pq+i\epsilon}\right]+(\omega\to-\omega)\nn
&=&\frac1{16\pi^2|\vec{q}|mc}\int_0^{k_F}dppn_p\left[\left(^{8m^2c^2}_{8(m\omega+p^2)+2\vec{q}^2}\right)\log\left|\frac{k+p}{k-p}\right|+8|\vec{q}|p\bigl(^{-1}_{~1}\bigr)\right] \!+ (\omega\to-\omega).
\eea
The momentum integrals can easily be carried out at vanishing temperature, where $n_p=\Theta(p^0)\Theta(k_F-|\vec{p}|)$, leading to the result
\bea\label{tsdiag}
\Re_x{\cal T}^{++}(q)&=&\frac{k_F^2mc}{2\pi^2|\vec{q}|}L_1(r)+(\omega\to-\omega),\nn
\Re_x{\cal S}^{++}(q)&=&\frac{k_F^2}{2\pi^2mc|\vec{q}|}\left[k_F^2L_3(r)+\left(m\omega+\frac{\vec{q}^2}4\right)L_1(r)\right]+\frac{k_F^3c}{6\pi^2m}+(\omega\to-\omega),
\eea
where we have defined
\bea\label{lfuncdef}
L_1(r)&=&\int_0^1dkk\log\left|\frac{r+k}{r-k}\right|=r+\hf(1-r^2)\log\left|\frac{r+1}{r-1}\right|,\nn
L_3(r)&=&\int_0^1dkk^3\log\left|\frac{r+k}{r-k}\right|=\frac{r}6+\frac{r^3}2+\frac14(1-r^4)\log\left|\frac{r+1}{r-1}\right|.
\eea

\begin{figure}
\includegraphics[scale=0.2]{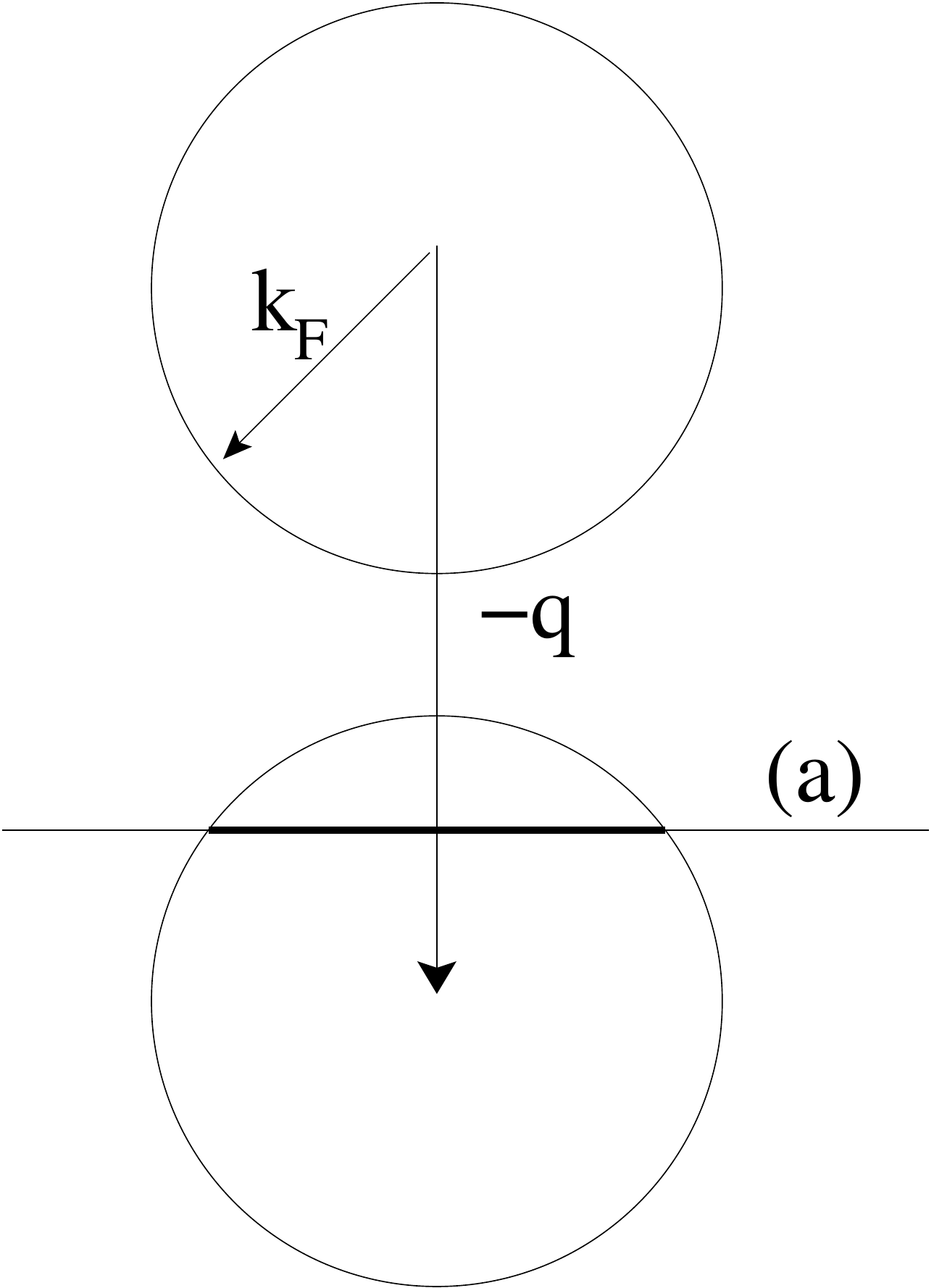}\caption{Domain of integration in Eq. \eq{tspm} for non-overlapping spheres, case (a), indicated by the solid line.}\label{imaga}
\end{figure}

\begin{figure}
\includegraphics[scale=0.2]{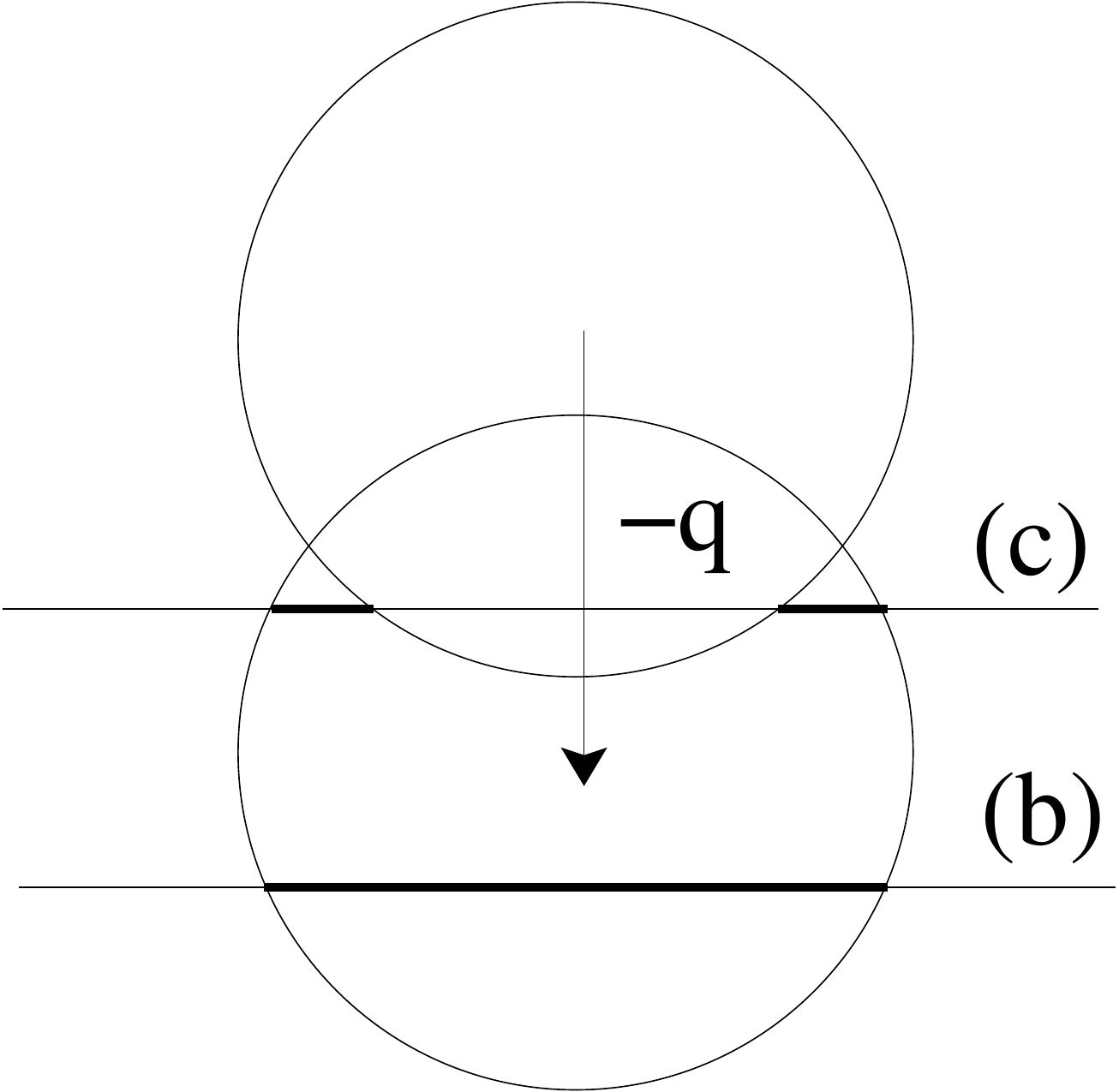}\caption{The same as Fig. \ref{imaga} except for overlapping spheres, cases (b) and (c).}\label{imagb}
\end{figure}

The next step is to compute the off-diagonal $+-$ CTP contributions to $\hat{\cal T}$ and $\hat{\cal{S}}$. At the leading order in the $1/c$ expansion, the off-diagonal CTP block is
\bea\label{tspm}
\!\!\!\!\!      \left(^{{\cal T}(q)}_{{\cal S}(q)}\right)^{+-}&=&-iI^{(2)}\left[\left(^{t'}_{s'}\right) \left[\Theta(p^0)-n_p \right]n_{q+p}\right]\nn
&=&-\frac{imc}{2\pi^2|\vec{q}|}\int d^3p\left(^1_{\frac1{m^2c^2}(m\omega+p^2+\frac{\vec{q}^2}4)}\right)\Theta(k_F-|\vec{p}+\vec{q}|)\Theta(|\vec{p}|-k_F)\delta(p_z-rk_F).
\eea
We choose the $z$-direction to be parallel to $\vec{q}$. This integral is over the region of a plane of the height $p_z=k_Fr$, which is shown in Figs. \ref{imaga} and \ref{imagb}. The integration thus needs to be taken over the points of the plane which are outside of the Fermi sphere, centred at the origin, and inside of another Fermi sphere, which is centred at $-\vec{q}$. It is advantageous to parametrise these integrals by the dimensionless variables $x=\frac{m \omega}{k_F |\vec{q}| }$ and $y=\frac{|\vec{q}|}{k_F}$. There are three different functional forms for these integrals, corresponding to the three cases shown in Figs. \ref{imaga} and \ref{imagb}: (a): $y>2$, $-y-1<r_+<-y+1$, (b): $y<2$, $-1-y<r_-<-1$, and (c): $y<2$, $1<r_+<-\frac{y}2$, where $r_\pm=\frac{q^2\pm2m\omega}{2|\vec{q}|k_F}\to\pm x-\frac{y}2$ in the non-relativistic limit. Straightforward integration yields
\bea\label{tsndiag}
{\cal T}^{+-}(q)&=&-\frac{imck_F^2}{2\pi|\vec{q}|}M_1(x,y),\nn
{\cal S}^{+-}(q)&=&-\frac{ik_F^2}{2\pi mc|\vec{q}|}\left[k_F^2M_3(x,y)+\left(\omega m+k_F^2r^2+\frac{\vec{q}^2}4\right)M_1(x,y)\right],
\eea
where
\bea\label{ilfuncdef}
M_1(x,y)&=&2\int_{p_1}^{p_2}dpp=\begin{cases}1-r_-^2&(a)\cr1-r_-^2&(b)\cr-2xy&(c)\end{cases},\nn
M_3(x,y)&=&2\int_{p_1}^{p_2}dpp^3=\hf\begin{cases}(1-r_-^2)^2&(a)\cr(1-r_-^2)^2&(b)\cr xy(y^2+4x^2-4)&(c)\end{cases}.
\eea
The integrals over the interval are given in terms of $p_2=\sqrt{1-r_-^2}$ and the relative locations of the two Fermi spheres are (a): $p_1=0$, (b): $p_1=0$, (c): $p_1=\sqrt{1-r^2_+}$. The real part of the off-diagonal CTP block is therefore
\be
\Re_x\left(^{{\cal T}(q)}_{{\cal S}(q)}\right)^{+-}=\hf\left[\left(^{{\cal T}(q)}_{{\cal S}(q)}\right)^{+-}-\left(^{{\cal T}(-q)}_{{\cal S}(-q)}\right)^{+-}\right].
\ee

\subsection{Low-temperature expansion}
In this part of the Appendix, we calculate the loop integrals \eq{ctpblfsp} at a small, non-zero temperature,  $T\ll\epsilon_F=\frac{k_F^2}{2m}$, to find the temperature-dependent corrections to ${\cal T}^{++}$, ${\cal S}^{++}$, ${\cal T}^{+-}$ and ${\cal S}^{+-}$. The temperature dependence of the vacuum contribution is strongly suppressed and will be ignored. We start with the ${++}$ CTP block, for which one needs to evaluate integrals of the type
\be\label{pptint}
I(T)=\int_0^\infty\frac{dxf(x)}{e^{\frac{x^2-\epsilon_F}T}+1}.
\ee
The low-temperature expansion involves the primitive function $F(x)$, where $F'(x)=f(x)$, and is introduced because of the partial integration,
\be
I(T)=\tilde I(T)-\frac{F(0)}{e^{-\frac{\epsilon_F}{T}}+1}
\approx\tilde I(T)-F(0),
\ee
where the exponentially small quantities are neglected and
\bea
\tilde I(T)&=&-\int_0^\infty dxF(x)\partial_x\frac1{e^{\frac{x^2-\epsilon_F}T}+1} = \int_{-\frac{\epsilon_F}T}^\infty dyF(\sqrt{Ty+\epsilon_F})\frac{e^y}{(e^y+1)^2}.
\eea
The next step is to expand $F(\sqrt{T y+\epsilon_F})$ around $y=0$, extend the integration over $-\infty<y<\infty$ and keep the even part of the integrand, while the odd part vanishes. By neglecting the $\ord{T^4}$ terms, the leading-order result becomes
\be\label{ITexp}
\tilde I(T)=\int_0^\infty dy\frac{e^y}{(e^y+1)^2}\left[2F(\sqrt{\epsilon_F})+\frac14\left(\frac{F''(\sqrt{\epsilon_F})}{{\epsilon_F}}-\frac{F'(\sqrt{\epsilon_F})}{{\epsilon_F}^{3/2}}\right)y^2T^2\right].
\ee
To compute the remaining integral, we write
\be
K_n=\int_0^\infty dy \frac{y^n e^y}{(e^y+1)^2},
\ee
and solve for $n=0$ and $n=2$, finding the values $K_0=\hf$ and $K_2=\frac{\pi^2}6$. By using the above expressions, we can find the leading-order result for the integral \eq{pptint}, which is
\be
I(T)=I(0)+\frac{\pi^2}{24}\frac{T^2}{{\epsilon_F}^2} \left[k_F^2\partial^2_{k_F}I(0)-k_F\partial_{k_F}I(0)\right],
\ee
showing that the low-temperature expansion is an expansion of the integrand around the Fermi surface. The final expression for the ++ CTP block, which includes an $\CO(T^2)$ correction of the zero-temperature result presented in Eqs. \eq{tsdiag}, is 
\bea\label{ltctpbl}
\Re_x{\cal T}^{++}(q)&=&\frac{k_F^2mc}{2\pi^2|\vec{q}|}\left[L_1(r)+\frac{\pi^2}{24}\frac{T^2}{{\epsilon_F}^2}[r^2L_1''(r)-rL'_1(r)]\right]+(\omega\to-\omega),\nn
\Re_x{\cal S}^{++}(q)&=&\frac{k_F^2}{2\pi^2mc|\vec{q}|}\biggl[k_F^2L_3(r)+\left(m\omega+\frac{\vec{q}^2}4\right)L_1(r)\nn
&&+\frac{\pi^2}{24}\frac{T^2}{{\epsilon_F}^2}\left(k_F^2[8L_3(r)-5rL'_3(r)+r^2L''_3(r)]+\left(m\omega+\frac{\vec{q}^2}4\right)[r^2L''_1(r)-rL'_1(r)]\right)\biggr]\nn
&&+\left(1+\frac{\pi^2}4\frac{T^2}{{\epsilon_F}^2}\right)\frac{k_F^3}{6\pi^2mc}+(\omega\to-\omega).
\eea

The $+-$ CTP block contains an integral of the type
\be\label{ipm}
I(T)=\int_0^\infty dxf(x)\frac{e^{\frac{x^2-\epsilon_1}T}}{\left(e^{\frac{x^2-\epsilon_1}T}+1\right)\left(e^{\frac{x^2-\epsilon_1}T}+1\right)},
\ee
which at low temperature receives non-zero contributions from the interval $\epsilon_1 < x^2 < \epsilon_2$, where
\begin{align}
\epsilon_1=\epsilon_F-\frac{k_F^2r^2}{2m},&&\epsilon_2=\epsilon_F-\frac{\vec{q}^2}{2m}-\frac{|\vec{q}|k_Fr}{m}-\frac{k_F^2r^2}{2m}.
\end{align}
After partial integration, we find
\be
I(T)\approx\begin{cases}\tilde I(T),&\epsilon_1\le0,\cr\tilde I(T)-\frac{F(\sqrt{{\epsilon_F}_1})}2,&\epsilon_1>0,\end{cases}
\ee
where
\be
\tilde I(T)=\int_0^\infty dyF\left(\sqrt{Ty}\right)N(y),
\ee
and
\be
N(y)=\frac1{(1+e^{\nu_1-y})^2(e^{y-\nu_2}+1)}+\frac{e^\nu}{(e^{y-\nu_1}+1)(1+e^{\nu_2-y})^2}-\frac1{(1+e^{\nu_1-y})(e^{y-\nu_2}+1)},
\ee
with $\nu_j=\frac{\epsilon_j}T$ and $\nu=\frac{\epsilon_2-\epsilon_1}T$. The function $N(y)$ in the integrand has a peak around $y=\nu_j$; hence,
\be
N(y)\approx\begin{cases}-\frac{e^{\nu_1-y}}{(1+e^{\nu_1-y})^2},&y\sim\nu_1,\cr
\frac{e^{\nu_2-y}}{(1+e^{\nu_2-y})^2},&y\sim\nu_2.\end{cases}
\ee
By expanding $F(\sqrt{Ty+\epsilon_j})$, the integral becomes
\be
\int_{-\Delta y}^{\Delta y}dyF(\sqrt{Ty+\epsilon_j})N(y+\nu_j)
\approx\int_0^{\Delta y}dy\left[2F(\sqrt{\epsilon_j})+\frac1{4\epsilon_j}\left(F''(\sqrt{\epsilon_j})-\frac{F'(\sqrt{\epsilon_j})}{\sqrt{\epsilon_j}}\right)y^2T^2\right]\frac{e^y}{(1+e^y)^2}.
\ee
We take $\Delta y$ to be large, but smaller than $1/T$, and find
\bea\label{ItildeTPM}
\tilde I(T)&=&F(\sqrt{\epsilon_2})-F(\sqrt{\epsilon_1}) \nn
&& +T^2\frac{\pi^2}{24}\left[\frac1{\epsilon_2}\left(F''(\sqrt{\epsilon_2})-\frac{F'(\sqrt{\epsilon_2})}{\sqrt{\epsilon_2}}\right)-\frac1{\epsilon_1}\left(F''(\sqrt{\epsilon_1})-\frac{F'(\sqrt{\epsilon_1})}{\sqrt{\epsilon_1}}\right)\right].
\eea
Now, the expressions for $M_n$ in Eqs. \eq{ilfuncdef} contain the functions $F_n(x)=2\pi(mx^2)^{\frac{n+1}2}$, with $n=1$ and $n=3$, when written in the form of Eq. \eq{ipm}, i.e. at finite temperature. By using Eq. \eqref{ItildeTPM} and the fact that only $n=1$ and $n=3$ are relevant, we recover
\bea
I_n(T)&=&F_n(\sqrt{\epsilon_2})-F_n(\sqrt{\epsilon_1})+T^2\frac{\pi^2}{24}\left[\frac1{\epsilon_2}\left(F_n''(\sqrt{\epsilon_2})-\frac{F_n'(\sqrt{\epsilon_2})}{\sqrt{\epsilon_2}}\right)-\frac1{\epsilon_1}\left(F''_n(\sqrt{\epsilon_1})-\frac{F'_n(\sqrt{\epsilon_1})}{\sqrt{\epsilon_1}}\right)\right]\nn
&=&I_n(0)+T^2\frac{\pi^3}3\frac{n+1}2\left(\frac{n+1}2-1\right)m^{\frac{n+1}2}\left(\epsilon_2^{\frac{n+1}2-2}-\epsilon_1^{\frac{n+1}2-2}\right) \nn
&=& I_n(0).
\eea
What this result implies for the low-temperature expansion of Eqs. \eq{tsndiag} is that the finite temperature dependence of the spectral functions ${\cal T}^{+-}$ and ${\cal S}^{+-}$ only appears at the $\ord{T^4}$ order. For the purposes of this calculation, we neglect the $\ord{T^4}$ contributions, which makes the temperature dependence of the $+-$ CTP blocks drop out altogether.

Finally, the retarded Green's function, $G_r=G_n+G_f=G_{++}-G_{+-}$, has longitudinal and transverse components, as in the parametrisation of Eq. \eq{parproj}. In the non-relativistic limit, the two sectors are defined by the expressions
\bea\label{retccgf}
G^r_\ell&=&\frac1c \, \Re_x \left[{\cal T}^{++}-{\cal T}^{+-}\right],\nn
G^r_t&=&\frac{c}2 \, \Re_x \left[({\cal T}^{++}-{\cal T}^{+-})\xi^2-{\cal S}^{++}+{\cal S}^{+-}\right].
\eea
The variable $\xi$ was defined above as $\xi=\omega/c|\vec{q}|$.

\newpage
\bibliography{HydroCTPRefsEGAS}

\begin{thebibliography}{43}%
\makeatletter
\providecommand \@ifxundefined [1]{%
 \@ifx{#1\undefined}
}%
\providecommand \@ifnum [1]{%
 \ifnum #1\expandafter \@firstoftwo
 \else \expandafter \@secondoftwo
 \fi
}%
\providecommand \@ifx [1]{%
 \ifx #1\expandafter \@firstoftwo
 \else \expandafter \@secondoftwo
 \fi
}%
\providecommand \natexlab [1]{#1}%
\providecommand \enquote  [1]{``#1''}%
\providecommand \bibnamefont  [1]{#1}%
\providecommand \bibfnamefont [1]{#1}%
\providecommand \citenamefont [1]{#1}%
\providecommand \href@noop [0]{\@secondoftwo}%
\providecommand \href [0]{\begingroup \@sanitize@url \@href}%
\providecommand \@href[1]{\@@startlink{#1}\@@href}%
\providecommand \@@href[1]{\endgroup#1\@@endlink}%
\providecommand \@sanitize@url [0]{\catcode `\\12\catcode `\$12\catcode
  `\&12\catcode `\#12\catcode `\^12\catcode `\_12\catcode `\%12\relax}%
\providecommand \@@startlink[1]{}%
\providecommand \@@endlink[0]{}%
\providecommand \url  [0]{\begingroup\@sanitize@url \@url }%
\providecommand \@url [1]{\endgroup\@href {#1}{\urlprefix }}%
\providecommand \urlprefix  [0]{URL }%
\providecommand \Eprint [0]{\href }%
\providecommand \doibase [0]{http://dx.doi.org/}%
\providecommand \selectlanguage [0]{\@gobble}%
\providecommand \bibinfo  [0]{\@secondoftwo}%
\providecommand \bibfield  [0]{\@secondoftwo}%
\providecommand \translation [1]{[#1]}%
\providecommand \BibitemOpen [0]{}%
\providecommand \bibitemStop [0]{}%
\providecommand \bibitemNoStop [0]{.\EOS\space}%
\providecommand \EOS [0]{\spacefactor3000\relax}%
\providecommand \BibitemShut  [1]{\csname bibitem#1\endcsname}%
\let\auto@bib@innerbib\@empty
%</preamble>
\bibitem [{\citenamefont {Landau}\ and\ \citenamefont
  {Lifshitz}(2013)}]{landau2013fluid}%
  \BibitemOpen
  \bibfield  {author} {\bibinfo {author} {\bibfnamefont {L.}~\bibnamefont
  {Landau}}\ and\ \bibinfo {author} {\bibfnamefont {E.}~\bibnamefont
  {Lifshitz}},\ }\href@noop {} {\emph {\bibinfo {title} {Fluid Mechanics:
  Landau and Lifshitz: Course of Theoretical Physics, Volume 6}}}\ (\bibinfo
  {publisher} {Elsevier Science},\ \bibinfo {year} {2013})\BibitemShut
  {NoStop}%
\bibitem [{\citenamefont {Forster}(1990)}]{forster}%
  \BibitemOpen
  \bibfield  {author} {\bibinfo {author} {\bibfnamefont {D.}~\bibnamefont
  {Forster}},\ }\href@noop {} {\emph {\bibinfo {title} {Hydrodynamic
  Fluctuations, Broken Symmetry, and Correlation Functions}}},\ Advanced book
  classics\ (\bibinfo  {publisher} {Perseus Books},\ \bibinfo {year}
  {1990})\BibitemShut {NoStop}%
\bibitem [{\citenamefont {Kovtun}(2012)}]{Kovtun:2012rj}%
  \BibitemOpen
  \bibfield  {author} {\bibinfo {author} {\bibfnamefont {P.}~\bibnamefont
  {Kovtun}},\ }\href {\doibase 10.1088/1751-8113/45/47/473001} {\bibfield
  {journal} {\bibinfo  {journal} {J.Phys.}\ }\textbf {\bibinfo {volume}
  {A45}},\ \bibinfo {pages} {473001} (\bibinfo {year} {2012})},\ \Eprint
  {http://arxiv.org/abs/1205.5040} {arXiv:1205.5040 [hep-th]} \BibitemShut
  {NoStop}%
%%CITATION = ARXIV:1205.5040;%%
\bibitem [{\citenamefont {Dubovsky}\ \emph {et~al.}(2006)\citenamefont
  {Dubovsky}, \citenamefont {Gregoire}, \citenamefont {Nicolis},\ and\
  \citenamefont {Rattazzi}}]{Dubovsky:2005xd}%
  \BibitemOpen
  \bibfield  {author} {\bibinfo {author} {\bibfnamefont {S.}~\bibnamefont
  {Dubovsky}}, \bibinfo {author} {\bibfnamefont {T.}~\bibnamefont {Gregoire}},
  \bibinfo {author} {\bibfnamefont {A.}~\bibnamefont {Nicolis}}, \ and\
  \bibinfo {author} {\bibfnamefont {R.}~\bibnamefont {Rattazzi}},\ }\href
  {\doibase 10.1088/1126-6708/2006/03/025} {\bibfield  {journal} {\bibinfo
  {journal} {JHEP}\ }\textbf {\bibinfo {volume} {0603}},\ \bibinfo {pages}
  {025} (\bibinfo {year} {2006})},\ \Eprint
  {http://arxiv.org/abs/hep-th/0512260} {arXiv:hep-th/0512260 [hep-th]}
  \BibitemShut {NoStop}%
%%CITATION = HEP-TH/0512260;%%
\bibitem [{\citenamefont {Dubovsky}\ \emph {et~al.}(2012)\citenamefont
  {Dubovsky}, \citenamefont {Hui}, \citenamefont {Nicolis},\ and\ \citenamefont
  {Son}}]{Dubovsky:2011sj}%
  \BibitemOpen
  \bibfield  {author} {\bibinfo {author} {\bibfnamefont {S.}~\bibnamefont
  {Dubovsky}}, \bibinfo {author} {\bibfnamefont {L.}~\bibnamefont {Hui}},
  \bibinfo {author} {\bibfnamefont {A.}~\bibnamefont {Nicolis}}, \ and\
  \bibinfo {author} {\bibfnamefont {D.~T.}\ \bibnamefont {Son}},\ }\href
  {\doibase 10.1103/PhysRevD.85.085029} {\bibfield  {journal} {\bibinfo
  {journal} {Phys.Rev.}\ }\textbf {\bibinfo {volume} {D85}},\ \bibinfo {pages}
  {085029} (\bibinfo {year} {2012})},\ \Eprint {http://arxiv.org/abs/1107.0731}
  {arXiv:1107.0731 [hep-th]} \BibitemShut {NoStop}%
%%CITATION = ARXIV:1107.0731;%%
\bibitem [{\citenamefont {Bhattacharya}\ \emph {et~al.}(2013)\citenamefont
  {Bhattacharya}, \citenamefont {Bhattacharyya},\ and\ \citenamefont
  {Rangamani}}]{Bhattacharya:2012zx}%
  \BibitemOpen
  \bibfield  {author} {\bibinfo {author} {\bibfnamefont {J.}~\bibnamefont
  {Bhattacharya}}, \bibinfo {author} {\bibfnamefont {S.}~\bibnamefont
  {Bhattacharyya}}, \ and\ \bibinfo {author} {\bibfnamefont {M.}~\bibnamefont
  {Rangamani}},\ }\href {\doibase 10.1007/JHEP02(2013)153} {\bibfield
  {journal} {\bibinfo  {journal} {JHEP}\ }\textbf {\bibinfo {volume} {1302}},\
  \bibinfo {pages} {153} (\bibinfo {year} {2013})},\ \Eprint
  {http://arxiv.org/abs/1211.1020} {arXiv:1211.1020 [hep-th]} \BibitemShut
  {NoStop}%
%%CITATION = ARXIV:1211.1020;%%
\bibitem [{\citenamefont {Haehl}\ \emph {et~al.}(2014)\citenamefont {Haehl},
  \citenamefont {Loganayagam},\ and\ \citenamefont
  {Rangamani}}]{Haehl:2013hoa}%
  \BibitemOpen
  \bibfield  {author} {\bibinfo {author} {\bibfnamefont {F.~M.}\ \bibnamefont
  {Haehl}}, \bibinfo {author} {\bibfnamefont {R.}~\bibnamefont {Loganayagam}},
  \ and\ \bibinfo {author} {\bibfnamefont {M.}~\bibnamefont {Rangamani}},\
  }\href {\doibase 10.1007/JHEP03(2014)034} {\bibfield  {journal} {\bibinfo
  {journal} {JHEP}\ }\textbf {\bibinfo {volume} {1403}},\ \bibinfo {pages}
  {034} (\bibinfo {year} {2014})},\ \Eprint {http://arxiv.org/abs/1312.0610}
  {arXiv:1312.0610 [hep-th]} \BibitemShut {NoStop}%
%%CITATION = ARXIV:1312.0610;%%
\bibitem [{\citenamefont {Haehl}\ and\ \citenamefont
  {Rangamani}(2013)}]{Haehl:2013kra}%
  \BibitemOpen
  \bibfield  {author} {\bibinfo {author} {\bibfnamefont {F.~M.}\ \bibnamefont
  {Haehl}}\ and\ \bibinfo {author} {\bibfnamefont {M.}~\bibnamefont
  {Rangamani}},\ }\href {\doibase 10.1007/JHEP10(2013)074} {\bibfield
  {journal} {\bibinfo  {journal} {JHEP}\ }\textbf {\bibinfo {volume} {1310}},\
  \bibinfo {pages} {074} (\bibinfo {year} {2013})},\ \Eprint
  {http://arxiv.org/abs/1305.6968} {arXiv:1305.6968 [hep-th]} \BibitemShut
  {NoStop}%
%%CITATION = ARXIV:1305.6968;%%
\bibitem [{\citenamefont {Grozdanov}\ and\ \citenamefont
  {Polonyi}(2013)}]{Grozdanov:2013dba}%
  \BibitemOpen
  \bibfield  {author} {\bibinfo {author} {\bibfnamefont {S.}~\bibnamefont
  {Grozdanov}}\ and\ \bibinfo {author} {\bibfnamefont {J.}~\bibnamefont
  {Polonyi}},\ }\href@noop {} {\  (\bibinfo {year} {2013})},\ \Eprint
  {http://arxiv.org/abs/1305.3670} {arXiv:1305.3670 [hep-th]} \BibitemShut
  {NoStop}%
%%CITATION = ARXIV:1305.3670;%%
\bibitem [{\citenamefont {Endlich}\ \emph {et~al.}(2013)\citenamefont
  {Endlich}, \citenamefont {Nicolis}, \citenamefont {Porto},\ and\
  \citenamefont {Wang}}]{Endlich:2012vt}%
  \BibitemOpen
  \bibfield  {author} {\bibinfo {author} {\bibfnamefont {S.}~\bibnamefont
  {Endlich}}, \bibinfo {author} {\bibfnamefont {A.}~\bibnamefont {Nicolis}},
  \bibinfo {author} {\bibfnamefont {R.~A.}\ \bibnamefont {Porto}}, \ and\
  \bibinfo {author} {\bibfnamefont {J.}~\bibnamefont {Wang}},\ }\href {\doibase
  10.1103/PhysRevD.88.105001} {\bibfield  {journal} {\bibinfo  {journal}
  {Phys.Rev.}\ }\textbf {\bibinfo {volume} {D88}},\ \bibinfo {pages} {105001}
  (\bibinfo {year} {2013})},\ \Eprint {http://arxiv.org/abs/1211.6461}
  {arXiv:1211.6461 [hep-th]} \BibitemShut {NoStop}%
%%CITATION = ARXIV:1211.6461;%%
\bibitem [{\citenamefont {Kovtun}\ \emph {et~al.}(2014)\citenamefont {Kovtun},
  \citenamefont {Moore},\ and\ \citenamefont {Romatschke}}]{Kovtun:2014hpa}%
  \BibitemOpen
  \bibfield  {author} {\bibinfo {author} {\bibfnamefont {P.}~\bibnamefont
  {Kovtun}}, \bibinfo {author} {\bibfnamefont {G.~D.}\ \bibnamefont {Moore}}, \
  and\ \bibinfo {author} {\bibfnamefont {P.}~\bibnamefont {Romatschke}},\
  }\href {\doibase 10.1007/JHEP07(2014)123} {\bibfield  {journal} {\bibinfo
  {journal} {JHEP}\ }\textbf {\bibinfo {volume} {1407}},\ \bibinfo {pages}
  {123} (\bibinfo {year} {2014})},\ \Eprint {http://arxiv.org/abs/1405.3967}
  {arXiv:1405.3967 [hep-ph]} \BibitemShut {NoStop}%
%%CITATION = ARXIV:1405.3967;%%
\bibitem [{\citenamefont {Galley}\ \emph {et~al.}(2014)\citenamefont {Galley},
  \citenamefont {Tsang},\ and\ \citenamefont {Stein}}]{Galley:2014wla}%
  \BibitemOpen
  \bibfield  {author} {\bibinfo {author} {\bibfnamefont {C.~R.}\ \bibnamefont
  {Galley}}, \bibinfo {author} {\bibfnamefont {D.}~\bibnamefont {Tsang}}, \
  and\ \bibinfo {author} {\bibfnamefont {L.~C.}\ \bibnamefont {Stein}},\
  }\href@noop {} {\  (\bibinfo {year} {2014})},\ \Eprint
  {http://arxiv.org/abs/1412.3082} {arXiv:1412.3082 [math-ph]} \BibitemShut
  {NoStop}%
%%CITATION = ARXIV:1412.3082;%%
\bibitem [{\citenamefont {Schwinger}(1961)}]{schw}%
  \BibitemOpen
  \bibfield  {author} {\bibinfo {author} {\bibfnamefont {J.~S.}\ \bibnamefont
  {Schwinger}},\ }\href {\doibase 10.1063/1.1703727} {\bibfield  {journal}
  {\bibinfo  {journal} {J.Math.Phys.}\ }\textbf {\bibinfo {volume} {2}},\
  \bibinfo {pages} {407} (\bibinfo {year} {1961})}\BibitemShut {NoStop}%
%%CITATION = JMAPA,2,407;%%
\bibitem [{\citenamefont {Schwinger}(1998)}]{schwbooks}%
  \BibitemOpen
  \bibfield  {author} {\bibinfo {author} {\bibfnamefont {J.}~\bibnamefont
  {Schwinger}},\ }\href@noop {} {\emph {\bibinfo {title} {Particles, Sources,
  And Fields, vol. I., II., and III.}}},\ Advanced Book Classics\ (\bibinfo
  {publisher} {Advanced Book Program, Perseus Books},\ \bibinfo {year}
  {1998})\BibitemShut {NoStop}%
\bibitem [{\citenamefont {Bakshi}\ and\ \citenamefont
  {Mahanthappa}(1963)}]{Bakshi:1962dv}%
  \BibitemOpen
  \bibfield  {author} {\bibinfo {author} {\bibfnamefont {P.~M.}\ \bibnamefont
  {Bakshi}}\ and\ \bibinfo {author} {\bibfnamefont {K.~T.}\ \bibnamefont
  {Mahanthappa}},\ }\href {\doibase 10.1063/1.1703883} {\bibfield  {journal}
  {\bibinfo  {journal} {J.Math.Phys.}\ }\textbf {\bibinfo {volume} {4}},\
  \bibinfo {pages} {1} (\bibinfo {year} {1963})}\BibitemShut {NoStop}%
%%CITATION = JMAPA,4,1;%%
\bibitem [{\citenamefont {Mahanthappa}(1962)}]{Mahanthappa:1962ex}%
  \BibitemOpen
  \bibfield  {author} {\bibinfo {author} {\bibfnamefont {K.~T.}\ \bibnamefont
  {Mahanthappa}},\ }\href {\doibase 10.1103/PhysRev.126.329} {\bibfield
  {journal} {\bibinfo  {journal} {Phys.Rev.}\ }\textbf {\bibinfo {volume}
  {126}},\ \bibinfo {pages} {329} (\bibinfo {year} {1962})}\BibitemShut
  {NoStop}%
%%CITATION = PHRVA,126,329;%%
\bibitem [{\citenamefont {Keldysh}(1964)}]{keldysh}%
  \BibitemOpen
  \bibfield  {author} {\bibinfo {author} {\bibfnamefont {L.}~\bibnamefont
  {Keldysh}},\ }\href@noop {} {\bibfield  {journal} {\bibinfo  {journal}
  {Zh.Eksp.Teor.Fiz.}\ }\textbf {\bibinfo {volume} {47}},\ \bibinfo {pages}
  {1515} (\bibinfo {year} {1964})}\BibitemShut {NoStop}%
%%CITATION = ZETFA,47,1515;%%
\bibitem [{\citenamefont {Pomeau}\ and\ \citenamefont
  {Resibois}(1975)}]{Pomeau:1974hg}%
  \BibitemOpen
  \bibfield  {author} {\bibinfo {author} {\bibfnamefont {Y.}~\bibnamefont
  {Pomeau}}\ and\ \bibinfo {author} {\bibfnamefont {P.}~\bibnamefont
  {Resibois}},\ }\href@noop {} {\bibfield  {journal} {\bibinfo  {journal}
  {Physics Reports}\ }\textbf {\bibinfo {volume} {19}},\ \bibinfo {pages} {63 }
  (\bibinfo {year} {1975})}\BibitemShut {NoStop}%
%%CITATION = SACLAY-DPH-T-74-87 ETC.;%%
\bibitem [{\citenamefont {Kovtun}\ and\ \citenamefont
  {Yaffe}(2003)}]{Kovtun:2003vj}%
  \BibitemOpen
  \bibfield  {author} {\bibinfo {author} {\bibfnamefont {P.}~\bibnamefont
  {Kovtun}}\ and\ \bibinfo {author} {\bibfnamefont {L.~G.}\ \bibnamefont
  {Yaffe}},\ }\href {\doibase 10.1103/PhysRevD.68.025007} {\bibfield  {journal}
  {\bibinfo  {journal} {Phys.Rev.}\ }\textbf {\bibinfo {volume} {D68}},\
  \bibinfo {pages} {025007} (\bibinfo {year} {2003})},\ \Eprint
  {http://arxiv.org/abs/hep-th/0303010} {arXiv:hep-th/0303010 [hep-th]}
  \BibitemShut {NoStop}%
%%CITATION = HEP-TH/0303010;%%
\bibitem [{\citenamefont {Caron-Huot}\ and\ \citenamefont
  {Saremi}(2010)}]{CaronHuot:2009iq}%
  \BibitemOpen
  \bibfield  {author} {\bibinfo {author} {\bibfnamefont {S.}~\bibnamefont
  {Caron-Huot}}\ and\ \bibinfo {author} {\bibfnamefont {O.}~\bibnamefont
  {Saremi}},\ }\href {\doibase 10.1007/JHEP11(2010)013} {\bibfield  {journal}
  {\bibinfo  {journal} {JHEP}\ }\textbf {\bibinfo {volume} {1011}},\ \bibinfo
  {pages} {013} (\bibinfo {year} {2010})},\ \Eprint
  {http://arxiv.org/abs/0909.4525} {arXiv:0909.4525 [hep-th]} \BibitemShut
  {NoStop}%
%%CITATION = ARXIV:0909.4525;%%
\bibitem [{\citenamefont {Toda}\ \emph {et~al.}(1991)\citenamefont {Toda},
  \citenamefont {Kubo}, \citenamefont {Sait{\=o}},\ and\ \citenamefont
  {Hashitsume}}]{kubo}%
  \BibitemOpen
  \bibfield  {author} {\bibinfo {author} {\bibfnamefont {M.}~\bibnamefont
  {Toda}}, \bibinfo {author} {\bibfnamefont {R.}~\bibnamefont {Kubo}}, \bibinfo
  {author} {\bibfnamefont {N.}~\bibnamefont {Sait{\=o}}}, \ and\ \bibinfo
  {author} {\bibfnamefont {N.}~\bibnamefont {Hashitsume}},\ }\href@noop {}
  {\emph {\bibinfo {title} {Statistical Physics II: Nonequilibrium Statistical
  Mechanics}}},\ Series C, English Authors\ (\bibinfo  {publisher} {Springer
  Berlin Heidelberg},\ \bibinfo {year} {1991})\BibitemShut {NoStop}%
\bibitem [{\citenamefont {Moore}\ and\ \citenamefont
  {Sohrabi}(2011)}]{Moore:2010bu}%
  \BibitemOpen
  \bibfield  {author} {\bibinfo {author} {\bibfnamefont {G.~D.}\ \bibnamefont
  {Moore}}\ and\ \bibinfo {author} {\bibfnamefont {K.~A.}\ \bibnamefont
  {Sohrabi}},\ }\href {\doibase 10.1103/PhysRevLett.106.122302} {\bibfield
  {journal} {\bibinfo  {journal} {Phys.Rev.Lett.}\ }\textbf {\bibinfo {volume}
  {106}},\ \bibinfo {pages} {122302} (\bibinfo {year} {2011})},\ \Eprint
  {http://arxiv.org/abs/1007.5333} {arXiv:1007.5333 [hep-ph]} \BibitemShut
  {NoStop}%
%%CITATION = ARXIV:1007.5333;%%
\bibitem [{\citenamefont {Landau}\ \emph {et~al.}(1980)\citenamefont {Landau},
  \citenamefont {Lifshitz},\ and\ \citenamefont
  {Pitaevskii}}]{landau1980statistical}%
  \BibitemOpen
  \bibfield  {author} {\bibinfo {author} {\bibfnamefont {L.}~\bibnamefont
  {Landau}}, \bibinfo {author} {\bibfnamefont {E.}~\bibnamefont {Lifshitz}}, \
  and\ \bibinfo {author} {\bibfnamefont {L.}~\bibnamefont {Pitaevskii}},\
  }\href {https://books.google.co.uk/books?id=NaB7oAkon9MC} {\emph {\bibinfo
  {title} {Statistical Physics}}},\ \bibinfo {series} {Course of theoretical
  physics}\ No.\ \bibinfo {number} {pt. 2}\ (\bibinfo  {publisher} {Pergamon
  Press},\ \bibinfo {year} {1980})\BibitemShut {NoStop}%
\bibitem [{\citenamefont {Baym}\ and\ \citenamefont
  {Pethick}(2008)}]{baym2008landau}%
  \BibitemOpen
  \bibfield  {author} {\bibinfo {author} {\bibfnamefont {G.}~\bibnamefont
  {Baym}}\ and\ \bibinfo {author} {\bibfnamefont {C.}~\bibnamefont {Pethick}},\
  }\href {http://books.google.co.uk/books?id=xmiV4YSEjE4C} {\emph {\bibinfo
  {title} {Landau Fermi-Liquid Theory: Concepts and Applications}}}\ (\bibinfo
  {publisher} {Wiley},\ \bibinfo {year} {2008})\BibitemShut {NoStop}%
\bibitem [{\citenamefont {Polchinski}(1992)}]{Polchinski:1992ed}%
  \BibitemOpen
  \bibfield  {author} {\bibinfo {author} {\bibfnamefont {J.}~\bibnamefont
  {Polchinski}},\ }\href@noop {} {\  (\bibinfo {year} {1992})},\ \Eprint
  {http://arxiv.org/abs/hep-th/9210046} {arXiv:hep-th/9210046 [hep-th]}
  \BibitemShut {NoStop}%
%%CITATION = HEP-TH/9210046;%%
\bibitem [{\citenamefont {Polonyi}(2010)}]{idgas}%
  \BibitemOpen
  \bibfield  {author} {\bibinfo {author} {\bibfnamefont {J.}~\bibnamefont
  {Polonyi}},\ }\href {http://stacks.iop.org/0295-5075/91/i=6/a=67003}
  {\bibfield  {journal} {\bibinfo  {journal} {EPL (Europhysics Letters)}\
  }\textbf {\bibinfo {volume} {91}},\ \bibinfo {pages} {67003} (\bibinfo {year}
  {2010})}\BibitemShut {NoStop}%
\bibitem [{\citenamefont {Clemmow}(1995)}]{clemmow}%
  \BibitemOpen
  \bibfield  {author} {\bibinfo {author} {\bibfnamefont {P.~C.}\ \bibnamefont
  {Clemmow}},\ }\href@noop {} {\emph {\bibinfo {title} {Electrodynamics of
  Particles and Plasmas}}},\ Advanced Books Classics Series\ (\bibinfo
  {publisher} {Perseus Books Group},\ \bibinfo {year} {1995})\BibitemShut
  {NoStop}%
\bibitem [{\citenamefont {Kamenev}(2011)}]{kamenev}%
  \BibitemOpen
  \bibfield  {author} {\bibinfo {author} {\bibfnamefont {A.}~\bibnamefont
  {Kamenev}},\ }\href@noop {} {\emph {\bibinfo {title} {Field Theory of
  Non-Equilibrium Systems}}},\ Field Theory of Non-equilibrium Systems\
  (\bibinfo  {publisher} {Cambridge University Press},\ \bibinfo {year}
  {2011})\BibitemShut {NoStop}%
\bibitem [{\citenamefont {Calzetta}\ and\ \citenamefont {Hu}(2008)}]{calzetta}%
  \BibitemOpen
  \bibfield  {author} {\bibinfo {author} {\bibfnamefont {E.}~\bibnamefont
  {Calzetta}}\ and\ \bibinfo {author} {\bibfnamefont {B.}~\bibnamefont {Hu}},\
  }\href@noop {} {\emph {\bibinfo {title} {Nonequilibrium Quantum Field
  Theory}}},\ Cambridge Monographs on Mathematical Physics\ (\bibinfo
  {publisher} {Cambridge University Press},\ \bibinfo {year}
  {2008})\BibitemShut {NoStop}%
\bibitem [{\citenamefont {Bloch}\ and\ \citenamefont {Nordsieck}(1937)}]{bn}%
  \BibitemOpen
  \bibfield  {author} {\bibinfo {author} {\bibfnamefont {F.}~\bibnamefont
  {Bloch}}\ and\ \bibinfo {author} {\bibfnamefont {A.}~\bibnamefont
  {Nordsieck}},\ }\href {\doibase 10.1103/PhysRev.52.54} {\bibfield  {journal}
  {\bibinfo  {journal} {Phys.Rev.}\ }\textbf {\bibinfo {volume} {52}},\
  \bibinfo {pages} {54} (\bibinfo {year} {1937})}\BibitemShut {NoStop}%
%%CITATION = PHRVA,52,54;%%
\bibitem [{\citenamefont {Kinoshita}(1962)}]{kin}%
  \BibitemOpen
  \bibfield  {author} {\bibinfo {author} {\bibfnamefont {T.}~\bibnamefont
  {Kinoshita}},\ }\href {\doibase 10.1063/1.1724268} {\bibfield  {journal}
  {\bibinfo  {journal} {J.Math.Phys.}\ }\textbf {\bibinfo {volume} {3}},\
  \bibinfo {pages} {650} (\bibinfo {year} {1962})}\BibitemShut {NoStop}%
%%CITATION = JMAPA,3,650;%%
\bibitem [{\citenamefont {Lee}\ and\ \citenamefont {Nauenberg}(1964)}]{ln}%
  \BibitemOpen
  \bibfield  {author} {\bibinfo {author} {\bibfnamefont {T.}~\bibnamefont
  {Lee}}\ and\ \bibinfo {author} {\bibfnamefont {M.}~\bibnamefont
  {Nauenberg}},\ }\href {\doibase 10.1103/PhysRev.133.B1549} {\bibfield
  {journal} {\bibinfo  {journal} {Phys.Rev.}\ }\textbf {\bibinfo {volume}
  {133}},\ \bibinfo {pages} {B1549} (\bibinfo {year} {1964})}\BibitemShut
  {NoStop}%
%%CITATION = PHRVA,133,B1549;%%
\bibitem [{\citenamefont {Yennie}\ \emph {et~al.}(1961)\citenamefont {Yennie},
  \citenamefont {Frautschi},\ and\ \citenamefont {Suura}}]{yennie}%
  \BibitemOpen
  \bibfield  {author} {\bibinfo {author} {\bibfnamefont {D.}~\bibnamefont
  {Yennie}}, \bibinfo {author} {\bibfnamefont {S.~C.}\ \bibnamefont
  {Frautschi}}, \ and\ \bibinfo {author} {\bibfnamefont {H.}~\bibnamefont
  {Suura}},\ }\href {\doibase 10.1016/0003-4916(61)90151-8} {\bibfield
  {journal} {\bibinfo  {journal} {Annals Phys.}\ }\textbf {\bibinfo {volume}
  {13}},\ \bibinfo {pages} {379} (\bibinfo {year} {1961})}\BibitemShut
  {NoStop}%
%%CITATION = APNYA,13,379;%%
\bibitem [{\citenamefont {Polonyi}(2012)}]{arrow}%
  \BibitemOpen
  \bibfield  {author} {\bibinfo {author} {\bibfnamefont {J.}~\bibnamefont
  {Polonyi}},\ }\href@noop {} {\  (\bibinfo {year} {2012})},\ \Eprint
  {http://arxiv.org/abs/1206.5781} {arXiv:1206.5781 [hep-th]} \BibitemShut
  {NoStop}%
%%CITATION = ARXIV:1206.5781;%%
\bibitem [{\citenamefont {Feynman}\ and\ \citenamefont {{Vernon,
  Jr.}}(1963)}]{Feynman:1963:TGQ}%
  \BibitemOpen
  \bibfield  {author} {\bibinfo {author} {\bibfnamefont {R.~P.}\ \bibnamefont
  {Feynman}}\ and\ \bibinfo {author} {\bibfnamefont {F.~L.}\ \bibnamefont
  {{Vernon, Jr.}}},\ }\href {\doibase
  http://dx.doi.org/10.1016/0003-4916(63)90068-X} {\bibfield  {journal}
  {\bibinfo  {journal} {Annals of Physics}\ }\textbf {\bibinfo {volume} {24}},\
  \bibinfo {pages} {118} (\bibinfo {year} {1963})}\BibitemShut {NoStop}%
\bibitem [{\citenamefont {Polonyi}(2014{\natexlab{a}})}]{PhysRevD.90.065010}%
  \BibitemOpen
  \bibfield  {author} {\bibinfo {author} {\bibfnamefont {J.}~\bibnamefont
  {Polonyi}},\ }\href {\doibase 10.1103/PhysRevD.90.065010} {\bibfield
  {journal} {\bibinfo  {journal} {Phys. Rev. D}\ }\textbf {\bibinfo {volume}
  {90}},\ \bibinfo {pages} {065010} (\bibinfo {year}
  {2014}{\natexlab{a}})}\BibitemShut {NoStop}%
\bibitem [{\citenamefont {Polonyi}(2014{\natexlab{b}})}]{effthpch}%
  \BibitemOpen
  \bibfield  {author} {\bibinfo {author} {\bibfnamefont {J.}~\bibnamefont
  {Polonyi}},\ }\href {\doibase 10.1016/j.aop.2014.01.008} {\bibfield
  {journal} {\bibinfo  {journal} {Annals Phys.}\ }\textbf {\bibinfo {volume}
  {342}},\ \bibinfo {pages} {239} (\bibinfo {year} {2014}{\natexlab{b}})},\
  \Eprint {http://arxiv.org/abs/1302.3864} {arXiv:1302.3864 [hep-th]}
  \BibitemShut {NoStop}%
%%CITATION = ARXIV:1302.3864;%%
\bibitem [{\citenamefont {Polonyi}(2011)}]{irrev}%
  \BibitemOpen
  \bibfield  {author} {\bibinfo {author} {\bibfnamefont {J.}~\bibnamefont
  {Polonyi}},\ }\href {\doibase 10.1103/PhysRevD.84.105021} {\bibfield
  {journal} {\bibinfo  {journal} {Phys.Rev.}\ }\textbf {\bibinfo {volume}
  {D84}},\ \bibinfo {pages} {105021} (\bibinfo {year} {2011})},\ \Eprint
  {http://arxiv.org/abs/1109.2228} {arXiv:1109.2228 [hep-th]} \BibitemShut
  {NoStop}%
%%CITATION = ARXIV:1109.2228;%%
\bibitem [{\citenamefont {Zeh}(1970)}]{zeh}%
  \BibitemOpen
  \bibfield  {author} {\bibinfo {author} {\bibfnamefont {H.}~\bibnamefont
  {Zeh}},\ }\href {\doibase 10.1007/BF00708656} {\bibfield  {journal} {\bibinfo
   {journal} {Found.Phys.}\ }\textbf {\bibinfo {volume} {1}},\ \bibinfo {pages}
  {69} (\bibinfo {year} {1970})}\BibitemShut {NoStop}%
%%CITATION = FNDPA,1,69;%%
\bibitem [{\citenamefont {Zurek}(1991)}]{zurek}%
  \BibitemOpen
  \bibfield  {author} {\bibinfo {author} {\bibfnamefont {W.~H.}\ \bibnamefont
  {Zurek}},\ }\href {\doibase 10.1063/1.881293} {\bibfield  {journal} {\bibinfo
   {journal} {Phys.Today}\ }\textbf {\bibinfo {volume} {44N10}},\ \bibinfo
  {pages} {36} (\bibinfo {year} {1991})}\BibitemShut {NoStop}%
%%CITATION = PHTOA,44N10,36;%%
\bibitem [{\citenamefont {Griffiths}(2003)}]{grif}%
  \BibitemOpen
  \bibfield  {author} {\bibinfo {author} {\bibfnamefont {R.}~\bibnamefont
  {Griffiths}},\ }\href@noop {} {\emph {\bibinfo {title} {Consistent Quantum
  Theory}}}\ (\bibinfo  {publisher} {Cambridge University Press},\ \bibinfo
  {year} {2003})\BibitemShut {NoStop}%
\bibitem [{\citenamefont {Polonyi}(2006)}]{ed}%
  \BibitemOpen
  \bibfield  {author} {\bibinfo {author} {\bibfnamefont {J.}~\bibnamefont
  {Polonyi}},\ }\href {\doibase 10.1103/PhysRevD.74.065014} {\bibfield
  {journal} {\bibinfo  {journal} {Phys.Rev.}\ }\textbf {\bibinfo {volume}
  {D74}},\ \bibinfo {pages} {065014} (\bibinfo {year} {2006})},\ \Eprint
  {http://arxiv.org/abs/hep-th/0605218} {arXiv:hep-th/0605218 [hep-th]}
  \BibitemShut {NoStop}%
%%CITATION = HEP-TH/0605218;%%
\bibitem [{\citenamefont {Maldacena}\ and\ \citenamefont
  {Zhiboedov}(2013)}]{Maldacena:2011jn}%
  \BibitemOpen
  \bibfield  {author} {\bibinfo {author} {\bibfnamefont {J.}~\bibnamefont
  {Maldacena}}\ and\ \bibinfo {author} {\bibfnamefont {A.}~\bibnamefont
  {Zhiboedov}},\ }\href {\doibase 10.1088/1751-8113/46/21/214011} {\bibfield
  {journal} {\bibinfo  {journal} {J.Phys.}\ }\textbf {\bibinfo {volume}
  {A46}},\ \bibinfo {pages} {214011} (\bibinfo {year} {2013})},\ \Eprint
  {http://arxiv.org/abs/1112.1016} {arXiv:1112.1016 [hep-th]} \BibitemShut
  {NoStop}%
%%CITATION = ARXIV:1112.1016;%%
\end{thebibliography}%

\end{document}